\documentclass[11pt]{article}

\usepackage{epstopdf,float}
\usepackage{multirow}
\usepackage{graphics}
\usepackage{amsmath,amsthm,amsfonts}
\usepackage{subfigure}
\usepackage{color, setspace, multirow}
\usepackage[T1]{fontenc}
\usepackage[utf8]{inputenc}
\usepackage{authblk} 
\usepackage{natbib} 
\usepackage{graphicx}
\usepackage{array}
\usepackage{mwe,hyperref}
\usepackage{graphbox, lineno}
\usepackage[export]{adjustbox}
\usepackage[normalem]{ulem}
 \usepackage{enumitem,rotating}
\usepackage{subcaption,caption}

\newcommand{\bfD}{{\bf D}}

\newcommand{\bfX}{{\bf X}}

\newcommand{\bftheta}{\mbox{\boldmath $\theta$}}

\newcommand{\bfphi}{\mbox{\boldmath $\phi$}}

\newcommand{\bfbeta}{\mbox{\boldmath $\beta$}}

\newcommand{\bfgamma}{\mbox{\boldmath $\gamma$}}

\newcommand{\bfZ}{{\bf Z}}
\newcommand{\bfY}{{\bf Y}}

\newcommand{\bfA}{{\bf A}}

\newcommand\redsout{\bgroup\markoverwith{\textcolor{red}{\rule[0.5ex]{2pt}{0.4pt}}}\ULon}

\newcommand{\blind}{1}

\def\spacingset#1{\renewcommand{\baselinestretch}%
	{#1}\small\normalsize} \spacingset{1}

\date{}

\begin{document}

\if1\blind
{
	\title{Three Case Studies of Two-stage MCMC for Fast Bayesian Inference of Large Spatio-temporal Data}
	\author{Mostafa Shams$^{\dagger}$, Robert Erhardt, Staci Hepler\\
		Department of Statistical Sciences, Wake Forest University, U.S.
	}
	\maketitle
} \fi

\if0\blind
{
	\title{\bf A }
	\maketitle
} \fi

\bigskip
\begin{abstract}
High dimensional spatio-temporal models can quickly run up against computational limitations.  Large numbers of observations in either space or time, large numbers of model parameters, inversion of dense matrices, poor mixing within Markov chain Monte Carlo (MCMC), low acceptance rates for necessary Metropolis-Hastings steps, and an inability to use parallel computing due to a lack of independence in the model can all combine to produce unreasonable computational costs.  Recursive or multi-stage Bayesian algorithms are one way to address this computational challenge. Here we consider an increasingly used two-stage algorithm.  The first stage models locations independently and in parallel across space. Resampling particles from this stage-one model with Metropolis-within-Gibbs methods and suitably computed acceptance ratios can impose spatial dependence across locations at a low computational cost, while targeting the same posterior distribution as the single-stage MCMC algorithm at a fraction of the overall computational cost. In this paper we show three complete examples of how two-stage MCMC algorithms can be used to efficiently fit Bayesian spatio-temporal models to large datasets.  Specifically, we consider: a spatio-temporal self-exciting count model on areal data utilizing intrinsic conditional autoregressive (ICAR) prior distributions; a spatio-temporal model of point-referenced data using a stationary, isotropic distance-based covariance function; and a binary model fit to data on a spatial lattice utilizing a latent Gaussian process to capture all spatio-temporal dependence. In each example, we define the model and describe the corresponding two-stage MCMC approach. We then compare the resulting posterior samples and computational efficiency with those obtained using a standard single-stage MCMC algorithm. While varying across the three examples and numerous parameters, the two-stage approach often achieves roughly an order of magnitude higher computational efficiency, with the two-stage and single-stage algorithms producing closely agreeing posterior samples. 
\end{abstract}

\noindent%
{\it Keywords: Two-stage MCMC; spatial and temporal dependence; Bayesian hierarchical modeling; areal and point-referenced data. \\}  
\vfill

$\dagger$ Corresponding author.  Address 1834 Wake Forest Road, Winston-Salem NC, 27109, US.  Email \texttt{shamsm@wfu.edu}

\newpage
\spacingset{1.45} 

\section{Introduction}

Spatio-temporal models are widely used in settings where data exhibit both spatial dependence across locations as well as temporal dependence across time periods. Application areas include ecology, astronomy, climate and meteorology, spatial epidemiology, insurance, and many others. Spatial support can be point-referenced (geostatistical), such as weather monitoring stations with specific latitude and longitude coordinates, or areal, such as counts of some random phenomena within a prescribed area such as a US county. Given the potential for large numbers of spatial locations, a large number of repeated measurements over time, and potentially many variables recorded at each location and time period, spatio-temporal datasets are known to grow very large in many applications. 

A common tool for analyzing spatio-temporal data is the Gaussian process, a well-defined stochastic process by which every finite subset of variables has a multivariate Gaussian distribution. Despite its elegance, \cite{heaton2019case} illustrated how the Gaussian process quickly becomes infeasible for large spatio-temporal datasets. This is primarily the result of the dense, complex correlation structure required to capture spatio-temporal dependencies.  They provide an overview of numerous commonly used approaches to reduce the computational burden to fitting the Gaussian process, which includes: 

\begin{itemize}
\item Reduced rank or basis function approaches that approximate large spatial processes by projecting onto a lower dimensional latent field, including fixed rank kriging \citep{cressie2008fixedrank} and predictive processes \citep{banerjee2008gaussian}; 

\item Methods that assume sparsity in either the covariance or the precision, such as covariance tapering methods \citep{furrer2006covariance}, Vecchia approximations \citep{katzfuss2021general}, and nearest neighbor Gaussian process methods \citep{datta2016hierarchical}; and 

\item Methods which permit parallel computing such as \cite{katzfuss2017parallel} which shows how parallel computing can be used with low-rank models to further improve computation, and spatial partitioning approaches \citep{peruzzi2022highly}.
\end{itemize}

Each of the above-mentioned methods for reducing the computational burden fundamentally involves approximating the likelihood. However, some recursive or multi-stage approaches have been developed that can permit computationally efficient Markov chain Monte Carlo (MCMC) for some spatio-temporal models, without having to approximate the likelihood. \cite{lunn2013fully} developed a two-stage MCMC approach originally for meta analysis. They first used a computationally efficient model, called the stage one model, to approximate the posterior, and then used samples from the stage one model as proposed values in a Metropolis-Hastings algorithm for the full model. \cite{hooten2021making} tied this idea to recursive Bayesian inference. Papers that have studied and applied these two-stage methods include: \cite{wei2019parallel}, which performed simulation studies and implemented the method for a Bayesian hierarchical model; \cite{hepler2025two}, which used the two-stage approach to model spatio-temporal ordinal data; \cite{hooten2022bayesian}, which used the approach on capture-recapture data from ecology; \cite{taylor2025generative}, which used these methods for when data arrive continuously with an application to ecological time series count data; and \cite{leach2022recursive}, which used these recursive methods to help with optimal design in ecological studies. Each of these shows the success in applying the two-stage approach to sample from the posterior distribution without resorting to approximating the likelihood, all at a fraction of the computational cost a single-stage MCMC model fit would require. While previous studies have successfully demonstrated the two-stage approach in specific application settings, in this paper we illustrate its broad utility and flexibility across a variety of spatio-temporal models and data types through three distinct case studies.

We first describe the two-stage methodology in the context of spatio-temporal data in Section~\ref{sec:section2-general}. Section~\ref{sec:section3-data} describes our data, and Section~\ref{sec:section4-application} shows the method's specific use in three distinct case studies: a count response on areal spatial support for a disease application; a continuous response for point-referenced spatial data with a weather application; and a binary response on areal support from a drought application. In each study, we demonstrate the agreement between the computationally expensive single-stage MCMC model fit and the computationally efficient two-stage MCMC model fit, and compare their effective sample sizes per hour of compute time. Our results show that the two-stage approach can be used for many spatio-temporal applications, at a fraction of the computational cost of a single-stage algorithm. We conclude with a discussion in Section~\ref{sec:discussion}.

\section{Methodology} \label{sec:section2-general}
The two-stage methodology described in this section follows the approach developed by \cite{lunn2013fully}, which we restate here within Bayesian hierarchical modeling framework for the spatio-temporal settings considered in this paper. We first describe a spatio-temporal model whose posterior distribution is assumed to be too computationally expensive to sample from. The posterior distribution of this model is the target of the proposed two-stage algorithm, which is described after. To clarify terminology throughout the paper, we refer to a standard MCMC algorithm of the full spatio-temporal model posterior distribution as the \emph{single-stage} MCMC algorithm. The two components of the proposed \emph{two-stage} MCMC algorithm are referred to as \emph{stage one} and \emph{stage two}.

\subsection{The Full Model}
Define $Y_{i,t}$ to be the observed response variable for locations $i=1, \ldots, I$ and time periods $t=1, \ldots, T$.  This response can be count, continuous, binary, ordinal, etc.  We define the observed covariates as $X_{i,t,p}$ where $p=1, \ldots, P$ indexes the $p$th observed covariate and $\bfX_{i,t}$ is a $(P+1)$-dimensional vector of covariates including a 1 for an intercept parameter.  We assume that the distribution of $Y_{i,t}$ depends on covariates $X_{i,t,p}$ with parameter vector $\bftheta_{Y,i}$.  With no loss of generality, assume that $\bftheta_{Y,i}$ has length $P+1+Q$, where the first $P+1$ elements correspond to the $P$ covariates in $\bfX_{i,t}$ along with an intercept, and the remaining $Q$ elements capture any additional parameters not tied to covariates such as variance terms or temporal dependence terms.

It is convenient to write all of the above in vector format by combining over locations $I$.  Call $\bfY_{1:T}$ as the $IT$-dimensional vector of the observed response, and $\bfX_{1:T}$ the $IT \times (P+1)$ dimensional design matrix of covariates.  Observe that this matrix has a column of all 1s for the intercept. Call $\bftheta_Y$ the $I(P+1+Q)$ dimensional vector of location-specific parameters. The observed responses are assumed to be conditionally independent across locations given $\bftheta_Y$ and $\bfX_{1:T}$.

The full spatio-temporal model may also depend on hyperparameters. Let $\bfphi$ be the vector of spatially static hyperparameters.  In most cases, these govern the spatial dependence across locations and are therefore referred to as spatial dependence hyperparameters. The full parameter vector $\left( \bftheta_Y, \bfphi \right)$ consists of all site-specific parameters together with the hyperparameters. We consider Bayesian hierarchical models of the form
\begin{equation}
\begin{aligned}
        \text{Data Model: } & \bfY_{1:T} \mid \bftheta_Y, \bfX_{1:T} \sim f\left( \bfY_{1:T} \mid \bftheta_Y, \bfX_{1:T} \right) \\
        \text{Prior Models: } & \bftheta_Y, \bfphi \sim \pi\left(\bftheta_Y \mid \bfphi\right) \times \pi\left(\bfphi\right),
    \end{aligned}
    \label{eq:BHM}
    \end{equation}
where $\pi\left(\bftheta_Y\mid\bfphi\right)$ denotes the conditional prior density of the site-specific parameters given the hyperparameters, which induces spatial dependence among locations, and $\pi\left(\bfphi\right)$ denotes the prior density of the hyperparameters. The posterior distribution is proportional to
\begin{equation}
   \pi\left(\bftheta_Y, \bfphi \mid \bfY_{1:T}, \bfX_{1:T} \right)  \propto  f\left(\bfY_{1:T} \mid \bftheta_Y, \bfX_{1:T}\right) \pi\left(\bftheta_Y \mid \bfphi\right) \pi\left(\bfphi\right).
   \label{eq:singlestagepost}
\end{equation}

Sampling from the posterior distribution in equation~\eqref{eq:singlestagepost} using standard single-stage MCMC is often computationally impractical due to any combination of a large number of locations $I$, a large number of time periods $T$, a large dimension of the parameter space $\left( \bftheta_Y, \bfphi \right)$, the computational cost to evaluate the data likelihood $f\left( \bfY_{1:T} \mid \bftheta_Y, \bfX_{1:T} \right)$ due to the spatio-temporal model selected for $\bfY_{1:T}$, and other reasons. The purpose of the two-stage methodology is to sample from equation \eqref{eq:singlestagepost} efficiently by breaking the MCMC into two distinct stages.  The first permits parallelization across site-specific parameter posterior distributions, and the second stage uses the stage one posterior as a proposal density for Metropolis-Hastings updates of the site-specific parameters in the full model. We describe both stages below.

\subsection{Stage One}
To improve computational efficiency, the stage one model modifies the prior distribution to assume spatial independence across locations, and the hyperparameters are temporarily excluded, so the full parameter vector consists only of $\bftheta_Y$. The data model in equation~\eqref{eq:BHM} remains unchanged from the full hierarchical model.

Let $\tilde{\pi}\left(\bftheta_Y\right)$ be the prior distribution assigned to $\bftheta_Y$ in stage one, and let $\tilde{\pi}\left(\bftheta_Y \mid \bfY_{1:T}, \bfX_{1:T}\right)$ denote the resulting stage one posterior distribution. Due to the assumed independence of locations during stage one, the stage one posterior can be written as
\begin{equation}
        \begin{aligned}
            \tilde{\pi}\left(\bftheta_Y \mid \bfY_{1:T}, \bfX_{1:T}\right) &\propto f\left(\bfY_{1:T} \mid \bftheta_Y, \bfX_{1:T}\right) \tilde{\pi}\left(\bftheta_Y\right) \\
            &= \prod_{i=1}^I f\left(\bfY_{i,1:T} \mid \bftheta_{Y,i}, \bfX_{i,1:T}\right) \tilde{\pi}\left(\bftheta_{Y,i}\right) \\
            &\propto \prod_{i=1}^I  \tilde{\pi}\left(\bftheta_{Y,i} \mid \bfY_{i,1:T}, \bfX_{i,1:T}\right),
            \label{eq:stageonepost}
        \end{aligned}
\end{equation}
where $\bftheta_{Y,i}$ is a location-specific parameter vector for location $i$. Observe that the stage one posterior factors into a product of $I$ independent site-specific posterior densities. This factorization allows for sampling from the posterior distribution in parallel by separately sampling from each of the $I$ site-specific posterior densities.

\subsection{Stage Two}\label{sec:stage-two-general}
The second stage implements a Metropolis-within-Gibbs algorithm targeting the posterior distribution of the full model. The spatial dependence hyperparameters $\bfphi$, which appear only in the full model, are updated from their full conditional distributions using Gibbs steps when direct sampling is possible and Metropolis-Hastings steps otherwise. The location-specific parameters $\bftheta_{Y,i}$, which appear in both the full model and the stage one model, are updated jointly for each location using Metropolis-Hastings steps, where proposed values for each location are drawn from the posterior sample generated in stage one. The $m$th iteration of the stage two MCMC algorithm proceeds as follows.

First, we update the spatial dependence hyperparameters, $\boldsymbol{\phi}$, one at a time from their full conditional distributions under the full hierarchical model. If a full conditional distribution is available in closed form, the corresponding hyperparameter can be sampled directly using a Gibbs update; otherwise, a Metropolis-Hastings update can be used. Let $\bfphi^{(m)}$ denote the updated value of the spatial dependence hyperparameters at iteration $m$.

Next, the location-specific parameters $\bftheta_{Y,i}$ are updated sequentially for $i=1,\ldots,I$ using Metropolis-Hastings steps. For each location, the proposed values are selected from the posterior sample generated for that location in stage one. The full conditional distribution for location $i$ implied by equation~\eqref{eq:singlestagepost}, and hence the target distribution for its Metropolis-Hastings update, denoted by $p\left(\bftheta_{Y,i}\right)$, is
\begin{equation}
\begin{aligned}
p\left(\bftheta_{Y,i}\right)
\equiv \pi\left(\bftheta_{Y,i}\mid\bftheta_{Y,-i},\boldsymbol{\phi},\bfY_{1:T},\bfX_{1:T}\right) 
\propto
f\left(\bfY_{i,1:T}\mid\bftheta_{Y,i},\bfX_{i,1:T}\right)
\pi\left(\bftheta_{Y,i}\mid\bftheta_{Y,-i},\boldsymbol{\phi}\right),
\end{aligned}
\label{eq:fullpost-target}
\end{equation}
where $\bftheta_{Y,-i}$ denotes the collection of location-specific parameter vectors for all locations other than $i$. The proposed candidate when updating location $i$ is $\bftheta_{Y,i}^*$, a random draw from the stage one posterior. The independence of the stage one posterior shown in equation \eqref{eq:stageonepost} implies the proposal distribution is
\begin{equation}
 \begin{aligned}
     q\left(\bftheta_{Y,i}^* \right) 
     \equiv  \tilde{\pi}\left(\bftheta_{Y,i}^* \mid \bfY_{i,1:T}, \bfX_{i,1:T}\right) 
     \propto f(\bfY_{i,1:T} \mid \bftheta_{Y,i}^*, \bfX_{i,1:T}) \tilde{\pi}(\bftheta_{Y,i}^*).
 \end{aligned}
\label{eq:proposal}
\end{equation}
The proposed candidate is accepted with Metropolis-Hastings acceptance probability $\min\left(1,R_i\right)$, where $R_i$ denotes the acceptance ratio. If the proposal is accepted, then $\bftheta_{Y,i}^{(m)} = \bftheta_{Y,i}^*$; otherwise, $\bftheta_{Y,i}^{(m)} = \bftheta_{Y,i}^{(m-1)}$. 

Let $\bftheta_{Y,j}^{(m^*)}$ denote the current value of the site-specific parameters for location $j$ when location $i$ is updated during iteration $m$ of the stage two MCMC algorithm. Because location-specific parameters are updated in sequence, $m^* = m$ for locations $j=1,\ldots,i-1$ (which have already been updated in the current iteration) and $m^* = m-1$ for locations $j=i+1,\ldots,I$ (which have not yet been updated). Based on the target and proposal distributions in equations~\eqref{eq:fullpost-target} and~\eqref{eq:proposal}, the acceptance ratio for the update at location $i$ is
\begin{equation}
\begin{aligned}
R_i
&=
\frac{p\left(\bftheta_{Y,i}^*\right)}
     {p\left(\bftheta_{Y,i}^{(m-1)}\right)} \times
\frac{q\left(\bftheta_{Y,i}^{(m-1)}\right)}
     {q\left(\bftheta_{Y,i}^*\right)} \\[4pt]
&=
\frac{
f\left(\bfY_{i,1:T}\mid\bftheta_{Y,i}^*,\bfX_{i,1:T}\right)
\pi\left(\bftheta_{Y,i}^*\mid\bftheta_{Y,-i}^{(m^*)},\bfphi^{(m)}\right)}
{
f\left(\bfY_{i,1:T}\mid\bftheta_{Y,i}^{(m-1)},\bfX_{i,1:T}\right)
\pi\left(\bftheta_{Y,i}^{(m-1)}\mid\bftheta_{Y,-i}^{(m^*)},\bfphi^{(m)}\right)}
\times
\frac{
f\left(\bfY_{i,1:T}\mid\bftheta_{Y,i}^{(m-1)},\bfX_{i,1:T}\right)
\tilde{\pi}\left(\bftheta_{Y,i}^{(m-1)}\right)}
{
f\left(\bfY_{i,1:T}\mid\bftheta_{Y,i}^*,\bfX_{i,1:T}\right)
\tilde{\pi}\left(\bftheta_{Y,i}^*\right)} \\[4pt]
&=
\frac{
\pi\left(\bftheta_{Y,i}^*\mid\bftheta_{Y,-i}^{(m^*)},\bfphi^{(m)}\right)
\tilde{\pi}\left(\bftheta_{Y,i}^{(m-1)}\right)}
{
\pi\left(\bftheta_{Y,i}^{(m-1)}\mid\bftheta_{Y,-i}^{(m^*)},\bfphi^{(m)}\right)
\tilde{\pi}\left(\bftheta_{Y,i}^*\right)}.
\end{aligned}
\label{eq:R0}
\end{equation}

Thus, the data-model cancels from the acceptance ratio. In particular, the acceptance ratio $R_i$ does not depend on the data likelihood $f\left(\bfY_{i,1:T} \mid \bftheta_{Y,i},\bfX_{i,1:T}\right)$. This cancellation substantially reduces the computational burden because evaluating this data density can be expensive, as in our applications. Consequently, the Metropolis-Hastings update depends only on the spatial conditional priors for $\bftheta_{Y,i}$ and the stage one priors. The resulting stage two Metropolis-within-Gibbs algorithm directly samples from the target posterior distribution in equation \eqref{eq:singlestagepost}.

\section{Data} \label{sec:section3-data}

\subsection{County-level Counts of Coccidioidomycosis}
In the first case study, the response variable is weekly, county-level reported cases of \textit{coccidioidomycosis}.  \textit{Coccidioidomycosis} (cocci) is caused by inhaling spores from the fungus \textit{Coccidioides}. This fungus lives in the soil primarily in the southwestern United States and is thought to thrive during wet winters, with spores becoming airborne during dry, windy periods that follow wet growth phases \citep{camponuri2025recent}. Accordingly, case counts are related to weather variables.  Case counts were provided by the Centers for Disease Control and Prevention using the National Notifiable Diseases Surveillance System (NNDSS), and cover the period from January 1, 2000 through June 30, 2022.  The date of any individual case is the ``event date'', or the earliest of onset, diagnosis, or specimen collection.  In the application, we focus on the 227 counties from seven states --- WA, OR, CA, NV, UT, AZ, and NM --- as the vast majority of cases occur in the American southwest and west.  There are a few important considerations with these data.  Cases in Washington weren't identified until roughly 2010, so essentially all WA cases post-date that discovery period.  Further, each state set its own rules and timeline for reporting case counts to NNDSS, with some states (such as NM) adopting reporting in the middle of this time period, in 2006.  Finally, the uniform criteria for what constitutes a case was modified once during this time period, in 2011 \footnote{\url{https://ndc.services.cdc.gov/case-definitions/coccidioidomycosis-2011/}}.

\subsection{Environmental Data}
Environmental data are used as covariates in all three case studies, and as the response variable in case study two. Specifically, we consider four measurements: accumulated precipitation ($kg \cdot m^{-2})$, potential evaporation ($kg \cdot m^{-2})$, soil moisture ($kg \cdot m^{-2})$, and soil temperature ($K$).  These were first obtained from the North America Land Data Assimilation System Phase 2 (NLDAS-2), an integrated observation and model reanalysis data set \citep{mitchell2004multi, xia2012continental}, and upscaled to a lattice of 0.5 degrees latitude by 0.5 degrees longitude covering the conterminous United States (CONUS). For case study one, all grid cells whose centroid was within a county border were averaged over the same week as the cocci case counts, and counties with no grid cells used the single closest grid cell's data. For the other case studies, these environmental data were averaged at the same weekly time to match the temporal support of the drought monitor. Environmental data came from a database published in \url{https://datadryad.org/dataset/doi:10.5061/dryad.g1jwstqw7}, and described in detail in \cite{erhardt2024homogenized}.  

\subsection{Binary Drought Data}
In the third case study, the response variable is a weekly, binary measure of drought at the 0.5 degree spatial scale.  Raw drought data were first obtained from the United States Drought Monitor (USDM, \cite{svoboda2002drought}, \url{https://www.drought.gov/data-maps-tools/us-drought-monitor}). This publicly available data product is updated weekly and classifies all US locations into an ordinal measure of drought with six levels. These categories range from 0 (no drought) to D0 (pre-drought), followed by D1, D2, D3, and D4, which represent progressively more severe drought conditions. We pre-processed these to binary by combining the top four levels (D1--D4) as ``drought'' and the bottom two (0 and D0) as ``no drought'', and extracted the data on grid cell centroids at the 0.5 degree spatial resolution.

\section{Three Case Studies} \label{sec:section4-application}
To demonstrate the flexibility and computational efficiency of the two-stage MCMC algorithm, we apply it to three spatio-temporal models for count, continuous, and binary responses in the following three subsections. These case studies illustrate how the general framework accommodates different likelihoods, temporal dependence structures, and spatial prior specifications. For each example, we first define the corresponding full hierarchical spatio-temporal model. We then describe the specific two-stage implementation by formulating the stage one model with spatially independent priors and deriving the application-specific stage two Metropolis-Hastings acceptance ratio.

\subsection{Case Study 1: Self-exciting Count Model on Areal Support}
Let $Y_{i,t}$ denote the observed case count of \textit{coccidioidomycosis} (cocci) in county $i$ during week $t$, for $i = 1, \ldots, I=227$ and $t = 1, \ldots, T=1174$. Because the count data exhibit both excess zeros and overdispersion, we assume that $Y_{i,t}$ follows a zero-inflated negative binomial (ZINB) distribution:
\[
Y_{i,t} \sim
\begin{cases}
0 & \text{with probability } \pi_{i}, \\[1ex]
\text{NB}(\mu_{i,t}, r_{i}) & \text{with probability } 1-\pi_{i},
\end{cases}
\]
where $\pi_i\in(0,1)$ denotes the county-specific zero-inflation probability, representing the probability of a structural zero, $\mu_{i,t} > 0$ is the mean of the negative binomial component, and $r_{i} > 0$ is the county-specific dispersion parameter.  In this case study, we do not model $\pi_i$ with covariates but instead allow it to vary by county.  One could include covariates for $\pi_i$ using this same methodology, though this would increase the dimension of the parameter space and could impact chain mixing. 

To account for temporal dependence, we model the mean on the log scale using an observation-driven first-order autoregressive structure. Specifically, we assume
\begin{equation}
\log(\mu_{i,t}) =
\begin{cases}
\log(P_{i,t}) + \mathbf{X}_{i,t}\boldsymbol{\beta}_i, & \text{for } t=1, \\[1ex]
\log(P_{i,t}) + \mathbf{X}_{i,t}\boldsymbol{\beta}_i
+ \rho_i \left(\log\bigl(Y_{i,t-1}^*\bigr) - \log(P_{i,t-1}) - \mathbf{X}_{i,t-1}\boldsymbol{\beta}_i \right),
& \text{for } t>1,
\end{cases}
\label{eq:logmu}
\end{equation}
where $P_{i,t}$ denotes the population for county $i$ at week $t$; $\rho_i \in (-1,1)$ is the county-specific autoregressive parameter that captures temporal dependence; $\mathbf{X}_{i,t}$ is the $(P+1)$-dimensional vector consisting of a $1$ for the intercept term and $P=3$ environmental covariates, which are weekly total precipitation, weekly average soil moisture content, and weekly average soil temperature; and $\boldsymbol{\beta}_i = (\beta_{0i}, \beta_{1i}, \ldots, \beta_{Pi})'$ is the county-specific $(P+1)$-dimensional vector of regression coefficients. These environmental covariates were originally available on a grid of 0.5 degrees latitude by 0.5 degrees longitude, but we took averages of all grid cells whose centroids were within a given county.  For any (small) counties with no gridded data with a centroid in the interior, we took the single nearest grid cell.  Accordingly, in this case study the response and all covariates had the same spatial support as counties. Prior to model fitting, all three environmental covariates were standardized to have mean zero and standard deviation one.

In addition, $Y_{i,t-1}^* = \max(Y_{i,t-1}, c)$, where $0 < c < 1$ is a small positive constant introduced to ensure that the logarithm is well defined when $Y_{i,t-1}=0$ \citep{angelakis2025modeling}. In this study, we set $c=0.5$, so that zero counts are replaced by a small positive value before applying the log transformation. 

For prior specification, we assign independent uniform priors to the zero-inflation parameters across counties, $\pi_i \sim \mathrm{U}(0,1)$ for $i=1,\ldots,I$, and independent gamma priors to the dispersion parameters, $r_{i} \sim \mathrm{Gamma}(2, 2)$ for $i=1,\ldots,I$. The gamma prior ensures support on the positive real line for the dispersion parameter. For each $p=0,\ldots,P$, let $\bfbeta_{(p)}=\left(\beta_{p1},\ldots,\beta_{pI}\right)'$ denote the vector of the $p$th regression coefficients across all $I$ counties. The parenthetical subscript $(p)$ distinguishes $\bfbeta_{(p)}$ from $\bfbeta_i$, where $\bfbeta_i$ denotes the vector of $P+1$ regression coefficients associated with county $i$. To capture spatial dependence across counties, we specify that the regression coefficient vectors, $\bfbeta_{(p)}$ for $p=0,\ldots,P$ follow mutually independent intrinsic conditional autoregressive (ICAR) prior distributions described below. Since $\rho_i$ is restricted to the interval $(-1,1)$, we first transform it to the real line by defining $\gamma_i = \frac{1}{2}\log\left(\frac{1+\rho_i}{1-\rho_i}\right)$, which implies $\rho_i = \frac{e^{2\gamma_i}-1}{e^{2\gamma_i}+1}$. We also specify that the transformed autoregressive parameter vector, $\bfgamma $, follows an ICAR prior distribution.

For the ICAR priors, we define the adjacency weights $a_{i\ell} = 1$ if locations $i$ and $\ell$ are neighbors, where neighborhood is determined using queen adjacency \citep{banerjee2003hierarchical}, and $a_{i\ell} = 0$ otherwise. Let $a_{i+} = \sum_{\ell} a_{i\ell}$ denote the number of neighbors for location $i$. Then, under the ICAR prior,
\begin{equation}
\gamma_i\mid\bfgamma_{-i},\sigma_\gamma^2
\sim
N\left(
\frac{1}{a_{i+}}\sum_{\ell=1}^I a_{i\ell}\gamma_\ell, \
\frac{\sigma_\gamma^2}{a_{i+}}
\right),
\label{eq:icar}
\end{equation}
where $\bfgamma_{-i} = \{\gamma_{\ell} : \ell \neq i\}$ and $\sigma^2_{\gamma}$ is a variance parameter. This conditional specification corresponds to the improper joint density $\pi(\bfgamma\mid\sigma_\gamma^2) \propto \exp \left(-\frac{1}{2\sigma^2_{\gamma}} \bfgamma' (\bfD - \bfA) \bfgamma \right)$, where $\bfA$ is the $I \times I$ symmetric adjacency matrix with entries $a_{i\ell}$ and $\bfD = \mathrm{diag}(a_{1+}, \ldots, a_{I+})$ is the diagonal matrix of neighbor counts \citep{banerjee2003hierarchical}. Similarly, each regression coefficient vector $\bfbeta_{(p)}$ is assigned an ICAR prior with variance parameter $\sigma^2_{(p)}$ for $p=0,\ldots,P$. This formulation induces spatial smoothing across neighboring counties in both the autoregressive and regression coefficients while ensuring that each $\rho_i$ remains in the interval $(-1,1)$. Finally, the variance parameters $\sigma_\gamma^2$ and $\{\sigma_{(p)}^2 : p = 0, \ldots, P\}$ are each assigned an independent weakly informative conjugate prior, $\mathrm{Inverse\text{-}Gamma}(0.5, 0.5)$.

We fit the (computationally intensive) single-stage count model using NIMBLE \citep{nimble-article:2017,nimble-software:2024,nimble-manual:2024}, which allows users to specify models in R using the BUGS language, but runs the MCMC in compiled C++ code for efficiency. Specifically, we ran an MCMC chain for 100,000 iterations, applying a burn-in period of 20,000 iterations and a thinning interval of 20, resulting in 4,000 posterior samples. We used NIMBLE's automated factor slice sampler \citep{tibbits2014automated,turek2017automated} to jointly update each block $(\bfbeta_i,\gamma_i)$, for $i=1,\ldots,I$. This block sampler was used to account for the strong posterior dependence between $\bfbeta_i$ and $\gamma_i$ and thereby improve the mixing of these parameters in the single-stage model. The total computation time (compiling and running) for the single-stage model was approximately 20.78 hours on one node of the DEAC high-performance computing cluster at Wake Forest University \citep{WakeHPC}. While we ran a single chain, we assessed Markov chain mixing and convergence to the stationary distribution using trace plots and autocorrelation function plots, some of which are included in the Appendix. These diagnostics indicated satisfactory mixing and convergence.  This holds for all case studies presented.

\subsubsection{Stage One}
The stage one model assumes independence across counties, and accordingly we replace the ICAR priors on $\bfbeta_{(p)}$ and $\bfgamma$ with independent priors across locations. The spatial ICAR priors on the regression coefficients are replaced with independent Gaussian priors. Because county population is included in the model as an offset, the intercept represents the baseline log expected case rate after adjusting for population size. Therefore, for each county $i$, the intercept is assigned the prior $\beta_{0i} \sim N(\mu_0,5^2)$, where $\mu_0$ is the overall log case rate estimated from the data, providing a weakly informative prior centered at a plausible baseline value. The remaining regression coefficients are assigned independent priors $\beta_{pi} \sim N(0,2^2)$ for $p=1,\ldots,P$ (recall that all covariates were first scaled to mean 0 and variance 1). Consequently, no spatial smoothing is imposed during stage one. 

Likewise, the transformed county-specific autoregressive parameters are assumed to be independent across counties. Recall that $\gamma_i=\frac{1}{2}\log\left(\frac{1+\rho_i}{1-\rho_i}\right)$, which maps $\rho_i \in (-1,1)$ to $\gamma_i \in \mathbb{R}$. In the stage one model, we assume $\gamma_i \sim \mathrm{Logistic}(0,0.5)$ for $i=1,\ldots,I$, where $\mathrm{Logistic}(0,0.5)$ denotes the logistic distribution with mean $0$ and scale parameter $0.5$. Under the inverse transformation $\rho_i=\frac{e^{2\gamma_i}-1}{e^{2\gamma_i}+1}$, this specification is equivalent to $\rho_i \sim \mathrm{U}(-1,1)$.

An important consequence of the stage one specification is that the joint posterior of the stage one model factorizes into a product of $I$ county-specific posterior densities as shown in equation~\eqref{eq:stageonepost}. As a result, parallel computing can be used to efficiently sample from the posterior by running the county-specific posterior samplers simultaneously. 

For each county $i$, we ran the stage one MCMC for 100,000 iterations, discarding the first 20,000 as burn-in and thinning by 20, which resulted in 4,000 draws of the stage one posteriors. We used NIMBLE's automated factor slice sampler to jointly update the parameter block $(\bfbeta_i,\gamma_i)$ to improve the mixing of these parameters in the stage one model. The average computation time was approximately 6.90 minutes per county. The resulting posterior draws served as proposal draws in stage two.

\subsubsection{Stage Two} \label{sec:stage2-count}
Stage two uses a Metropolis-within-Gibbs algorithm targeting the posterior distribution under the full spatial count model. At iteration $m$, we first update the ICAR variance parameters, $\bfphi = \left(\sigma^2_{\gamma}, \{\sigma^2_{(p)}, p=0,\ldots, P\}\right)$, which only appear in the full model. The priors on all ICAR variances, $\sigma_\gamma^2$ and $\sigma_{(p)}^2$ for $p=0,\ldots,P$, were independent $\mathrm{Inverse\text{-}Gamma}(0.5, 0.5)$, which permit Gibbs updates under conjugacy. The full conditional distribution of each ICAR variance parameter is an inverse-gamma distribution with shape parameter $a + I/2$. The corresponding scale parameter is $b + 0.5\sum_{i \sim j} \left(\gamma_i^{(m-1)} - \gamma_j^{(m-1)} \right)^2$ for $\sigma^{2}_{\gamma}$, and $b + 0.5\sum_{i \sim j} \left(\beta_{pi}^{(m-1)} - \beta_{pj}^{(m-1)} \right)^2$ for each $\sigma^{2}_{(p)}$, $p=0,\ldots,P$. Here, $a=0.5$ and $b=0.5$ are the shape and scale parameters of the prior distribution, and the notation $i \sim j$ indicates all pairs of neighboring locations defined by the adjacency matrix $\bfA$. Let $\boldsymbol{\phi}^{(m)}=\left(\sigma_{\gamma}^{2(m)},\sigma_{(0)}^{2(m)},\ldots,\sigma_{(P)}^{2(m)}\right)$ denote the updated values of the spatial variance parameters at iteration $m$.

For each county $i=1,\ldots,I$, the county-specific parameters, $\bftheta_{Y,i}=(\bfbeta_i,\gamma_i, r_{i}, \pi_{i})$, which appear in both the full model and the stage one model, are updated jointly using a Metropolis-Hastings step with proposals drawn from the stage one posterior samples. Following the general stage two methodology presented in Section~\ref{sec:stage-two-general} and equation~\eqref{eq:R0}, the Metropolis-Hastings acceptance ratio for county $i$ in the count model becomes 
\begin{equation}
\begin{aligned}
R_i
&=
\frac{
\pi\left(\gamma_i^*\mid\bfgamma_{-i}^{(m^*)},\sigma_\gamma^{2(m)}\right)
\prod_{p=0}^P
\pi\left(\beta_{pi}^*\mid\bfbeta_{p,-i}^{(m^*)},\sigma_{(p)}^{2(m)}\right)}
{
\pi\left(\gamma_i^{(m-1)}\mid\bfgamma_{-i}^{(m^*)},\sigma_\gamma^{2(m)}\right)
\prod_{p=0}^P
\pi\left(\beta_{pi}^{(m-1)}\mid\bfbeta_{p,-i}^{(m^*)},\sigma_{(p)}^{2(m)}\right)}
\times
\frac{
\widetilde{\pi}\left(\gamma_i^{(m-1)}\right)
\prod_{p=0}^P
\widetilde{\pi}\left(\beta_{pi}^{(m-1)}\right)}
{
\widetilde{\pi}\left(\gamma_i^*\right)
\prod_{p=0}^P
\widetilde{\pi}\left(\beta_{pi}^*\right)},
\end{aligned}
\label{eq:R0-count}
\end{equation}
where $\pi(\cdot\mid\cdot)$ denotes the full-model conditional Gaussian densities induced by the ICAR priors in equation~\eqref{eq:icar}, and $\widetilde{\pi}(\cdot)$ denotes the stage one prior densities. Because the priors assigned to the dispersion parameter $r_{i}$ and the zero-inflation parameter $\pi_{i}$ are identical in both the stage one and full models, their corresponding prior-density terms cancel from the acceptance ratio. The acceptance ratio in equation~\eqref{eq:R0-count} reduces to the expression given in equation~\eqref{eq:R-count} in the Appendix.

We ran the stage two Metropolis-within-Gibbs MCMC for 200,000 iterations. After discarding an initial burn-in of 50,000 iterations and applying a thinning interval of 50, this yielded 3,000 posterior draws. The stage two algorithm required approximately 0.37 hours to complete on a single node of the DEAC high-performance computing cluster.

\subsubsection{Results}
\begin{figure}
    \centering
    {\includegraphics[width=0.75\textwidth]{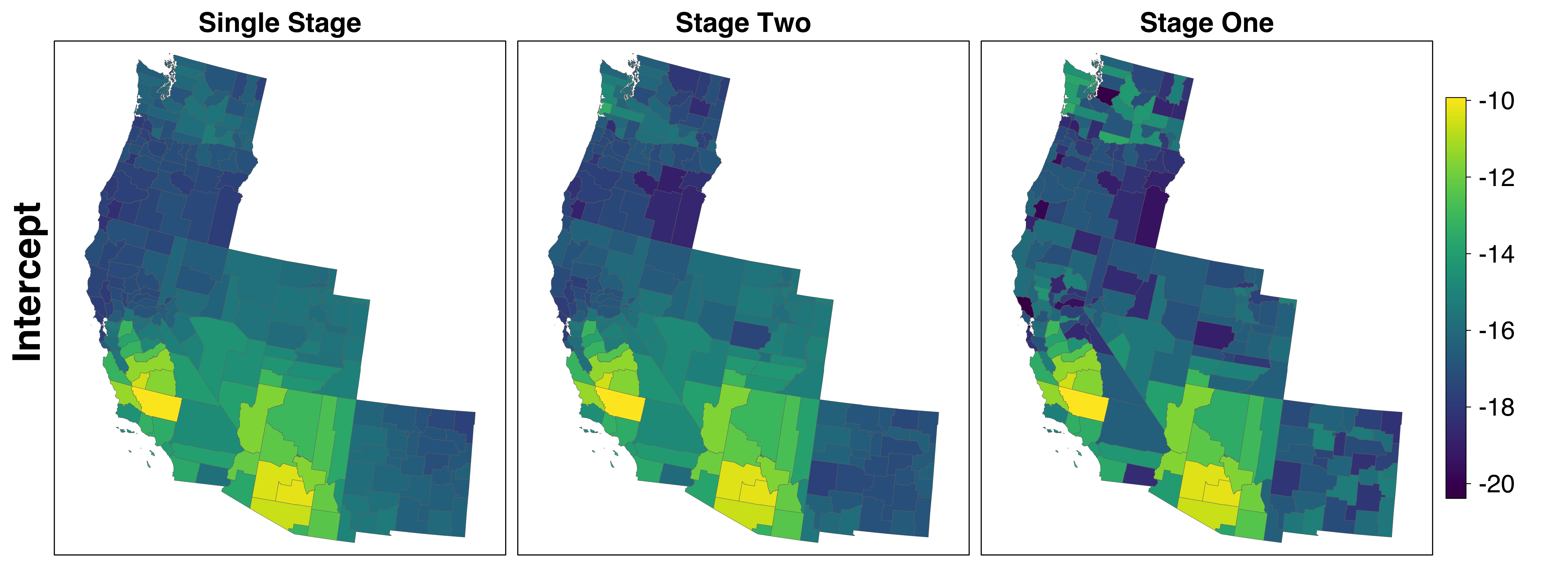}}  
    \par\vspace{1mm}
    {\includegraphics[trim={0 0 0 70mm},clip, width=0.75\textwidth]{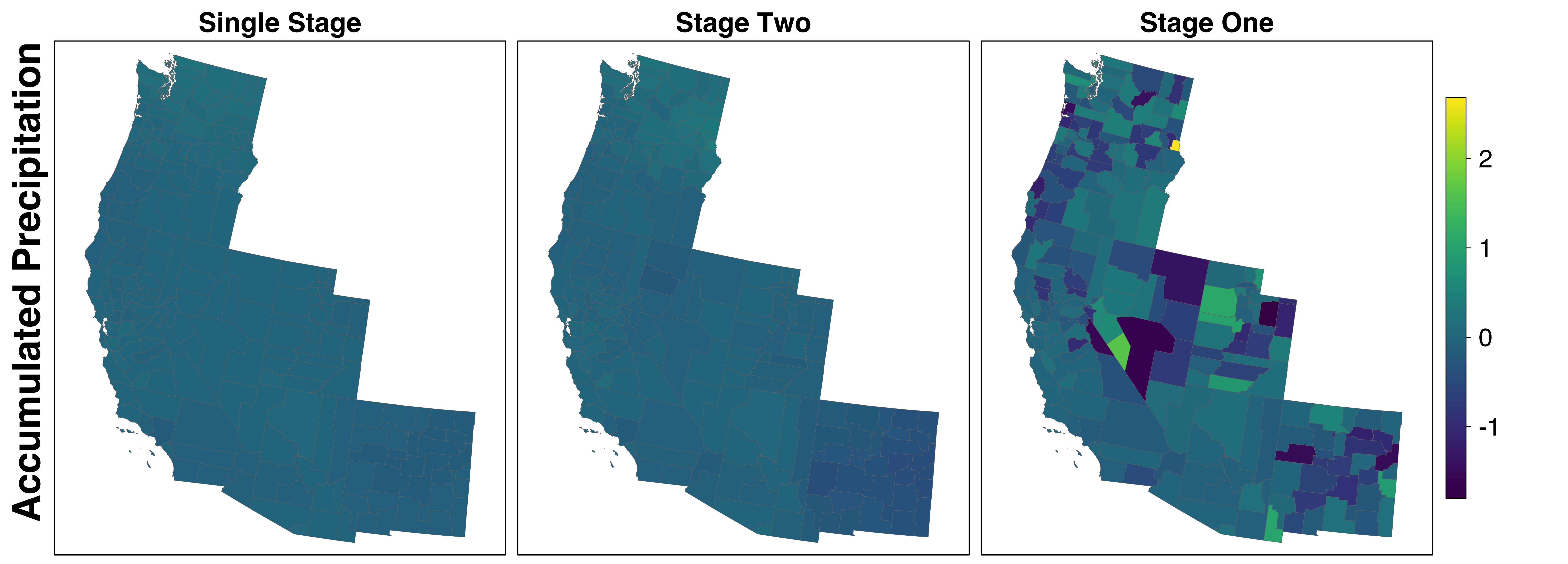}} 
    \par\vspace{1mm}
    {\includegraphics[trim={0 0 0 70mm},clip,width=0.75\textwidth]{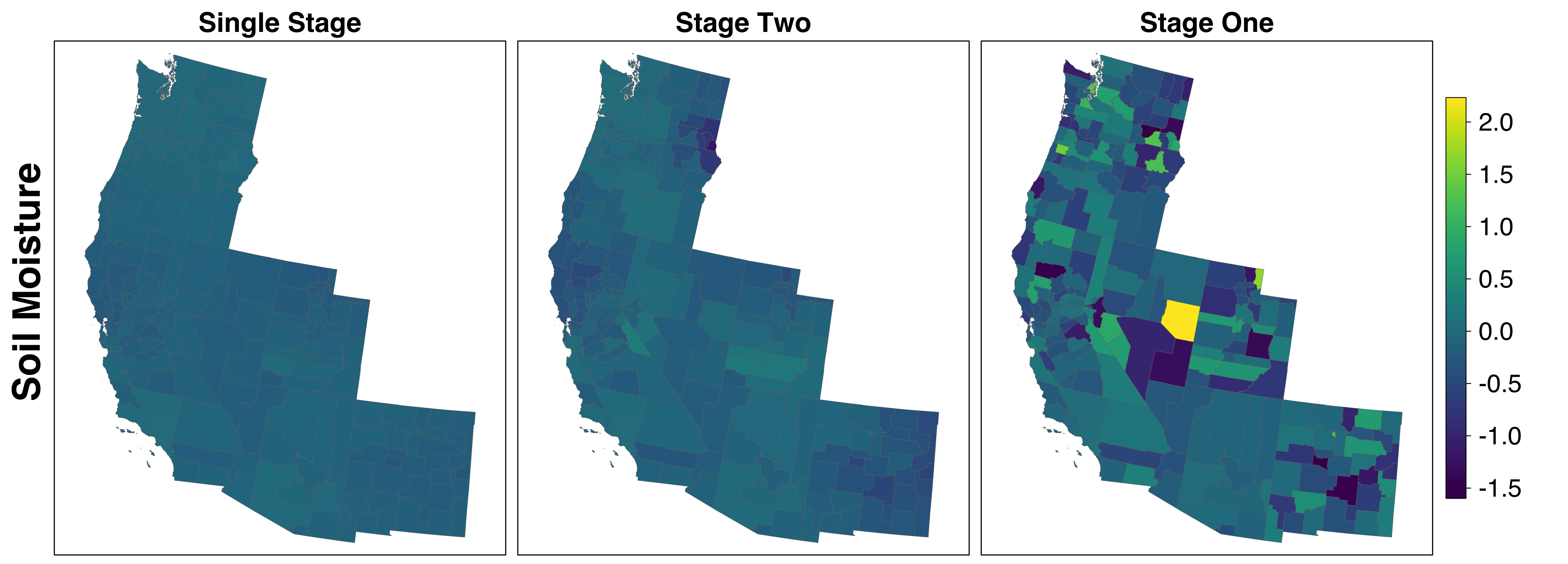}}
    \par\vspace{1mm}
    {\includegraphics[trim={0 0 0 70mm},clip,width=0.75\textwidth]{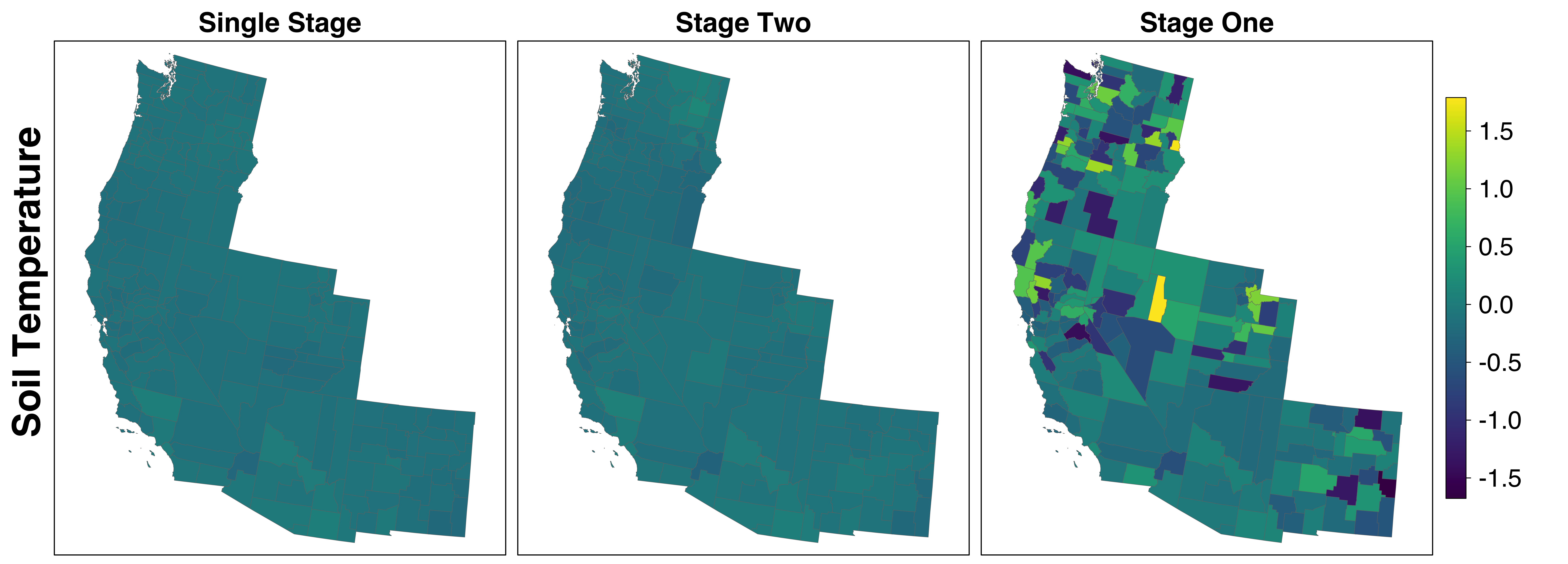}}
    \par\vspace{1mm}
    {\includegraphics[trim={0 0 0 70mm},clip,width=0.75\textwidth]{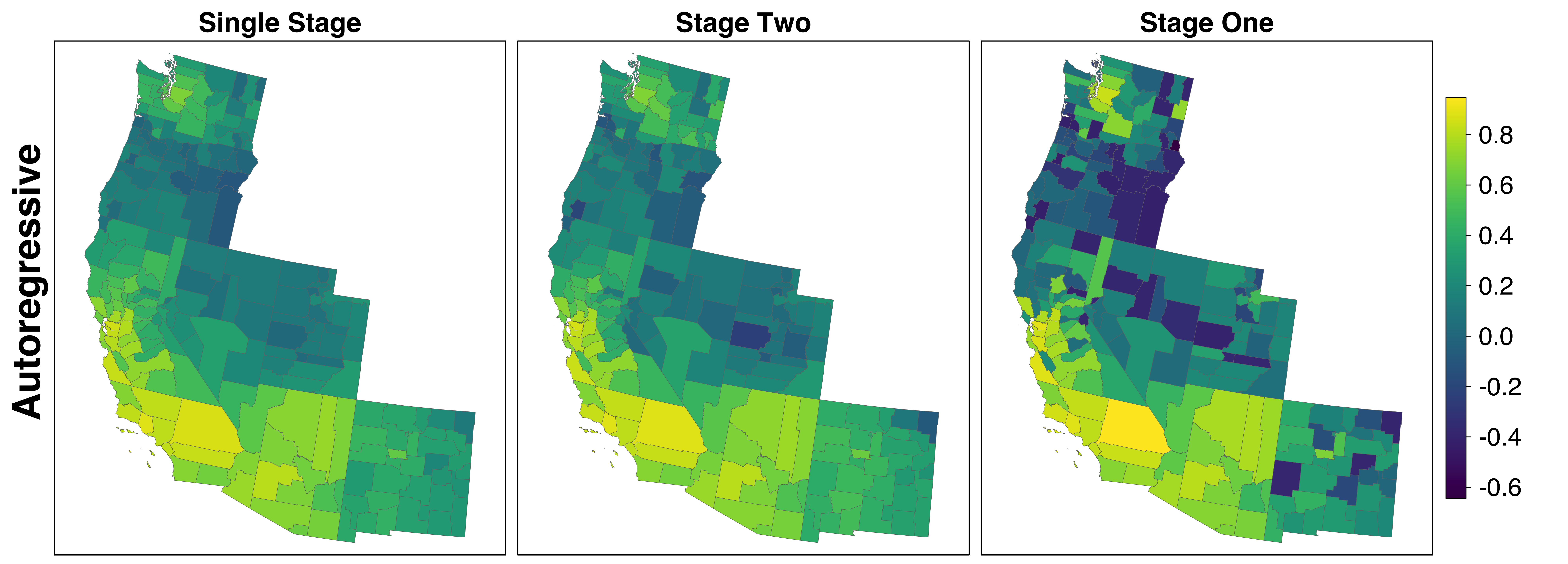}} 

    \caption{Posterior means of the county-specific parameters assigned spatial ICAR priors in the count model under the single-stage method (left column), the two-stage method (center column), and after stage one only (right column). From top to bottom, the rows show the intercept $\beta_0$, effect of accumulated precipitation $\beta_{\mathrm{apcp}}$, effect of soil moisture $\beta_{\mathrm{soilm}}$, effect of soil temperature $\beta_{\mathrm{tsoil}}$, and autoregressive parameter $\rho$.}
    \label{fig:count-mean}
\end{figure}

Figure~\ref{fig:count-mean} shows the posterior means of the county-specific parameters assigned spatial ICAR priors in the full model across the single-stage method, the two-stage method, and stage one alone. The Figure~\ref{fig:count-sd} in the Appendix shows the corresponding posterior standard deviations. 

The single-stage results in the left column are obtained by sampling from the target posterior distribution, so agreement between the left and center columns indicates that the two-stage algorithm is also correctly sampling from the same posterior distribution. Furthermore, comparing the right column (stage one) with the center column (stage two) highlights the effect of introducing the spatial ICAR priors. Because the stage one model assumes independence across counties, its estimates display greater county-to-county variation and more irregular spatial patterns. In the full model, information is shared across neighboring counties, resulting in spatially smoother posterior mean surfaces and a noticeable reduction in posterior uncertainty (Figure~\ref{fig:count-sd}). Overall, the close agreement between the single-stage and stage two maps provides visual evidence that both computational approaches are generating samples from the same target posterior distribution.

The regression coefficients $\beta_{\mathrm{apcp}}$, $\beta_{\mathrm{soilm}}$, and $\beta_{\mathrm{tsoil}}$ represent the effects of accumulated precipitation, soil moisture, and soil temperature, respectively, on the log expected cocci case count after accounting for the population offset and temporal dependence. Positive values (Figure \ref{fig:count-mean}) indicate that increases in the corresponding covariate are associated with higher expected case counts, whereas negative values indicate lower expected case counts. Because the population is included as an offset, the intercept $\beta_0$ represents the baseline log expected case rate. The autoregressive parameter $\rho$ captures temporal dependence, with larger values indicating stronger dependence from one week to the next. 

Figures~\ref{fig:count-mean-pi-r} and~\ref{fig:count-sd-pi-r} in the Appendix also present the posterior means and standard deviations, respectively, of the zero-inflation and dispersion parameters, which are not assigned spatial ICAR priors. Trace plots of the county-specific parameters for one county under the single-stage method, the two-stage method, and stage one alone are shown in Figure~\ref{fig:count-trace} in the Appendix.

We next compared the computational efficiency of the single-stage and two-stage count models using effective sample size (ESS) per hour. As shown in Table~\ref{tab:ess-count}, the two-stage model produced substantially larger ESS per hour for every county-specific parameter considered. Depending on the parameter, the two-stage ESS per hour was approximately 4 to 5 times larger than for    the single-stage model. The total run time was also substantially lower, dropping from approximately 20.78 hours for the single-stage model to less than one hour for the two-stage procedure. These results show that the two-stage procedure is more computationally efficient.

\begin{table}[!h]
    \centering
    \begin{tabular}{c|c|c}
        & \text{Single-stage model} & \text{Two-stage model} \\
        \hline
        Parameter & ESS per hour & ESS per hour \\
        \hline
        $\beta_0$                & 149.2 & 726.1 \\
        $\beta_{\mathrm{apcp}}$  & 175.4 & 730.8 \\
        $\beta_{\mathrm{soilm}}$ & 138.2 & 688.9 \\
        $\beta_{\mathrm{tsoil}}$ & 147.2 & 608.2 \\
        $\rho$                   & 154.1 & 753.0 \\
        $\pi$                    & 189.1 & 1030.8 \\
        $r$                      & 191.1 & 1042.9 
    \end{tabular}

    \caption{Mean effective sample sizes per hour of computing time for the county-specific regression coefficients, autoregressive, zero-inflation, and dispersion parameters under the single-stage and two-stage count models. Effective sample sizes per hour are averaged over all locations. The reported run times exclude model compilation time but include burn-in.}
    \label{tab:ess-count}
\end{table}


\subsection{Cast Study 2: Continuous, non-Gaussian Spatio-temporal Model with Point-referenced Support}
Let $Y_{i,t}$ be soil moisture observations at location $i$ during week $t$, for $i=1,\ldots,I=187$ and $t=1,\ldots,T=104$. Locations are available on evenly spaced grid cells every 0.5 degrees latitude and 0.5 degrees longitude. In this analysis, we represent each grid cell by its centroid and treat the corresponding soil moisture observations as point-referenced (geostatistical) data observed over a continuous spatial domain. Spatial dependence in the regression coefficients and transformed autoregressive parameters is modeled through Gaussian processes with stationary, isotropic exponential covariance functions defined in terms of the distances between grid cell centroids.

In this application, weekly average soil moisture observations, $Y_{i,t}$, are strictly positive, continuous, and right-skewed. We therefore model $Y_{i,t}$ using a gamma likelihood. Specifically, we assume
\[
Y_{i,t} \mid \mu_{i,t}, \alpha_i
\sim
\mathrm{Gamma}\left(\alpha_i, \frac{\alpha_i}{\mu_{i,t}}\right),
\]
where the gamma distribution is parameterized in terms of shape and rate. Under this parameterization, $E(Y_{i,t}\mid \mu_{i,t},\alpha_i)=\mu_{i,t}$ and $\mathrm{Var}(Y_{i,t}\mid \mu_{i,t},\alpha_i)=\mu_{i,t}^2/\alpha_i$. Thus, $\mu_{i,t}$ represents the conditional mean of the continuous response, while $\alpha_i$ controls the conditional variability at location $i$.

Let $\mathbf{X}_{i,t}$ denote the $(P+1)$-dimensional vector of covariates at location $i$ and week $t$. This vector consists of a $1$ for the intercept term and $P = 3$ environmental covariates: weekly total precipitation, weekly average potential evaporation, and weekly average soil temperature. Before model fitting, these covariates were standardized to have mean zero and unit variance. Let $\boldsymbol{\beta}_i = (\beta_{0i}, \beta_{1i}, \ldots, \beta_{Pi})'$ denote the corresponding location-specific vector of regression coefficients. Temporal dependence is introduced through an observation-driven, centered AR(1) structure:
\begin{equation*}
\log(\mu_{i,t})
=
\begin{cases}
\mathbf{X}_{i,1}\boldsymbol{\beta}_i, & \text{for } t=1, \\[1ex]
\mathbf{X}_{i,t}\boldsymbol{\beta}_i
+
\rho_i
\left(
\log(Y_{i,t-1})
-
\mathbf{X}_{i,t-1}\boldsymbol{\beta}_i
\right), & \text{for } t>1,
\end{cases}
\label{eq:continuous_log_mean}
\end{equation*}
where $\rho_i\in(-1,1)$ is a location-specific autoregressive parameter. The term $\log(Y_{i,t-1})-\mathbf{X}_{i,t-1}\boldsymbol{\beta}_i$ is the previous-week residual on the log scale, and $\rho_i$ determines how strongly this residual affects the current log-mean.

To capture spatial dependence across locations, we model each regression coefficient and the transformed autoregressive parameter as a Gaussian process with an exponential covariance function. Evaluated at the $I$ observed locations, these Gaussian processes induce multivariate normal (MVN) priors for the corresponding parameter vectors. For the $p$th regression coefficient, let $\boldsymbol{\beta}_{(p)}=(\beta_{p1},\ldots,\beta_{pI})'$ for $p=0,\ldots,P$ denote the vector of coefficients across all locations. We assume
\(
\boldsymbol{\beta}_{(p)}
\sim
\mathrm{MVN}
\left(
\mu_{(p)}\mathbf{1},
\sigma_{(p)}^2\mathbf{R}_{\phi_{(p)}}
\right),
\)
for $p=0,\ldots,P$, where $\mathbf{1}$ is an $I$-dimensional vector of ones and the spatial correlation matrix is specified using an exponential correlation function,
\(
\mathbf{R}_{\phi_{(p)}}(i,j)
=
\exp(-\phi_{(p)}d_{ij}).
\)

Spatial dependence is also introduced for the transformed autoregressive parameters. To ensure that $\rho_i$ lies in the interval $(-1,1)$, we define $\gamma_i=\frac{1}{2}\log\left(\frac{1+\rho_i}{1-\rho_i}\right)$, which implies $\rho_i=\tanh(\gamma_i)=\frac{e^{2\gamma_i}-1}{e^{2\gamma_i}+1}$, where $\gamma_{i}\in\mathbb{R}$ is an unconstrained transformed autoregressive parameter. Let $\boldsymbol{\gamma}=(\gamma_{1},\ldots,\gamma_{I})'$. We assume
\(
\boldsymbol{\gamma}
\sim
\mathrm{MVN}
\left(
\mu_{\gamma}\mathbf{1},
\sigma_{\gamma}^2\mathbf{R}_{\phi_{\gamma}}
\right),
\)
where
\(
\mathbf{R}_{\phi_{\gamma}}(i,j)
=
\exp(-\phi_{\gamma}d_{ij}).
\)
In these specifications, $d_{ij}$ denotes the distance between locations $i$ and $j$. The parameters $\phi_{(p)}>0$ and $\phi_{\gamma}>0$ are the spatial decay parameters that control the rate at which correlation decreases with distance.

For prior specification, we assign $\mu_{(p)}\sim \mathrm{N}(0,10^2)$, $\sigma_{(p)}^2\sim \mathrm{Inverse\text{-}Gamma}(2,1)$, and $\phi_{(p)}\sim \mathrm{U}(\phi_L,\phi_U)$, for $p=0,\ldots,P$. Similarly, we assign $\mu_{\gamma}\sim \mathrm{N}(0,10^2)$, \\$\sigma_{\gamma}^2\sim \mathrm{Inverse\text{-}Gamma}(2,1)$, and $\phi_{\gamma}\sim \mathrm{U}(\phi_L,\phi_U)$. The lower and upper bounds for the spatial decay parameters are chosen based on the observed pairwise distances in the spatial domain. Specifically, we define $\phi_L = 0.01/\mathrm{maxD}$ and $\phi_U = 10/\mathrm{minD}$, where $\mathrm{maxD}$ and $\mathrm{minD}$ denote the maximum and minimum observed pairwise distances, respectively. For the gamma shape parameters, we use weakly informative independent gamma priors, $\alpha_i \sim \mathrm{Gamma}(0.01,0.01)$, for $i=1,\ldots,I$. 

We implemented the full single-stage continuous model in NIMBLE. An MCMC chain was run for 10,000 iterations, with the first 4,000 iterations discarded as burn-in and a thinning interval of 2. We used NIMBLE's automated factor slice sampler \citep{tibbits2014automated,turek2017automated} to jointly update each regression coefficient vector, $\boldsymbol{\beta}_{(p)}$ for $p=0,\ldots,P$, and the transformed autoregressive parameter vector, $\boldsymbol{\gamma}$, to improve MCMC mixing of these parameters in the single-stage model. The MCMC run required approximately 48.30 hours (compiling and running) on one node of the DEAC high-performance computing cluster at Wake Forest University.

\subsubsection{Stage One}
The stage one model replaces the Gaussian process priors for the spatially dependent parameters with normally distributed priors that are independent across locations. Specifically, the spatial MVN prior on each regression coefficient vector $\boldsymbol{\beta}_{(p)}$ is replaced with independent normal priors, $\beta_{pi} \sim \mathrm{N}(0,10^2)$, for $p=0,\ldots,P$ and $i=1,\ldots,I$. Similarly, the transformed autoregressive parameters $\boldsymbol{\gamma}$ are modeled independently. In stage one, we assign $\gamma_i \sim \mathrm{N}(0,10^2)$ for $i=1,\ldots,I$. The priors for the gamma shape parameters $\alpha_i$ remain unchanged in stage one. In particular, $\alpha_i \sim \mathrm{Gamma}(0.01,0.01)$ for $i=1,\ldots,I$, exactly as in the full model.

For each location $i$, the stage one MCMC was run for 200,000 iterations, with the first 50,000 iterations discarded as burn-in and a thinning interval of 50. This produced 3,000 posterior draws for each location. The average computation time was approximately 1.27 minutes per location.

\subsubsection{Stage Two}

Stage two uses a Metropolis-within-Gibbs algorithm targeting the posterior distribution under the full continuous spatial model. We first update the spatial hyperparameters that appear in the full model but not in the stage one model. The spatial means, $\mu_{\gamma}$ and $\mu_{(p)}$, and spatial variances, $\sigma_{\gamma}^2$ and $\sigma_{(p)}^2$, are updated using Gibbs steps, whereas the spatial decay parameters, $\phi_{\gamma}$ and $\phi_{(p)}$, are updated using random-walk Metropolis-Hastings steps, for $p=0,\ldots,P$. Details of these full conditional distributions and Metropolis-Hastings updates are provided in Section~\ref{sec:continuous-hyperparameters-updates} of the Appendix. 

Let $\boldsymbol{\phi}^{(m)}=\left((\mu_{\gamma}^{(m)},\sigma^{2(m)}_{\gamma},\phi_{\gamma}^{(m)}), \{(\mu_{(p)}^{(m)},\sigma^{2(m)}_{(p)},\phi_{(p)}^{(m)}), p=0,\ldots, P\}\right)$ denote the updated spatial hyperparameters at iteration $m$. Following the general stage two procedure in Section~\ref{sec:stage-two-general}, the location-specific parameters $\bftheta_{Y,i}=(\bfbeta_i,\gamma_i,\alpha_i)$ are updated jointly using Metropolis--Hastings proposals drawn from the corresponding stage one posterior samples. In the continuous model, the gamma likelihood cancels from the acceptance ratio in equation~\eqref{eq:R0}, and because the gamma shape parameter $\alpha_i$ is assigned the same prior in the full and stage one models, its prior-density terms also cancel. Therefore, the acceptance ratio depends only on the conditional densities of the MVN priors from the full model and the independent normal priors from stage one.

Let $q_{\gamma,ii}$ denote the $i$th diagonal element of $\mathbf{Q}_{\gamma}$ and let $\mathbf{q}_{\gamma,i,-i}$ denote the remaining elements in row $i$. Under the MVN spatial prior, the full conditional distribution of $\gamma_i$ is
\begin{equation}
\gamma_i\mid\boldsymbol{\gamma}_{-i},\mu_{\gamma},\sigma_{\gamma}^2,\phi_{\gamma}
\sim
\mathrm{N}\left(
m_{\gamma,i},
v_{\gamma,i}
\right),
\label{eq:continuous-gamma-conditional}
\end{equation}
where
\[
m_{\gamma,i}
=
\mu_{\gamma}
-
\frac{\mathbf{q}_{\gamma,i,-i}}{q_{\gamma,ii}}
\left(
\boldsymbol{\gamma}_{-i}-\mu_{\gamma}\mathbf{1}
\right),
\qquad
v_{\gamma,i}
=
\frac{\sigma_{\gamma}^2}{q_{\gamma,ii}}.
\]
Similarly, if $q_{(p),ii}$ is the $i$th diagonal element of $\mathbf{Q}_{(p)}$ and $\mathbf{q}_{(p),i,-i}$ denotes the remaining elements of row $i$, then, for each $p=0,\ldots,P$,
\begin{equation}
\beta_{pi}\mid\boldsymbol{\beta}_{(p),-i},\mu_{(p)},\sigma_{(p)}^2,\phi_{(p)}
\sim
\mathrm{N}\left(
m_{(p),i},
v_{(p),i}
\right),
\label{eq:continuous-beta-conditional}
\end{equation}
where
\[
m_{(p),i}
=
\mu_{(p)}
-
\frac{\mathbf{q}_{(p),i,-i}}{q_{(p),ii}}
\left(
\boldsymbol{\beta}_{(p),-i}-\mu_{(p)}\mathbf{1}
\right),
\qquad
v_{(p),i}
=
\frac{\sigma_{(p)}^2}{q_{(p),ii}}.
\]
The proposed location-specific vector is accepted with probability $\min(1,R_i)$, where
\begin{equation}
\begin{aligned}
R_i
&=
\frac{
\pi\left(\gamma_i^*\mid\boldsymbol{\gamma}_{-i}^{(m^*)},\mu_{\gamma}^{(m)},\sigma_{\gamma}^{2(m)},\phi_{\gamma}^{(m)}\right)
\prod_{p=0}^{P}
\pi\left(\beta_{pi}^*\mid\boldsymbol{\beta}_{(p),-i}^{(m^*)},\mu_{(p)}^{(m)},\sigma_{(p)}^{2(m)},\phi_{(p)}^{(m)}\right)
}{
\pi\left(\gamma_i^{(m-1)}\mid\boldsymbol{\gamma}_{-i}^{(m^*)},\mu_{\gamma}^{(m)},\sigma_{\gamma}^{2(m)},\phi_{\gamma}^{(m)}\right)
\prod_{p=0}^{P}
\pi\left(\beta_{pi}^{(m-1)}\mid\boldsymbol{\beta}_{(p),-i}^{(m^*)},\mu_{(p)}^{(m)},\sigma_{(p)}^{2(m)},\phi_{(p)}^{(m)}\right)
}
\\
&\quad\times
\frac{
\widetilde{\pi}\left(\gamma_i^{(m-1)}\right)
\prod_{p=0}^{P}\widetilde{\pi}\left(\beta_{pi}^{(m-1)}\right)
}{
\widetilde{\pi}\left(\gamma_i^*\right)
\prod_{p=0}^{P}\widetilde{\pi}\left(\beta_{pi}^*\right)
}.
\end{aligned}
\label{eq:R0-continuous}
\end{equation}
Here, the full-model conditional densities, $\pi(\cdot \mid \cdot)$, are the normal densities given in equations~\eqref{eq:continuous-gamma-conditional} and~\eqref{eq:continuous-beta-conditional}, while the stage one priors, $\widetilde{\pi}(\cdot)$, are $\gamma_i\sim\mathrm{N}(0,10^2)$ and $\beta_{pi}\sim\mathrm{N}(0,10^2)$. Thus, under the continuous model specification, the acceptance ratio in equation~\eqref{eq:R0-continuous} reduces to the expression given in equation~\eqref{eq:R-continuous} in the Appendix.

For stage two, the Metropolis-within-Gibbs MCMC was run for 200,000 iterations, with the first 50,000 iterations discarded as burn-in and a thinning interval of 50. This resulted in 3,000 posterior draws. The stage two computation required approximately 0.71 hours on a single node of the DEAC high-performance computing cluster.

\subsubsection{Results}

\begin{figure}
    \centering
    {\includegraphics[width=0.99\textwidth]{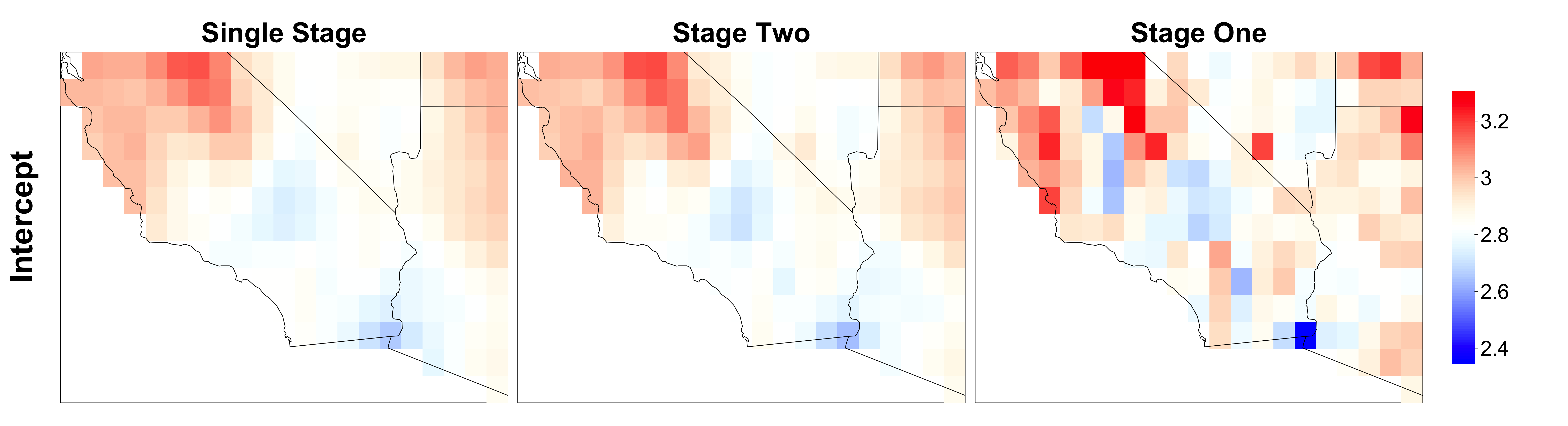}}  
    \par\vspace{1mm}
    {\includegraphics[trim={0 0 0 85mm},clip, width=0.99\textwidth]{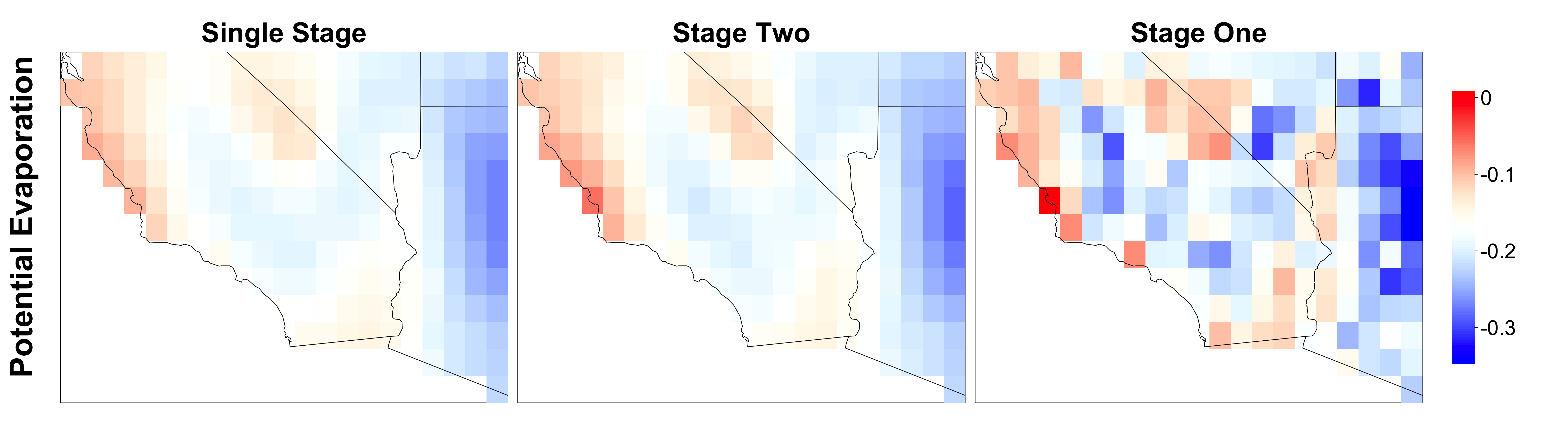}} 
    \par\vspace{1mm}
    {\includegraphics[trim={0 0 0 85mm},clip,width=0.99\textwidth]{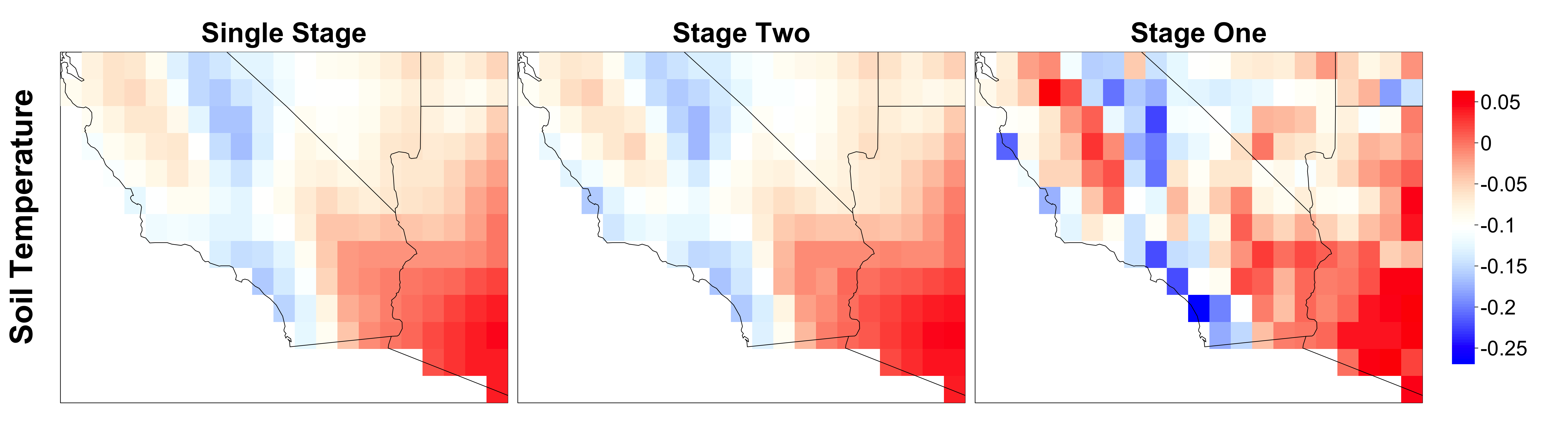}}
    \par\vspace{1mm}
    {\includegraphics[trim={0 0 0 85mm},clip,width=0.99\textwidth]{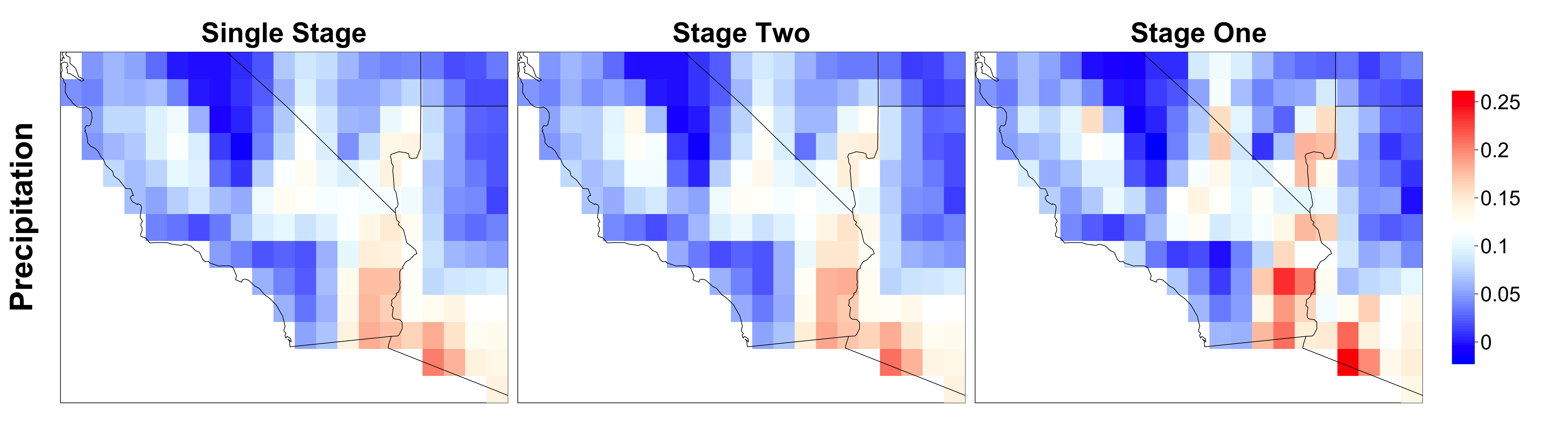}}
    \par\vspace{1mm}
    {\includegraphics[trim={0 0 0 85mm},clip,width=0.99\textwidth]{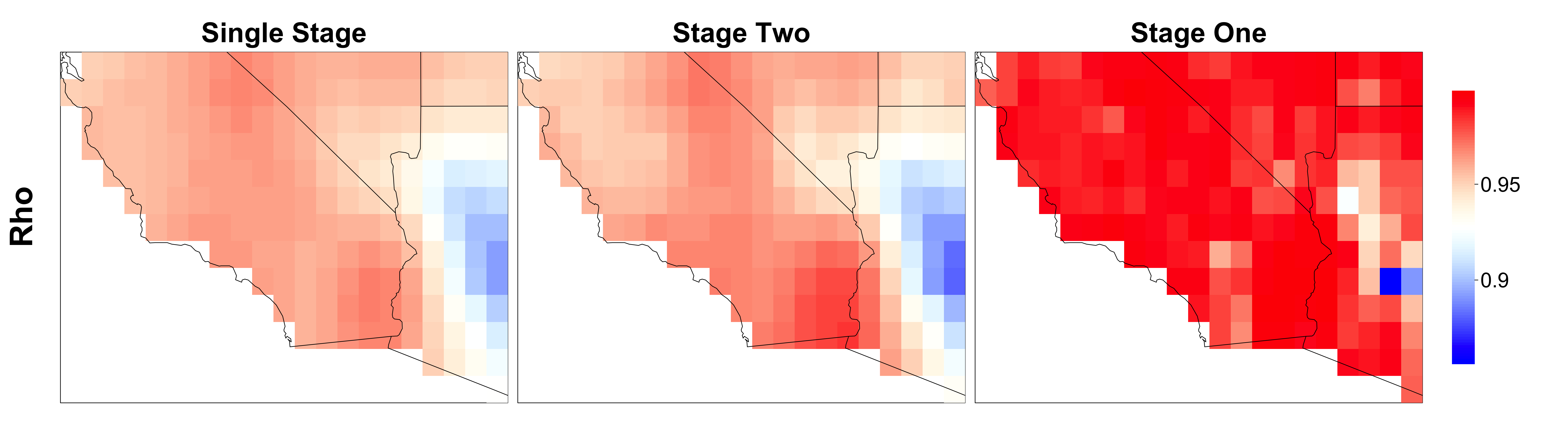}} 

    \caption{Posterior means of the location-specific parameters assigned spatial MVN priors in the continuous model under the single-stage method (left column), the two-stage method (center column), and after stage one only (right column). From top to bottom, the rows show the intercept $\beta_0$, potential evaporation effect $\beta_{\mathrm{pevap}}$, soil temperature effect $\beta_{\mathrm{tsoil}}$, precipitation effect $\beta_{\mathrm{apcp}}$, and autoregressive parameter $\rho$.}
    \label{fig:continuous-mean}
\end{figure}

For the continuous-model application, we considered grid cells located west of $112^\circ$ longitude and between $31^\circ$ and $38^\circ$ latitude. The study period extended from January 2019 through December 2020. The resulting dataset contained $I=187$ locations observed over $T=104$ weeks. Figure~\ref{fig:continuous-mean} compares the posterior means of the location-specific parameters assigned spatial MVN priors in the full continuous model across the single-stage method, the two-stage method, and stage one alone, whereas Figure~\ref{fig:continuous-sd} in the Appendix compares the corresponding posterior standard deviations. 

The single-stage posterior summaries, shown in the left column, are based on samples from the target posterior distribution, and their close agreement with the stage two results in the center column provides evidence that the two-stage algorithm is sampling from that same posterior distribution. The stage two posterior mean surfaces for the intercept, covariate effects, and autoregressive parameter show nearly identical spatial patterns to those from the single-stage model. Furthermore, the similarity in the posterior standard deviation maps (Figure~\ref{fig:continuous-sd}) demonstrates that the two-stage method closely reproduces the uncertainty of the full spatial model. Comparing the stage one and stage two results illustrates the effect of incorporating the spatial MVN priors, with stage two producing smoother posterior mean surfaces and lower posterior uncertainty. Overall, the strong agreement between the single-stage and stage two results demonstrates that both computational approaches are sampling from the same target posterior distribution, while the two-stage procedure avoids direct fitting of the computationally intensive full continuous model.

The coefficients $\beta_{\mathrm{pevap}}$, $\beta_{\mathrm{tsoil}}$, and $\beta_{\mathrm{apcp}}$ quantify how potential evaporation, soil temperature, and precipitation are related to the log expected soil moisture after adjusting for temporal dependence. Positive coefficient values (Figure \ref{fig:continuous-mean}) indicate that higher values of the corresponding covariate are associated with greater expected soil moisture, whereas negative values indicate lower expected soil moisture. The intercept $\beta_0$ represents the baseline log expected soil moisture. The autoregressive parameter $\rho$ measures the strength of week-to-week dependence, with larger values indicating stronger temporal dependence.

Furthermore, Figures~\ref{fig:continuous-mean-alpha} and~\ref{fig:continuous-sd-alpha} in the Appendix show the posterior means and posterior standard deviations, respectively, of the gamma shape parameter, which is not assigned a spatial MVN prior. Trace plots of location-specific parameters for one grid cell from the single-stage method, the two-stage method, and stage one alone are presented in Figure~\ref{fig:continuous-trace} in the Appendix.

We further evaluated the computational performance of the single-stage and two-stage continuous models using effective sample size (ESS) per hour. Table~\ref{tab:ess-continuous} shows that the two-stage method achieved higher ESS per hour for all regression coefficients and the autoregressive parameter. The ESS per hour improved by a factor of approximately 5.8 to 9.4 times. In addition, the overall computation time decreased from approximately 48.30 hours for the single-stage model to about 0.74 hours for the two-stage implementation. Thus, the two-stage approach provides substantially more effective posterior draws per unit of computing time while maintaining close agreement with the single-stage posterior results.

\begin{table}[!h]
    \centering
    \begin{tabular}{c|c|c}
        & \text{Single-stage model} & \text{Two-stage model} \\
        \hline
        Parameter & ESS per hour & ESS per hour \\
        \hline
        $\beta_0$                & 37.2 & 319.1 \\
        $\beta_{\mathrm{pevap}}$ & 31.4 & 183.2 \\
        $\beta_{\mathrm{tsoil}}$ & 32.7 & 222.5 \\
        $\beta_{\mathrm{apcp}}$  & 52.4 & 492.7 \\
        $\rho$                   & 29.1 & 167.6 \\
        $\alpha$                 & 26.5 & 528.1 
    \end{tabular}

    \caption{Mean effective sample sizes per hour of computing time for the location-specific regression coefficients, autoregressive parameter, and gamma shape parameter under the single-stage and two-stage continuous models. Effective sample sizes per hour are averaged over all locations. The reported run times exclude model compilation time but include burn-in.}
    \label{tab:ess-continuous}
\end{table}


\subsection{Cast Study 3: Binary Model on a Spatial Lattice with a Latent Gaussian Process}
Let $Y_{i,t}\in\{0,1\}$ denote the binary response at grid cell $i=1,\ldots,I=1354$ and time period $t=1,\ldots,T=104$, where $Y_{i,t}=1$ indicates a drought and $Y_{i,t}=0$ indicates no drought. Here, we define location $i$ as experiencing drought during time period $t$ if its category of the United States Drought Monitor (USDM) is D1, D2, D3, or D4, and as experiencing no drought if its category is 0 or D0. We assume that the distribution of $Y_{i,t}$ depends on a latent, continuous, unobserved random variable $Z_{i,t}$ where $Y_{i,t} = I(Z_{i,t} > 0)$, with
\begin{equation} 
Z_{i,t} = \begin{cases} 
 \bfX_{i,t} \bfbeta_{i} + \epsilon_{i,t} & \text{ for } t=1 \\
 \bfX_{i,t} \bfbeta_{i} + \rho_i \left(Z_{i, t-1} -\bfX_{i,t-1} \bfbeta_{i} \right) + \epsilon_{i,t}
 & \text{ for } t > 1.
\end{cases}
\label{eq:Zmod}
\end{equation}
We define $\bfX_{i,t}$ as the $(P+1)$-dimensional vector of covariates, which includes the $1$ for an intercept term and $P=3$ environmental covariates: weekly total precipitation, weekly mean soil moisture content, and weekly mean soil temperature. Before fitting the models, each covariate was standardized to have a mean of zero and unit variance. Let $\bfbeta_{i} = (\beta_{0i}, \beta_{1i}, \ldots, \beta_{Pi})'$ be the corresponding location-specific vector of regression coefficients. In this model, temporal dependence is captured through an AR(1) term with parameter $\rho_i \in (0,1)$. Observe the centering in equation $\eqref{eq:Zmod}$, where the previous time period's latent variable $Z_{i,t-1}$ is adjusted by $\bfX_{i,t-1} \bfbeta_{i}$ before being multiplied by $\rho_i$.  This model treats exceedances or shortfalls of $Z_{i,t-1}$ relative to its expected value as influencing future time periods.  A practical interpretation of this is that autoregression captures time-dependence after covariates are considered.  Finally, the errors are assumed to be normally distributed and independent as $\epsilon_{i,t} \overset{ind}{\sim} N(0, 1)$.  This error term implies that, for $t>1$, the latent variable $Z_{i,t}$ follows a Gaussian distribution with mean $\bfX_{i,t} \bfbeta_{i} + \rho_i \left(Z_{i, t-1} -\bfX_{i,t-1} \bfbeta_{i} \right)$ and variance $1$. Fixing the latent variance at $1$ results in no loss of generality because $Z_{i,t}$ is a latent unobserved variable and the response is binary, depending only on whether $Z_{i,t}$ lies above or below zero.

We define the $p$th regression coefficient across all locations as $\bfbeta_{(p)} = (\beta_{p1}, \ldots, \beta_{pI})'$ for $p = 0, \ldots, P$. In addition, since $\rho_i$ is restricted to the interval $(0,1)$, we transform it to the real line by defining $ \gamma_i = \log\left(\rho_i/(1-\rho_i)\right)$, which implies $\rho_i = e^{\gamma_i}/(1+e^{\gamma_i})$. To capture spatial dependence across locations, we assign independent ICAR priors, as defined for the count model in equation~\eqref{eq:icar}, to each regression coefficient vector $\bfbeta_{(p)}=(\beta_{p1},\ldots,\beta_{pI})'$, for $p=0,\ldots,P$, and to the transformed autoregressive parameter vector $\bfgamma=(\gamma_1,\ldots,\gamma_I)’$. Specifically, each $\bfbeta_{(p)}$ is assigned an ICAR prior with variance parameter $\sigma^2_{(p)}$ and the vector $\bfgamma$ is assigned an ICAR prior with variance parameter $\sigma_\gamma^2$. As in the count model, independent weakly informative conjugate priors, $\mathrm{Inverse\text{-}Gamma}(0.5,0.5)$, were assigned to the ICAR variance parameters $\sigma_\gamma^2$ and $\sigma_{(p)}^2$, for $p=0,\ldots,P$. 

Let $\bfZ_{1:T}$ denote the $IT$-dimensional vector of latent variables, and let $\bftheta_Z$ denote the vector of location-specific parameters. The full parameter vector $\left( \bftheta_Z, \bfphi \right)$ consists of all location-specific parameters together with the spatial dependence hyperparameters. For the binary model, we write $\bftheta_Z=\{\bftheta_{Z,i}:i=1,\ldots,I\}$, where $\bftheta_{Z,i}=(\bfbeta_i,\gamma_i)$. The spatial dependence hyperparameter vector contains the variance parameters associated with the ICAR priors and is given by $\bfphi = \left(\sigma^2_{\gamma}, \{\sigma^2_{(p)}, p=0,\ldots, P\}\right)$. 

Following the general formulation in equations~\eqref{eq:BHM} and~\eqref{eq:singlestagepost}, and accounting for the latent process $\bfZ_{1:T}$ in the binary model, the posterior distribution is proportional to
\begin{equation}
   \pi\left(\bfZ_{1:T}, \bftheta_Z, \bfphi \mid \bfY_{1:T}, \bfX_{1:T} \right)  \propto f\left(\bfY_{1:T} \mid \bfZ_{1:T}\right) \pi\left(\bfZ_{1:T} \mid \bftheta_Z, \bfX_{1:T}\right) \pi\left(\bftheta_Z \mid \bfphi\right) \pi\left(\bfphi\right).
   \label{eq:singlestagepost-binary}
\end{equation}

We fit the computationally intensive single-stage model using NIMBLE. Specifically, we ran one chain for 50,000 iterations, discarding the first 10,000 draws as a burn-in and thinning by 5. Total compute time (compiling and running) for these data on one node of the DEAC high-performance computing cluster was 101.18 hours.

\subsubsection{Stage One}
The stage one model treats locations as independent. In this stage, the hyperparameters are temporarily excluded, so the full parameter vector consists only of $\bftheta_Z$. The data and latent process models remain unchanged from the full hierarchical model; only the prior specification for $\bftheta_Z$ is modified to remove spatial dependence.

For this stage only, the spatially varying regression coefficients are assigned independent priors $\beta_{pi} \sim N(0,3^2)$, for $p=0,\ldots, P$ and $i=1,\ldots, I$, rather than ICAR priors. As a result, no spatial smoothing is introduced at this stage. Similarly, the transformed location-specific autoregressive parameters $\gamma_i$ are assumed to be independent across locations. Recall that $\gamma_i=\log\left(\rho_i/(1-\rho_i)\right)$, which maps $\rho_i \in (0,1)$ to $\gamma_i \in \mathbb{R}$. In the stage one model, we assume $\gamma_i \sim \mathrm{Logistic}(0,1)$ for $i=1,\ldots,I$. Under the inverse transformation $\rho_i = e^{\gamma_i}/(1+e^{\gamma_i})$, this specification is equivalent to $\rho_i \sim \mathrm{U}(0,1)$.

These modeling changes in stage one eliminate all spatial dependence across locations. Let $\tilde{\pi}\left(\bftheta_Z\right)$ be the prior distribution assigned to $\bftheta_Z$ in stage one, and let $\tilde{\pi}\left(\bftheta_Z, \bfZ_{1:T} \mid \bfY_{1:T}, \bfX_{1:T}\right)$ denote the resulting stage one posterior distribution. Under the stage one specification, the joint posterior separates into a product of $I$ location-specific posterior densities as
\begin{equation}
        \begin{aligned}
            \tilde{\pi}\left(\bftheta_Z, \bfZ_{1:T} \mid \bfY_{1:T}, \bfX_{1:T}\right) \propto \prod_{i=1}^I f\left(\bfY_{i,1:T} \mid \bfZ_{i,1:T}\right) \pi\left(\bfZ_{i,1:T} \mid \bftheta_{Z,i}, \bfX_{i,1:T}\right) \tilde{\pi}\left(\bftheta_{Z,i}\right).
            \label{eq:stageonepost-binary}
        \end{aligned}
\end{equation}

This factorization allows efficient parallel posterior sampling across locations. For each location $i$, we ran the stage one MCMC with 100,000 draws, discarding the first 50,000 as a burn-in, and thinning by 10, resulting in 5,000 draws of the stage one posteriors for each location.  Computing time was around 6 minutes per location.

\subsubsection{Stage Two} \label{sec:stage2-binary}
Stage two uses a Metropolis-within-Gibbs algorithm to target the full binary model posterior. The spatial hyperparameters $\bfphi$ are updated using Gibbs steps, whereas the location-specific latent variables and parameters $\left(\bfZ_{i,1:T},\bftheta_{Z,i}\right)$ are updated jointly for each location using a Metropolis-Hastings proposal drawn from the corresponding stage one posterior sample. The steps at iteration $m$ are described below.

First, update the ICAR variance hyperparameters $\boldsymbol{\phi}$ by Gibbs sampling from their conjugate full conditional distributions. Because $\sigma_{\gamma}^{2}$ and $\sigma_{(p)}^{2}$, for $p=0,\ldots,P$, have independent $\mathrm{Inverse\text{-}Gamma}(0.5,0.5)$ priors, these updates follow those described in Section~\ref{sec:stage2-count}. Let $\boldsymbol{\phi}^{(m)}=\left(\sigma_{\gamma}^{2(m)},\sigma_{(0)}^{2(m)},\ldots,\sigma_{(P)}^{2(m)}\right)'$ denote the updated spatial variance hyperparameters at iteration $m$.

Next, update $\left(\bfZ_{i,1:T},\bftheta_{Z,i}\right)$ sequentially for $i=1,\ldots,I$. From equation \eqref{eq:singlestagepost-binary}, the full conditional distribution, and thus the Metropolis-Hastings target for location $i$, is
\begin{equation}
\begin{aligned}
\pi\left(\bfZ_{i,1:T},\bftheta_{Z,i}\mid\bftheta_{Z,-i},\boldsymbol{\phi},\bfY_{1:T},\bfX_{1:T}\right) 
\propto
f\left(\bfY_{i,1:T}\mid\bfZ_{i,1:T}\right)
\pi\left(\bfZ_{i,1:T}\mid\bftheta_{Z,i},\bfX_{i,1:T}\right)
\pi\left(\bftheta_{Z,i}\mid\bftheta_{Z,-i},\boldsymbol{\phi}\right),
\end{aligned}
\label{eq:fullpost-target-binary}
\end{equation}
For location $i$, propose $\left(\bfZ_{i,1:T}^*,\bftheta_{Z,i}^*\right)$ by drawing from the corresponding stage one posterior. By the factorization in equation~\eqref{eq:stageonepost-binary}, the proposal distribution is
\begin{equation}
 \begin{aligned}  
     \tilde{\pi}\left(\bfZ_{i,1:T}^*, \bftheta_{Z,i}^* \mid \bfY_{i,1:T}, \bfX_{i,1:T}\right) 
     \propto f(\bfY_{i,1:T} \mid \bfZ_{i,1:T}^*) \pi(\bfZ_{i,1:T}^* \mid \bftheta_{Z,i}^*, \bfX_{i,1:T}) \tilde{\pi}(\bftheta_{Z,i}^*).
 \end{aligned}
\label{eq:proposal-binary}
\end{equation}

Let $\left(\bfZ_{j,1:T}^{(m^*)},\bftheta_{Z,j}^{(m^*)}\right)$ denote the current latent variables and parameters at location $j$ when location $i$ is updated at iteration $m$. Because locations are updated sequentially, $m^*=m$ for $j<i$ and $m^*=m-1$ for $j > i$. From equations~\eqref{eq:fullpost-target-binary} and~\eqref{eq:proposal-binary}, the proposal at location $i$ is accepted with probability $\min\{1,R_i\}$, where
\begin{equation}
    \begin{aligned}
        R_i 
        &= \frac{
        \pi\left(\gamma_i^*\mid\bfgamma_{-i}^{(m^*)},\sigma_\gamma^{2(m)}\right)
        \prod_{p=0}^P
        \pi\left(\beta_{pi}^*\mid\bfbeta_{p,-i}^{(m^*)},\sigma_{(p)}^{2(m)}\right)}
        {
        \pi\left(\gamma_i^{(m-1)}\mid\bfgamma_{-i}^{(m^*)},\sigma_\gamma^{2(m)}\right)
        \prod_{p=0}^P
        \pi\left(\beta_{pi}^{(m-1)}\mid\bfbeta_{p,-i}^{(m^*)},\sigma_{(p)}^{2(m)}\right)}
        \times
        \frac{
        \widetilde{\pi}\left(\gamma_i^{(m-1)}\right)
        \prod_{p=0}^P
        \widetilde{\pi}\left(\beta_{pi}^{(m-1)}\right)}
        {
        \widetilde{\pi}\left(\gamma_i^*\right)
        \prod_{p=0}^P
        \widetilde{\pi}\left(\beta_{pi}^*\right)}
    \end{aligned}
    \label{eq:R0-binary}
\end{equation}
If the proposal is accepted, then $\left(\bfZ_{i,1:T}^{(m)},\bftheta_{Z,i}^{(m)}\right) = \left(\bfZ_{i,1:T}^*,\bftheta_{Z,i}^*\right)$; otherwise, $\left(\bfZ_{i,1:T}^{(m)},\bftheta_{Z,i}^{(m)}\right) = \left(\bfZ_{i,1:T}^{(m-1)},\bftheta_{Z,i}^{(m-1)}\right)$.

The data-model and latent-process terms cancel from $R_i$, leaving only the spatial conditional priors and stage one priors. In particular, evaluating the latent-process density $\pi\left(\bfZ_{i,1:T}\mid\bftheta_{Z,i},\bfX_{i,1:T}\right)$ is unnecessary, substantially reducing computation when $T$ is large. In equation~\eqref{eq:R0-binary}, $\pi(\cdot\mid\cdot)$ denotes the Gaussian ICAR conditional densities from equation~\eqref{eq:icar}, whereas $\widetilde{\pi}(\cdot)$ denotes the stage one priors: $N(0,3^2)$ for the regression coefficients and $\mathrm{Logistic}(0,1)$ for the transformed autoregressive parameters. Under these specifications, equation~\eqref{eq:R0-binary} reduces to equation~\eqref{eq:R-binary} in the Appendix.

We ran the stage two Metropolis-within-Gibbs MCMC for 100,000 draws, discarding the first 10,000 as a burn-in and thinning by 5. Using a single core on the DEAC high-performance computing cluster, this second stage took 2.47 hours to run.

\subsubsection{Results}

\begin{figure}
    \centering
    {\includegraphics[width=0.99\textwidth, height=4cm]{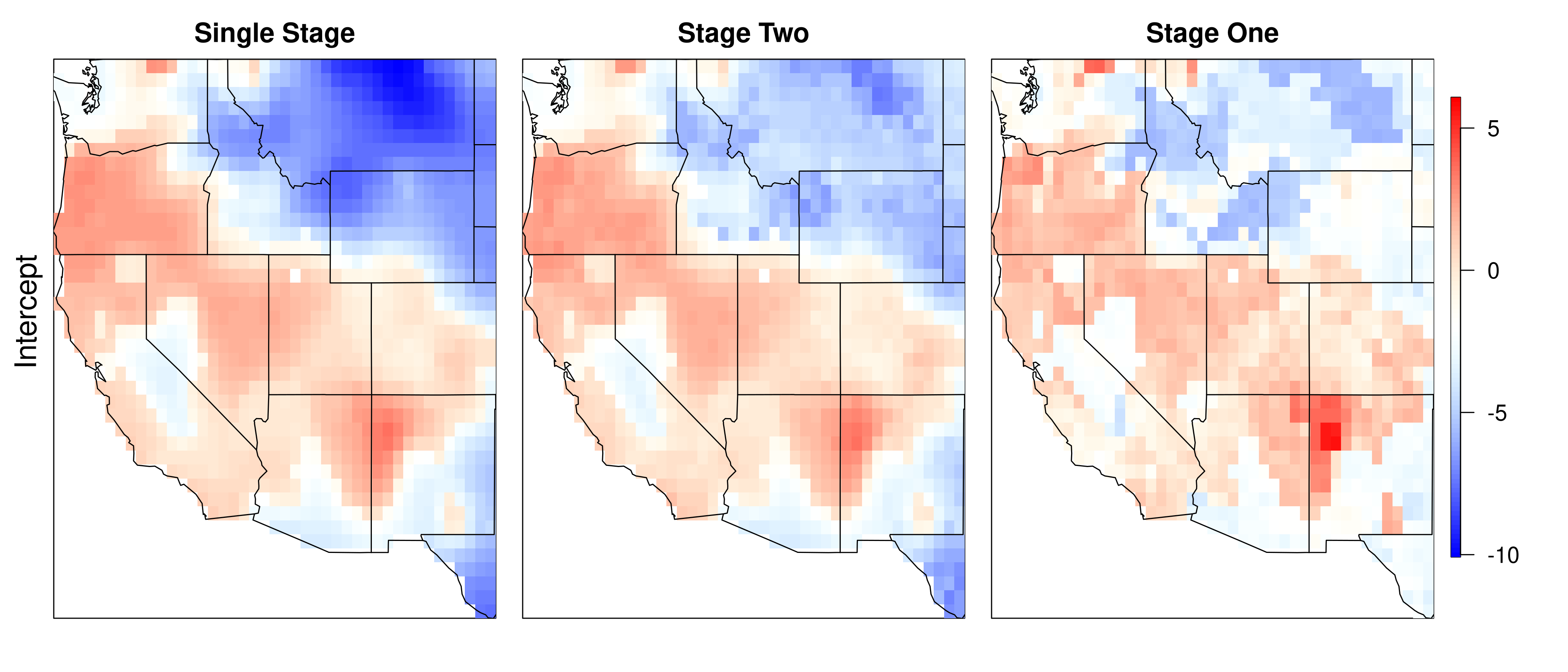}}  
    {\includegraphics[trim={0 0 0 11mm},clip, width=0.99\textwidth, height=4cm]{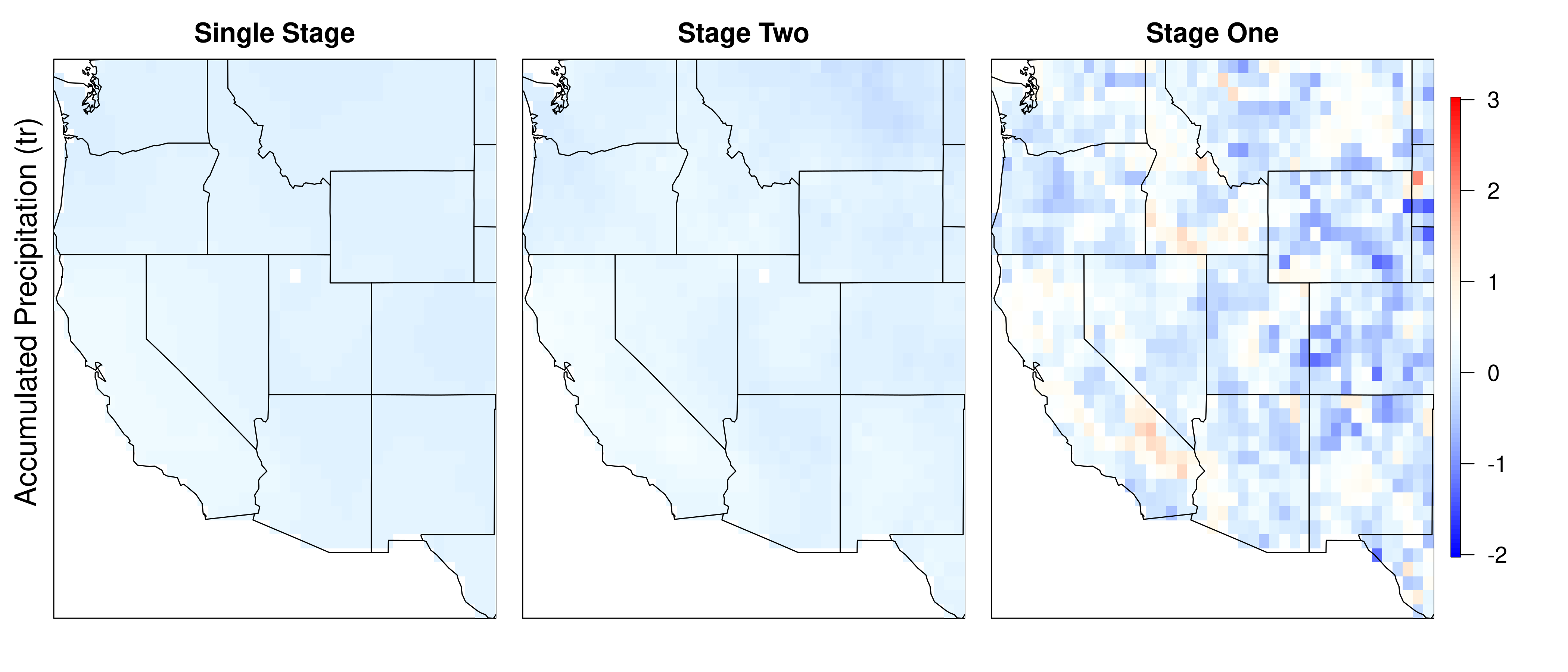}} 
    {\includegraphics[trim={0 0 0 11mm},clip,width=0.99\textwidth, height=4cm]{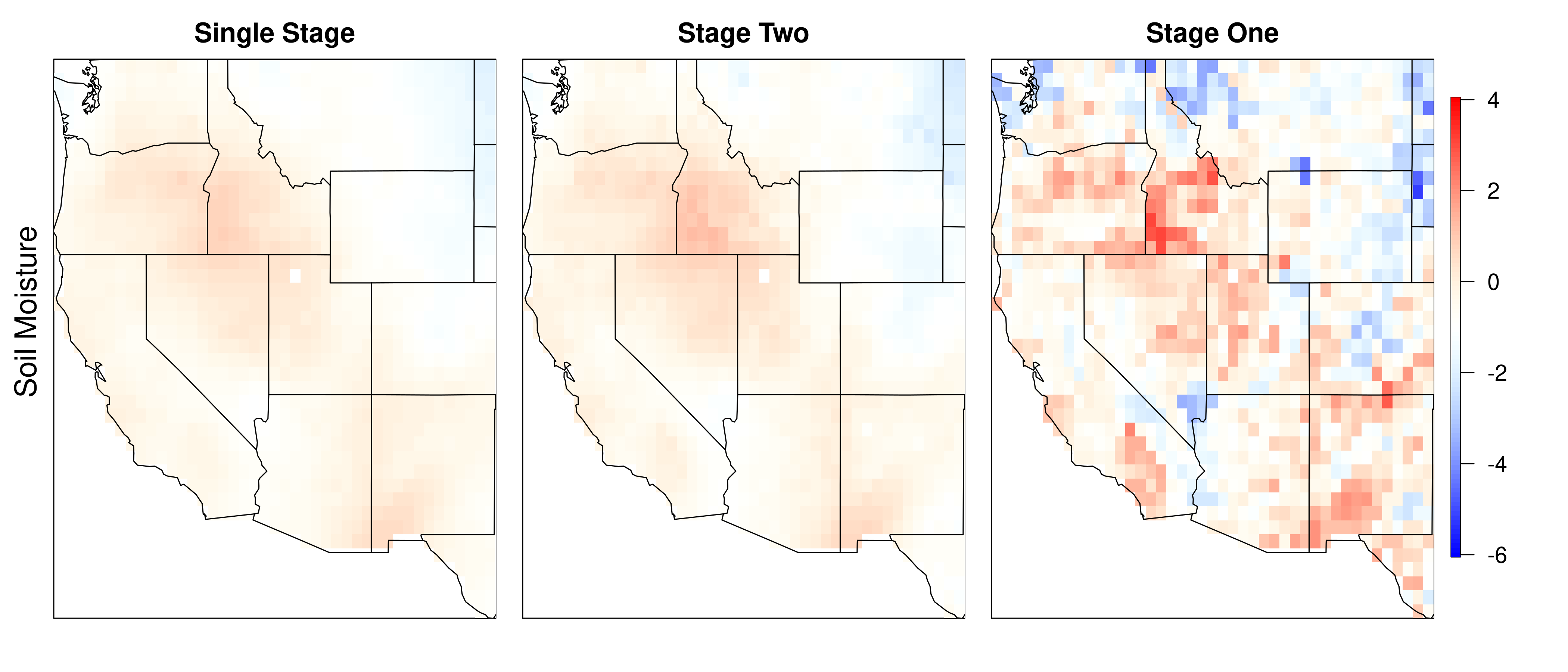}}
    {\includegraphics[trim={0 0 0 11mm},clip,width=0.99\textwidth, height=4cm]{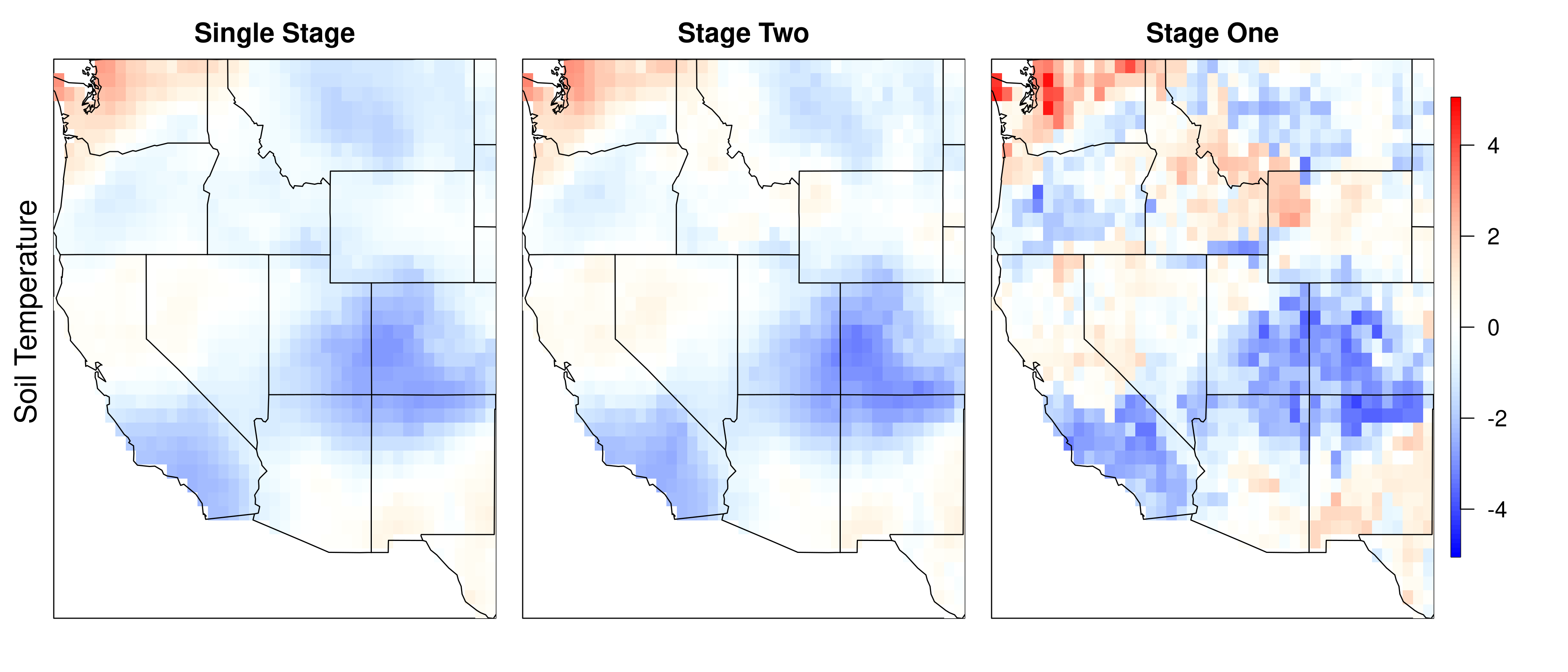}}
    {\includegraphics[trim={0 0 0 11mm},clip,width=0.99\textwidth, height=4cm]{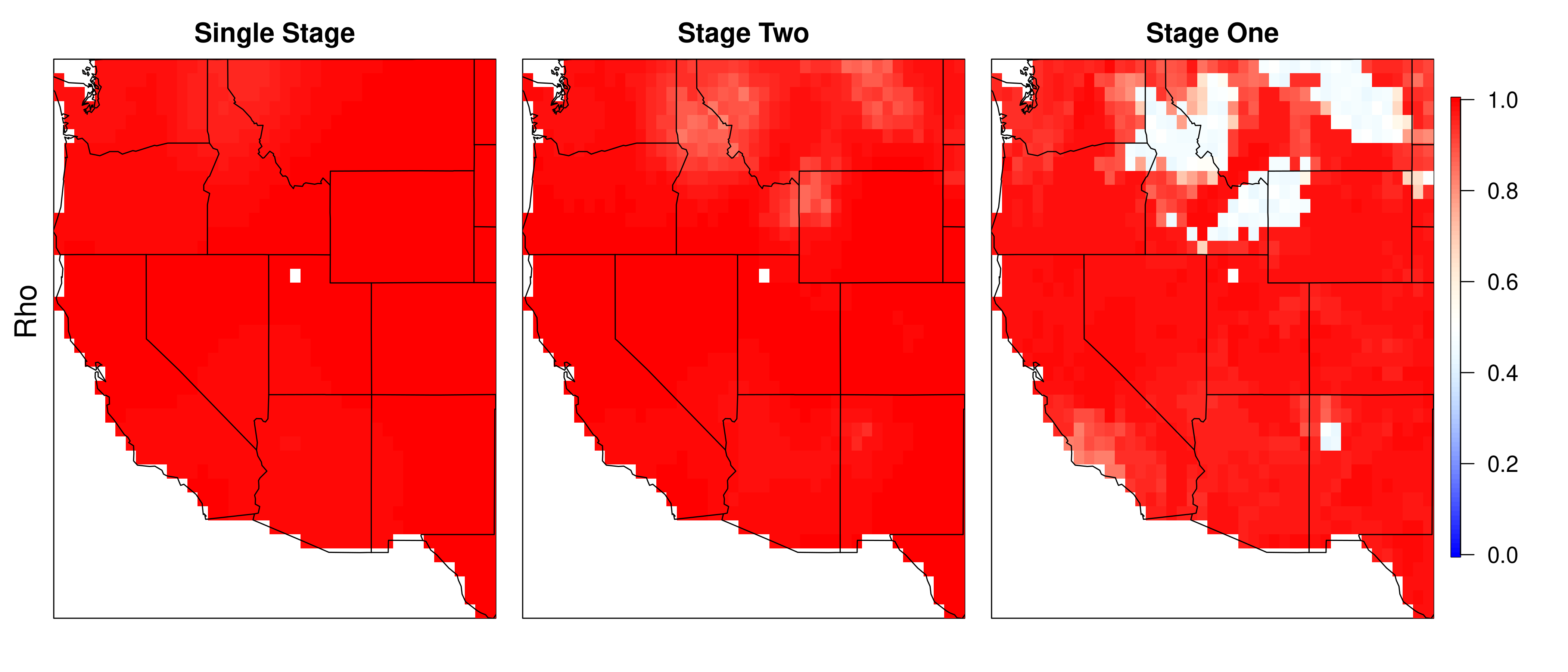}} 

    \caption{Posterior means of the location-specific parameters assigned spatial ICAR priors in the binary model taken from the single-stage method (left column), two-stage method (center column), and after stage one only (right column).  Parameters shown include: intercept $\beta_0$, effect of accumulated precipitation $\beta_{\mathrm{apcp}}$, effect of soil moisture $\beta_{\mathrm{soilm}}$, effect of soil temperature $\beta_{\mathrm{tsoil}}$, and  autoregression $\rho$.}.
    \label{fig:mean-binary}
\end{figure}

Figure \ref{fig:mean-binary} shows posterior means for the five model parameters, each obtained from the single-stage method (left), the two-stage method (center), and after stage one (right), whereas Figure \ref{fig:sd-binary} in the Appendix shows the corresponding posterior standard deviations. The close agreement between the center and left columns provides evidence that the two-stage and single-stage algorithms are sampling from the same target posterior distribution. Almost everywhere across all parameters, the two-stage and single-stage results show strong agreement. There are essentially no differences in posterior means or variances for any of the regression coefficients $\beta_{\mathrm{apcp}}$, $\beta_{\mathrm{soilm}}$, and $\beta_{\mathrm{tsoil}}$, while the posterior means and variances for $\rho$ and $\beta_0$ show overall agreement, with some mild localized differences discussed further in the Discussion and Appendix.

Turning next to compute times, we evaluate the single-stage and two-stage models by effective sample sizes (ESS) per hour. Table \ref{tab:ESS-binary} shows that ESS per hour is notably higher for the two-stage model. The total compute time is also much lower for the two-stage model (around 2.5 hours versus 100+ for single-stage); however, the mixing of stage two is worse than mixing in single-stage. This lowers the effective sample size due to autocorrelaiton and tempers the overall computational benefit somewhat.  Nevertheless, the two-stage model is often several times more efficient than the single-stage model when measured in ESS per hour over the single-stage, and achieves a very close approximation to the full single-stage posterior distribution.  

\begin{table}[!h]
    \centering
    \begin{tabular}{c|c|c}
    & {Single-stage model} & {Two-stage model}\\
    \hline
    Parameter & ESS per hour & ESS per hour\\
    \hline
        $\beta_0$ & 60.2 & 136.9 \\
        $\beta_{\mathrm{apcp}}$ & 23.4 & 95.0\\
        $\beta_{\mathrm{soilm}}$ & 38.5 & 109.7\\
        $\beta_{\mathrm{tsoil}}$ & 42.5 & 124.6\\
        $\rho$ & 15.0 & 131.1\\
    \end{tabular}   
    
    \caption{Effective sample sizes per hour of computing time for location-specific parameters in the binary model. Note these are averaged over all locations, and are run times only which do not include compile time but do include the burn-in time.}
    \label{tab:ESS-binary}
\end{table}

\section{Discussion} \label{sec:discussion}
In this paper, we presented three case studies of a two-stage framework for fitting spatio-temporal models to observed data.  The first stage fit independent models by location, and the second stage used proposals from the first stage to cancel out terms in the Metropolis-Hastings ratio, reducing the computational cost of fitting a Bayesian hierarchical model. We applied this two-stage approach across three settings, chosen to demonstrate its flexibility. These settings varied in the response variable (count, continuous, binary), the spatial support (areal, geostatistical), and the presence or absence of a latent variable and corresponding process layer in the Bayesian hierarchical model. The quality of the fit was assessed by visual comparisons of the single-stage and two-stage posterior distributions as well as trace plots, and computational gains were measured by improvements in effective sample sizes (ESS) per hour. These gains varied from roughly doubling to an order of magnitude improvement in ESS, depending on the parameter and the specific model (Tables~\ref{tab:ess-count},~\ref{tab:ess-continuous}, and~\ref{tab:ESS-binary}). 

A limitation of the two-stage approach is that the stage one models must be identifiable and must provide proposal distributions that are reasonably close to the corresponding full-model posterior distributions. If a stage one model is poorly informed by the data, its posterior may remain close to the prior, resulting in poor proposals during stage two and consequently slower mixing or lower acceptance rates. The most prominent example of this behavior can be seen in the binary case study for locations with no drought during the training period, where the autoregressive parameter $\rho$ was only weakly informed by the data in stage one. Some localized disagreement between the single-stage and stage two posterior means is visible around Idaho, Montana, and Wyoming, shown as the white region in the bottom right image of Figure \ref{fig:mean-binary}. In this region, several grid cells experienced no drought during the training period, and the stage one posterior for $\rho$ remained close to its prior, with a posterior mean near 0.5 (the prior mean). This led to poor proposal draws and weaker mixing during stage two. We explore this behavior in more detail for two representative grid cells in Section~\ref{sec:binary-grid-cells} of the Appendix: G21, located on the border between Idaho and Montana, and GG16, located in Southern California. This example demonstrates the importance of obtaining a reasonable stage one approximation for efficient performance of the two-stage methodology.

\section{Data and Code Availability}
All R code used to fit the models is available in a GitHub repository at \url{https://github.com/m0stafa-shams/two-stage-mcmc}. All drought and environmental data are freely and publicly available at \url{https://datadryad.org/dataset/doi:10.5061/dryad.g1jwstqw7}. \textit{Coccidioidomycosis} reported case counts were provided by the Centers for Disease Control and Prevention, through the National Notifiable Diseases Surveillance System (NNDSS), 2000 - 2022, as compiled by the National Center for Emerging and Zoonotic Infectious Diseases.  These data are governed under a data use agreement, and accordingly are not publicly available.  Please note that this means that readers can only replicate applications 2 and 3 from this paper.  

\section{Conflict of Interest}
The authors declare no conflicts of interest.

\section{Acknowledgments}
The authors acknowledge assistance from Dr. Mitsuru Toda at the National Center for Emerging and Zoonotic Infections Diseases at the Centers for Disease Control and Prevention, in providing raw case counts of cocci as well as oversight in the count model case study of section 4.1.  The authors also thank Ria Ghai and Seoyun Choe at the Mycotic Diseases Branch for helpful comments on this manuscript. 

\section{Funding Statement}
The authors acknowledge support from NSF award \#2151881.

{\small
\bibliographystyle{jasa}
\bibliography{ref} 
}


\clearpage
\appendix
\counterwithin{figure}{section}
\section{Appendix/Supplemental}

\subsection{Acceptance ratio for the count model} \label{sec:app-count-ratio}
The spatial ICAR priors in the full count model are assigned to $\bfbeta_{(p)}$, for $p=0,\ldots,P$, and $\bfgamma$. In stage one, these corresponding county-specific parameters are instead assigned the independent priors $\beta_{0i} \sim N(\mu_0,5^2)$, $\beta_{pi} \sim N(0,2^2)$, for $p=1,\ldots,P$, and $\gamma_i\sim\mathrm{Logistic}(0,0.5)$. Therefore, under the count model specification, the acceptance ratio in equation~\eqref{eq:R0-count} simplifies to
\begin{equation}
\begin{aligned}
R_i
&=
\frac{
\pi\left(\gamma_i^*\mid\bfgamma_{-i}^{(m^*)},\sigma_\gamma^{2(m)}\right)
\prod_{p=0}^P
\pi\left(\beta_{pi}^*\mid\bfbeta_{p,-i}^{(m^*)},\sigma_{(p)}^{2(m)}\right)}
{
\pi\left(\gamma_i^{(m-1)}\mid\bfgamma_{-i}^{(m^*)},\sigma_\gamma^{2(m)}\right)
\prod_{p=0}^P
\pi\left(\beta_{pi}^{(m-1)}\mid\bfbeta_{p,-i}^{(m^*)},\sigma_{(p)}^{2(m)}\right)}
\times
\frac{
\widetilde{\pi}\left(\gamma_i^{(m-1)}\right)
\prod_{p=0}^P
\widetilde{\pi}\left(\beta_{pi}^{(m-1)}\right)}
{
\widetilde{\pi}\left(\gamma_i^*\right)
\prod_{p=0}^P
\widetilde{\pi}\left(\beta_{pi}^*\right)}
\\[4pt]
&=
\frac{
\exp\left[
-\frac{a_{i+}}{2\sigma_\gamma^{2(m)}}
\left(\gamma_i^*-\bar{\gamma}_{j\sim i}^{(m^*)}\right)^2
\right]}
{
\exp\left[
-\frac{a_{i+}}{2\sigma_\gamma^{2(m)}}
\left(\gamma_i^{(m-1)}-\bar{\gamma}_{j\sim i}^{(m^*)}\right)^2
\right]}
\times
\frac{
\exp\left(-2\gamma_i^{(m-1)}\right)
\left[1+\exp\left(-2\gamma_i^*\right)\right]^2}
{
\left[1+\exp\left(-2\gamma_i^{(m-1)}\right)\right]^2
\exp\left(-2\gamma_i^*\right)}
\\[4pt]
&\qquad\times
\frac{
\prod_{p=0}^P
\exp\left[
-\frac{a_{i+}}{2\sigma_{(p)}^{2(m)}}
\left(\beta_{pi}^*-\bar{\beta}_{p,j\sim i}^{(m^*)}\right)^2
\right]}
{
\prod_{p=0}^P
\exp\left[
-\frac{a_{i+}}{2\sigma_{(p)}^{2(m)}}
\left(\beta_{pi}^{(m-1)}-\bar{\beta}_{p,j\sim i}^{(m^*)}\right)^2
\right]}
\\[4pt]
&\qquad\times
\frac{
\exp\left[
-\frac{1}{2(5^2)} \left(\beta_{0i}^{(m-1)} - \mu_0\right)^2
\right]
\prod_{p=1}^P
\exp\left[
-\frac{1}{2(2^2)} \left(\beta_{pi}^{(m-1)}\right)^2
\right]}
{
\exp\left[
-\frac{1}{2(5^2)} \left(\beta_{0i}^* - \mu_0\right)^2
\right]
\prod_{p=1}^P
\exp\left[
-\frac{1}{2(2^2)} \left(\beta_{pi}^*\right)^2
\right]},
\end{aligned}
\label{eq:R-count}
\end{equation}
where $j \sim i$ indexes the neighbors of county $i$, and $\bar{\gamma}_{j\sim i}^{(m^*)} = \frac{1}{a_{i+}}\sum_{\ell=1}^I a_{i\ell}\gamma_\ell^{(m^*)}$ and $\bar{\beta}_{p,j\sim i}^{(m^*)} = \frac{1}{a_{i+}}\sum_{\ell=1}^I a_{i\ell}\beta_{p\ell}^{(m^*)}$ denote the means of the current transformed autoregressive parameters and the $p$th regression coefficients, respectively, over the neighbors of county $i$.

\subsection{Stage Two Hyperparameter Updates for the Continuous Model} \label{sec:continuous-hyperparameters-updates}

Stage two targets the posterior distribution under the full continuous spatial model. At iteration $m$ of the MCMC algorithm, the spatial hyperparameters associated with the MVN priors, which appear only in the full model, are updated first. These updates are followed by sequential Metropolis-Hastings updates of the location-specific parameters. Unlike the count model, which uses ICAR priors, the continuous model uses MVN priors with unknown spatial means, variances, and decay parameters. Consequently, stage two requires updates of $\mu_{\gamma}$, $\sigma_{\gamma}^2$, and $\phi_{\gamma}$ for the transformed autoregressive parameter vector, together with $\mu_{(p)}$, $\sigma_{(p)}^2$, and $\phi_{(p)}$ for each regression coefficient vector, for $p=0,\ldots,P$.

Because conjugate normal and inverse-gamma priors are assigned to the spatial means and variances, respectively, these parameters can be updated directly using Gibbs sampling. For notational convenience, let $\mathbf{Q}_{\gamma}=\mathbf{R}_{\phi_{\gamma}}^{-1}$ and $\mathbf{Q}_{(p)}=\mathbf{R}_{\phi_{(p)}}^{-1}$. Since $\mu_{\gamma}\sim\mathrm{N}(0,10^2)$ and $\boldsymbol{\gamma}\sim\mathrm{MVN}(\mu_{\gamma}\mathbf{1},\sigma_{\gamma}^2\mathbf{R}_{\phi_{\gamma}})$, the full conditional distribution of $\mu_{\gamma}$ is
\begin{equation*}
\mu_{\gamma}\mid\boldsymbol{\gamma},\sigma_{\gamma}^2,\phi_{\gamma}
\sim
\mathrm{N}\left(m_{\mu_{\gamma}},V_{\mu_{\gamma}}\right),
\label{eq:continuous-mu-gamma-update}
\end{equation*}
where
\[
V_{\mu_{\gamma}}
=
\left(
\frac{1}{10^2}
+
\frac{\mathbf{1}'\mathbf{Q}_{\gamma}\mathbf{1}}{\sigma_{\gamma}^{2}}
\right)^{-1},
\qquad
m_{\mu_{\gamma}}
=
V_{\mu_{\gamma}}
\frac{\mathbf{1}'\mathbf{Q}_{\gamma}\boldsymbol{\gamma}}{\sigma_{\gamma}^{2}}.
\]
Similarly, for each $p=0,\ldots,P$,
\begin{equation*}
\mu_{(p)}\mid\boldsymbol{\beta}_{(p)},\sigma_{(p)}^2,\phi_{(p)}
\sim
\mathrm{N}\left(m_{\mu_{(p)}},V_{\mu_{(p)}}\right),
\label{eq:continuous-mu-beta-update}
\end{equation*}
with
\[
V_{\mu_{(p)}}
=
\left(
\frac{1}{10^2}
+
\frac{\mathbf{1}'\mathbf{Q}_{(p)}\mathbf{1}}{\sigma_{(p)}^{2}}
\right)^{-1},
\qquad
m_{\mu_{(p)}}
=
V_{\mu_{(p)}}
\frac{\mathbf{1}'\mathbf{Q}_{(p)}\boldsymbol{\beta}_{(p)}}{\sigma_{(p)}^{2}}.
\]

The spatial variance parameters have conjugate inverse-gamma full conditional distributions. Under the prior $\sigma_{\gamma}^2\sim\mathrm{Inverse\text{-}Gamma}(2,1)$,
\begin{equation*}
\sigma_{\gamma}^2
\mid
\boldsymbol{\gamma},\mu_{\gamma},\phi_{\gamma}
\sim
\mathrm{Inverse\text{-}Gamma}
\left(
2+\frac{I}{2},
1+\frac{1}{2}
(\boldsymbol{\gamma}-\mu_{\gamma}\mathbf{1})'
\mathbf{Q}_{\gamma}
(\boldsymbol{\gamma}-\mu_{\gamma}\mathbf{1})
\right).
\label{eq:continuous-sigma-gamma-update}
\end{equation*}
Likewise, for $p=0,\ldots,P$,
\begin{equation*}
\sigma_{(p)}^2
\mid
\boldsymbol{\beta}_{(p)},\mu_{(p)},\phi_{(p)}
\sim
\mathrm{Inverse\text{-}Gamma}
\left(
2+\frac{I}{2},
1+\frac{1}{2}
(\boldsymbol{\beta}_{(p)}-\mu_{(p)}\mathbf{1})'
\mathbf{Q}_{(p)}
(\boldsymbol{\beta}_{(p)}-\mu_{(p)}\mathbf{1})
\right).
\label{eq:continuous-sigma-beta-update}
\end{equation*}

Next, the spatial decay parameters, $\phi_{\gamma}$ and $\phi_{(p)}$ for $p=0,\ldots,P$, do not have standard full conditional distributions and are therefore updated using random-walk Metropolis-Hastings steps. Because each decay parameter is assigned a $\mathrm{U}(\phi_L,\phi_U)$ prior and is restricted to the interval $(\phi_L,\phi_U)$, we transform it to the real line before constructing the proposal. For a generic decay parameter $\phi$, define $z = \log\left(\frac{\phi-\phi_L}{\phi_U-\phi}\right)$, which implies $\phi = \frac{\phi_L+\phi_Ue^{z}}{1+e^{z}}$. At iteration $m$, we propose a candidate value using a Gaussian random walk, $z^* \sim \mathrm{N}(z^{(m-1)},s_{\phi}^2)$, and then transform it back to the original scale as $\phi^*=(\phi_L+\phi_Ue^{z^*})/(1+e^{z^*})$. The proposal variance $s_{\phi}^2$ is adaptively tuned during burn-in to achieve a suitable target acceptance rate. Specifically, the proposal scale is increased when the recent acceptance rate is above the target and decreased when it is below the target. This adaptation is performed separately for $\phi_{\gamma}$ and each $\phi_{(p)}$ and is discontinued after burn-in.

\subsection{Acceptance ratio for the continuous model} \label{sec:app-continuous-ratio}
Under the continuous model specification, the acceptance ratio in equation~\eqref{eq:R0-continuous} simplifies to
\begin{equation}
\begin{aligned}
R_i
&=
\frac{
\pi\left(\gamma_i^*\mid\boldsymbol{\gamma}_{-i}^{(m^*)},\mu_{\gamma}^{(m)},\sigma_{\gamma}^{2(m)},\phi_{\gamma}^{(m)}\right)
\prod_{p=0}^{P}
\pi\left(\beta_{pi}^*\mid\boldsymbol{\beta}_{(p),-i}^{(m^*)},\mu_{(p)}^{(m)},\sigma_{(p)}^{2(m)},\phi_{(p)}^{(m)}\right)
}{
\pi\left(\gamma_i^{(m-1)}\mid\boldsymbol{\gamma}_{-i}^{(m^*)},\mu_{\gamma}^{(m)},\sigma_{\gamma}^{2(m)},\phi_{\gamma}^{(m)}\right)
\prod_{p=0}^{P}
\pi\left(\beta_{pi}^{(m-1)}\mid\boldsymbol{\beta}_{(p),-i}^{(m^*)},\mu_{(p)}^{(m)},\sigma_{(p)}^{2(m)},\phi_{(p)}^{(m)}\right)
}
\\[4pt]
&\qquad\times
\frac{
\widetilde{\pi}\left(\gamma_i^{(m-1)}\right)
\prod_{p=0}^{P}\widetilde{\pi}\left(\beta_{pi}^{(m-1)}\right)
}{
\widetilde{\pi}\left(\gamma_i^*\right)
\prod_{p=0}^{P}\widetilde{\pi}\left(\beta_{pi}^*\right)
}
\\[4pt]
&=
\frac{
\exp\left[
-\frac{q_{\gamma,ii}^{(m)}}{2\sigma_{\gamma}^{2(m)}}
\left(\gamma_i^*-m_{\gamma,i}^{(m^*)}\right)^2
\right]}
{
\exp\left[
-\frac{q_{\gamma,ii}^{(m)}}{2\sigma_{\gamma}^{2(m)}}
\left(\gamma_i^{(m-1)}-m_{\gamma,i}^{(m^*)}\right)^2
\right]}
\times
\frac{
\prod_{p=0}^P
\exp\left[
-\frac{q_{(p),ii}^{(m)}}{2\sigma_{(p)}^{2(m)}}
\left(\beta_{pi}^*-m_{(p),i}^{(m^*)}\right)^2
\right]}
{
\prod_{p=0}^P
\exp\left[
-\frac{q_{(p),ii}^{(m)}}{2\sigma_{(p)}^{2(m)}}
\left(\beta_{pi}^{(m-1)}-m_{(p),i}^{(m^*)}\right)^2
\right]}
\\[4pt]
&\qquad\times
\frac{
\exp\left[
-\frac{1}{2(10^2)} \left(\gamma_i^{(m-1)}\right)^2
\right]
\prod_{p=0}^P
\exp\left[
-\frac{1}{2(10^2)} \left(\beta_{pi}^{(m-1)}\right)^2
\right]}
{
\exp\left[
-\frac{1}{2(10^2)} \left(\gamma_i^*\right)^2
\right]
\prod_{p=0}^P
\exp\left[
-\frac{1}{2(10^2)} \left(\beta_{pi}^*\right)^2
\right]},
\end{aligned}
\label{eq:R-continuous}
\end{equation}
where $m_{\gamma,i}^{(m^*)} = \mu_{\gamma}^{(m)} - \frac{\mathbf{q}_{\gamma,i,-i}^{(m)}}{q_{\gamma,ii}^{(m)}} \left( \boldsymbol{\gamma}_{-i}^{(m^*)}-\mu_{\gamma}^{(m)}\mathbf{1} \right)$ and $m_{(p),i}^{(m^*)} = \mu_{(p)}^{(m)} - \frac{\mathbf{q}_{(p),i,-i}^{(m)}}{q_{(p),ii}^{(m)}} \left( \boldsymbol{\beta}_{(p),-i}^{(m^*)}-\mu_{(p)}^{(m)}\mathbf{1} \right)$ denote the conditional means under the multivariate normal spatial priors for $\gamma_i$ and $\beta_{pi}$, evaluated at the current values of the parameters for all other locations.

\subsection{Acceptance ratio for the binary model} \label{sec:app-binary-ratio}
Under the binary model specification, the acceptance ratio in equation~\eqref{eq:R0-binary} simplifies to
\begin{equation}
\begin{aligned}
R_i
&=
\frac{
\pi\left(\gamma_i^*\mid\bfgamma_{-i}^{(m^*)},\sigma_\gamma^{2(m)}\right)
\prod_{p=0}^P
\pi\left(\beta_{pi}^*\mid\bfbeta_{p,-i}^{(m^*)},\sigma_{(p)}^{2(m)}\right)}
{
\pi\left(\gamma_i^{(m-1)}\mid\bfgamma_{-i}^{(m^*)},\sigma_\gamma^{2(m)}\right)
\prod_{p=0}^P
\pi\left(\beta_{pi}^{(m-1)}\mid\bfbeta_{p,-i}^{(m^*)},\sigma_{(p)}^{2(m)}\right)}
\times
\frac{
\widetilde{\pi}\left(\gamma_i^{(m-1)}\right)
\prod_{p=0}^P
\widetilde{\pi}\left(\beta_{pi}^{(m-1)}\right)}
{
\widetilde{\pi}\left(\gamma_i^*\right)
\prod_{p=0}^P
\widetilde{\pi}\left(\beta_{pi}^*\right)}
\\[4pt]
&=
\frac{
\exp\left[
-\frac{a_{i+}}{2\sigma_\gamma^{2(m)}}
\left(\gamma_i^*-\bar{\gamma}_{j\sim i}^{(m^*)}\right)^2
\right]}
{
\exp\left[
-\frac{a_{i+}}{2\sigma_\gamma^{2(m)}}
\left(\gamma_i^{(m-1)}-\bar{\gamma}_{j\sim i}^{(m^*)}\right)^2
\right]}
\times
\frac{
\exp\left(-\gamma_i^{(m-1)}\right)
\left[1+\exp\left(-\gamma_i^*\right)\right]^2}
{
\left[1+\exp\left(-\gamma_i^{(m-1)}\right)\right]^2
\exp\left(-\gamma_i^*\right)}
\\[4pt]
&\qquad\times
\frac{
\prod_{p=0}^P
\exp\left[
-\frac{a_{i+}}{2\sigma_{(p)}^{2(m)}}
\left(\beta_{pi}^*-\bar{\beta}_{p,j\sim i}^{(m^*)}\right)^2
\right]}
{
\prod_{p=0}^P
\exp\left[
-\frac{a_{i+}}{2\sigma_{(p)}^{2(m)}}
\left(\beta_{pi}^{(m-1)}-\bar{\beta}_{p,j\sim i}^{(m^*)}\right)^2
\right]}
\times
\frac{
\prod_{p=0}^P
\exp\left[
-\frac{1}{2(3^2)} \left(\beta_{pi}^{(m-1)}\right)^2
\right]}
{
\prod_{p=0}^P
\exp\left[
-\frac{1}{2(3^2)} \left(\beta_{pi}^*\right)^2
\right]}.
\end{aligned}
\label{eq:R-binary}
\end{equation}
where $j \sim i$ indexes the neighbors of location $i$, and $\bar{\gamma}_{j\sim i}^{(m^*)} = \frac{1}{a_{i+}}\sum_{\ell=1}^I a_{i\ell}\gamma_\ell^{(m^*)}$ and $\bar{\beta}_{p,j\sim i}^{(m^*)} = \frac{1}{a_{i+}}\sum_{\ell=1}^I a_{i\ell}\beta_{p\ell}^{(m^*)}$ denote the means of the current transformed autoregressive parameters and the $p$th regression coefficients, respectively, over the neighbors of location $i$.

\clearpage
\subsection{Posterior standard deviations}
\begin{figure}[H]
    \centering
    {\includegraphics[width=0.65\textwidth]{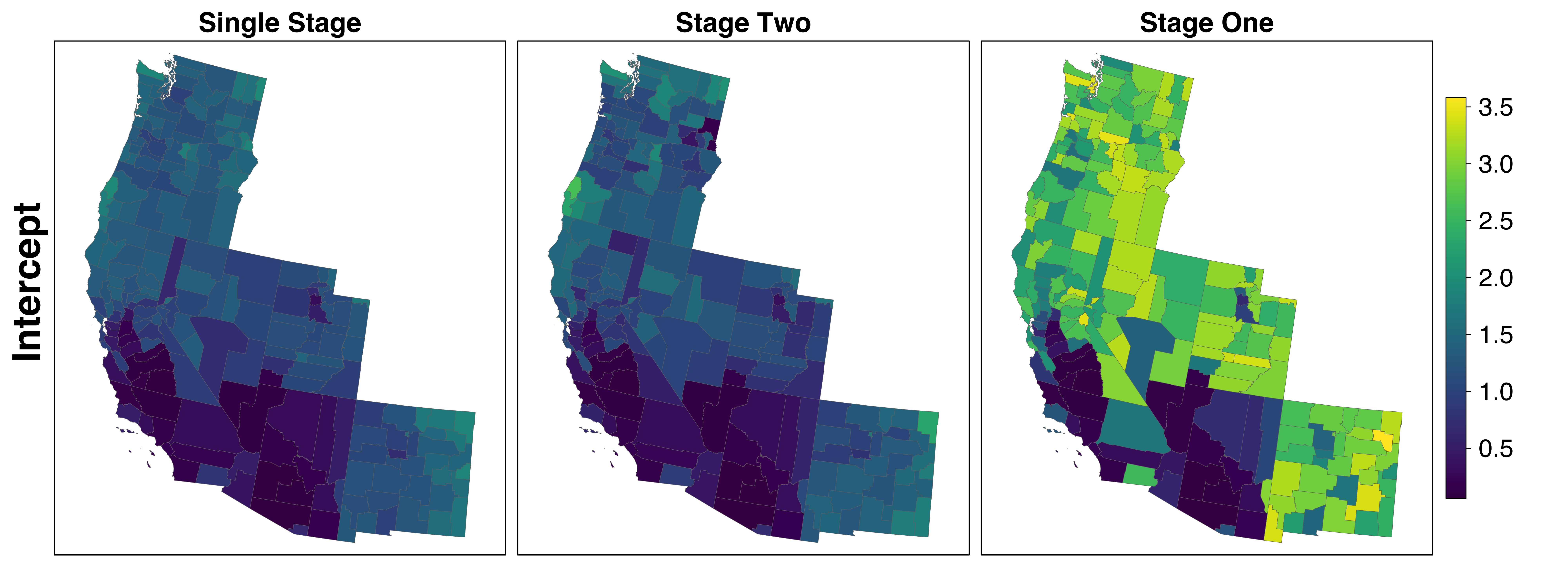}}  
    \par\vspace{1mm}
    {\includegraphics[trim={0 0 0 70mm},clip, width=0.65\textwidth]{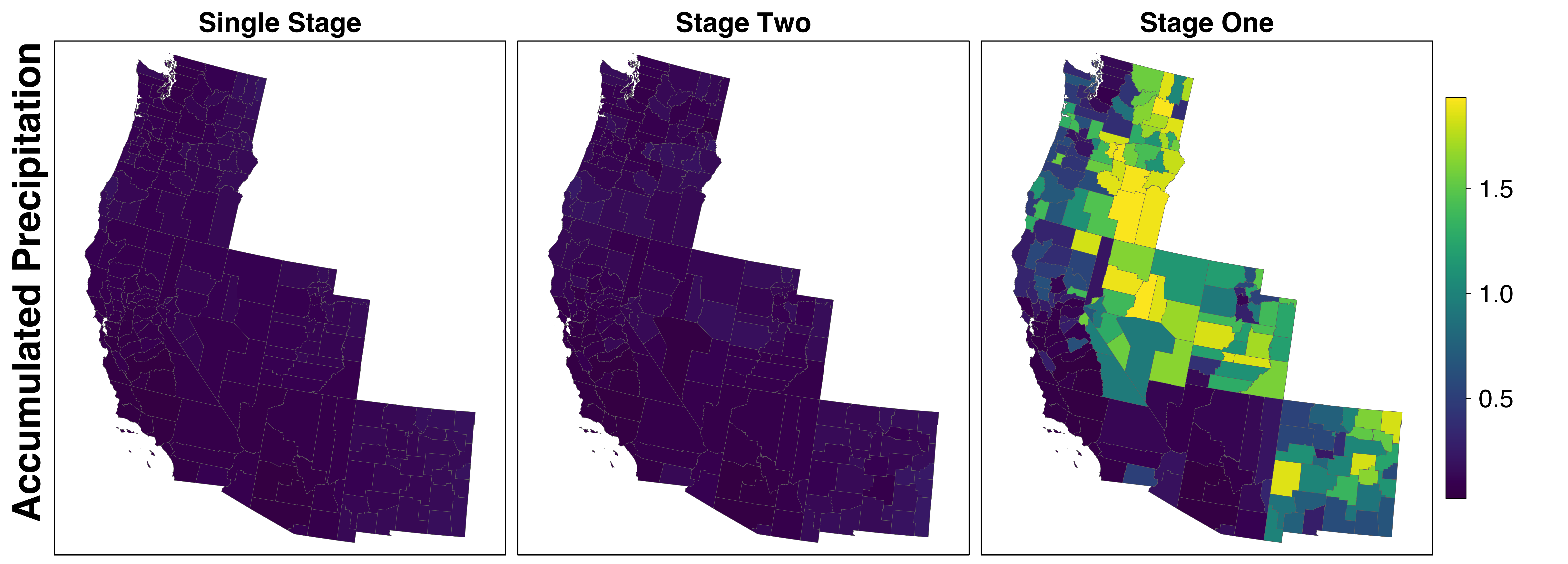}} 
    \par\vspace{1mm}
    {\includegraphics[trim={0 0 0 70mm},clip,width=0.65\textwidth]{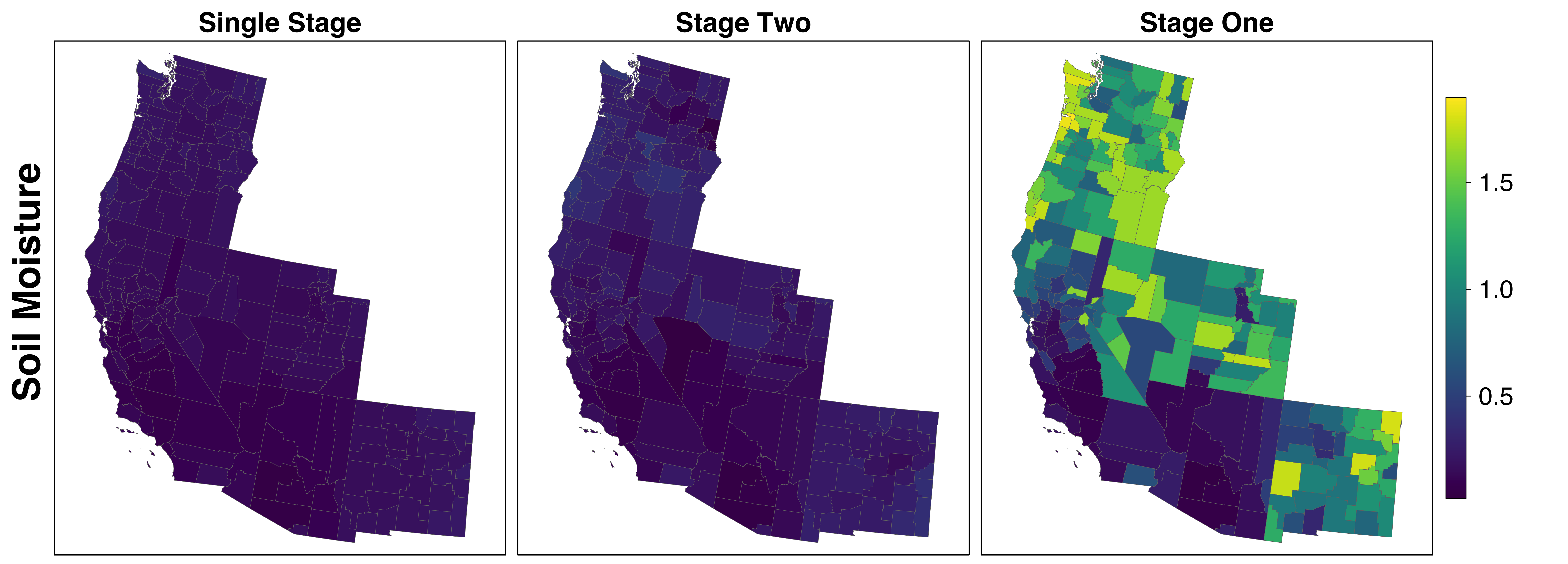}}
    \par\vspace{1mm}
    {\includegraphics[trim={0 0 0 70mm},clip,width=0.65\textwidth]{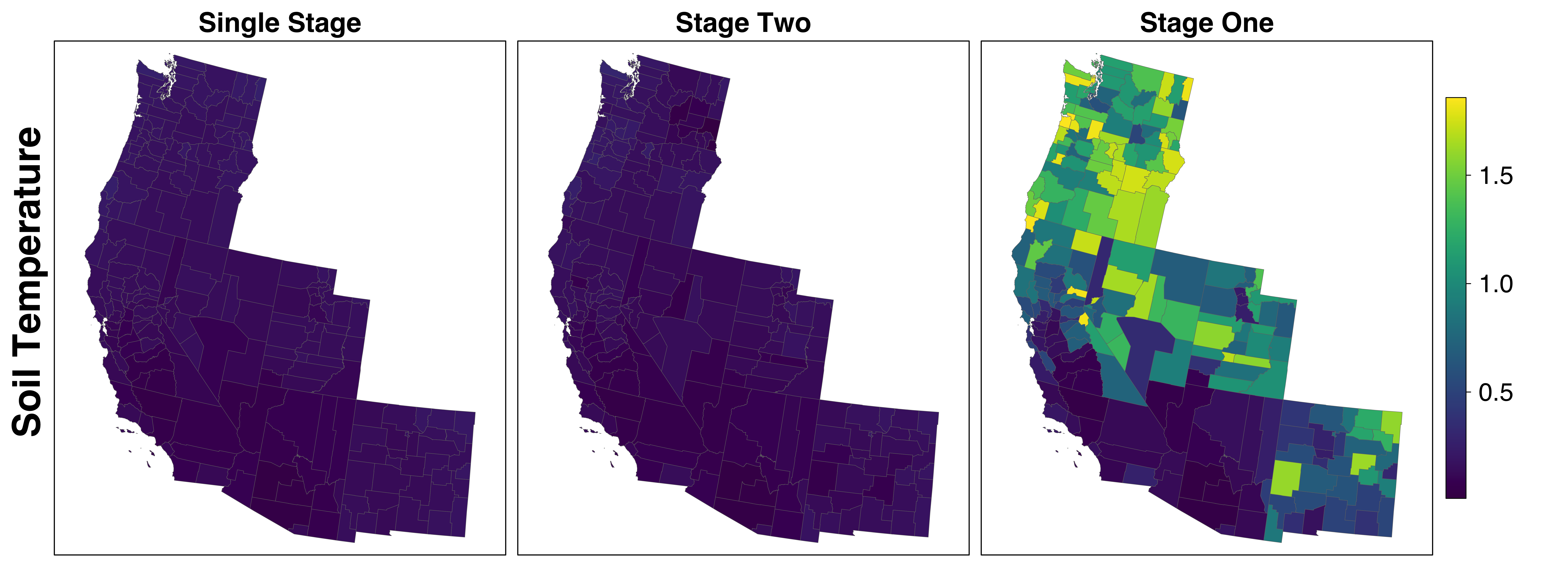}}
    \par\vspace{1mm}
    {\includegraphics[trim={0 0 0 70mm},clip,width=0.65\textwidth]{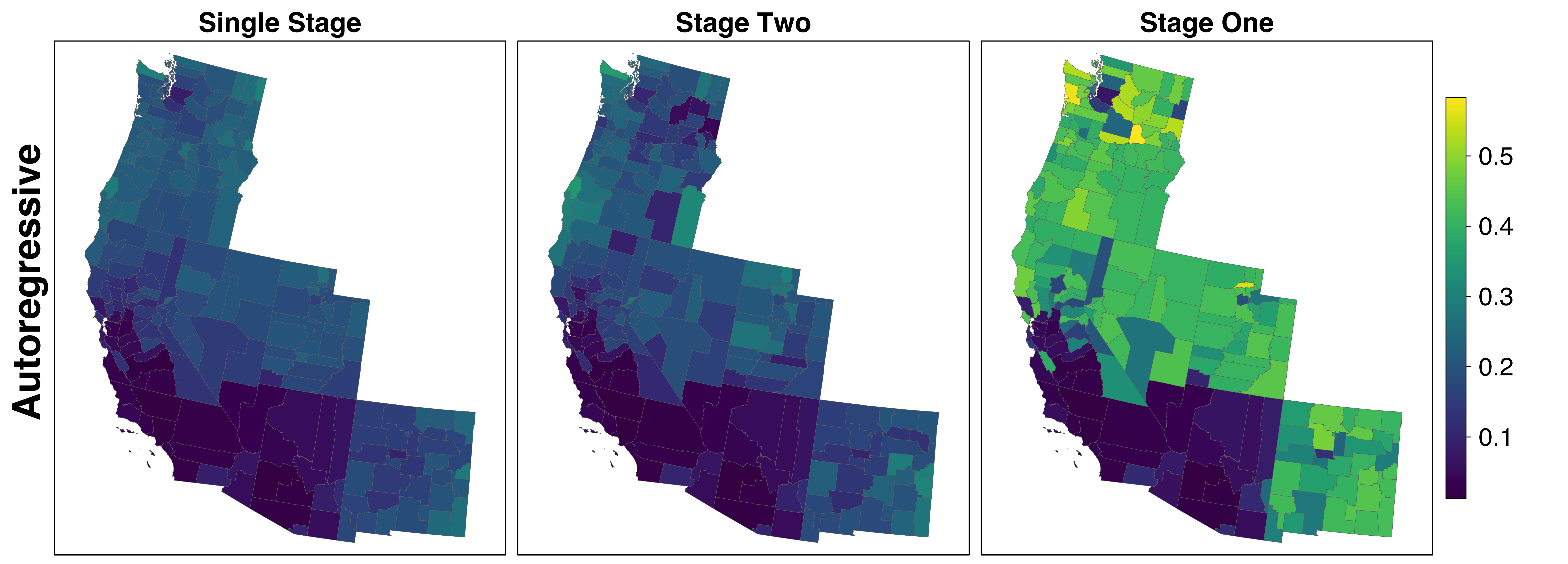}} 

    \caption{Posterior standard deviations of the county-specific parameters assigned spatial ICAR priors in the count model under the single-stage method (left column), the two-stage method (center column), and after stage one only (right column). From top to bottom, the rows show the intercept $\beta_0$, effect of accumulated precipitation $\beta_{\mathrm{apcp}}$, effect of soil moisture $\beta_{\mathrm{soilm}}$, effect of soil temperature $\beta_{\mathrm{tsoil}}$, and autoregressive parameter $\rho$.}
    \label{fig:count-sd}
\end{figure}

\begin{figure}[H]
    \centering
    {\includegraphics[width=0.99\textwidth]{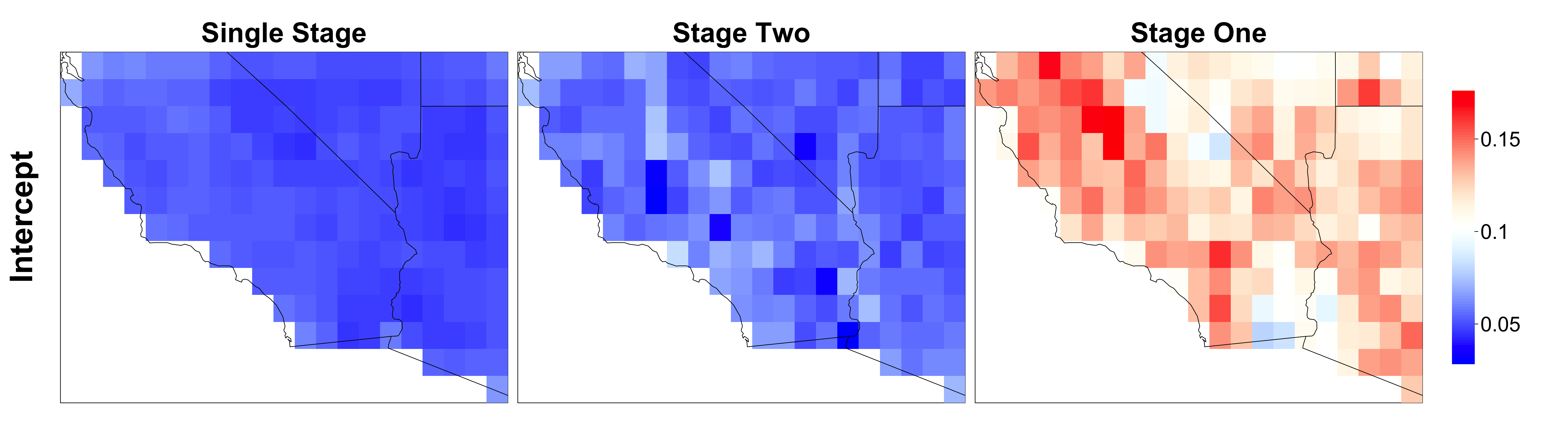}}  
    \par\vspace{1mm}
    {\includegraphics[trim={0 0 0 85mm},clip, width=0.99\textwidth]{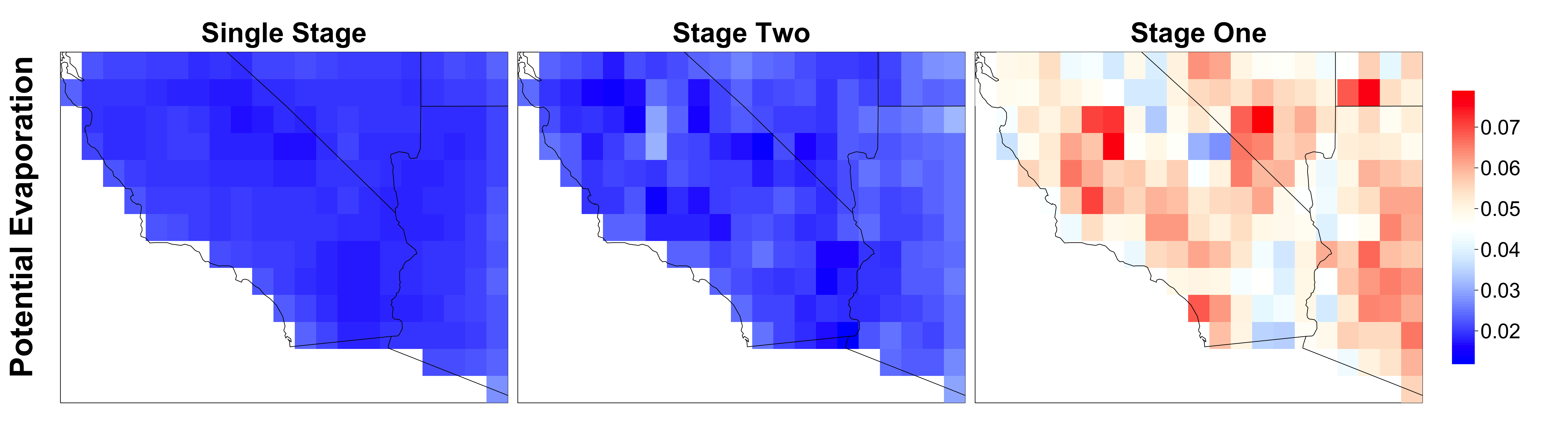}} 
    \par\vspace{1mm}
    {\includegraphics[trim={0 0 0 85mm},clip,width=0.99\textwidth]{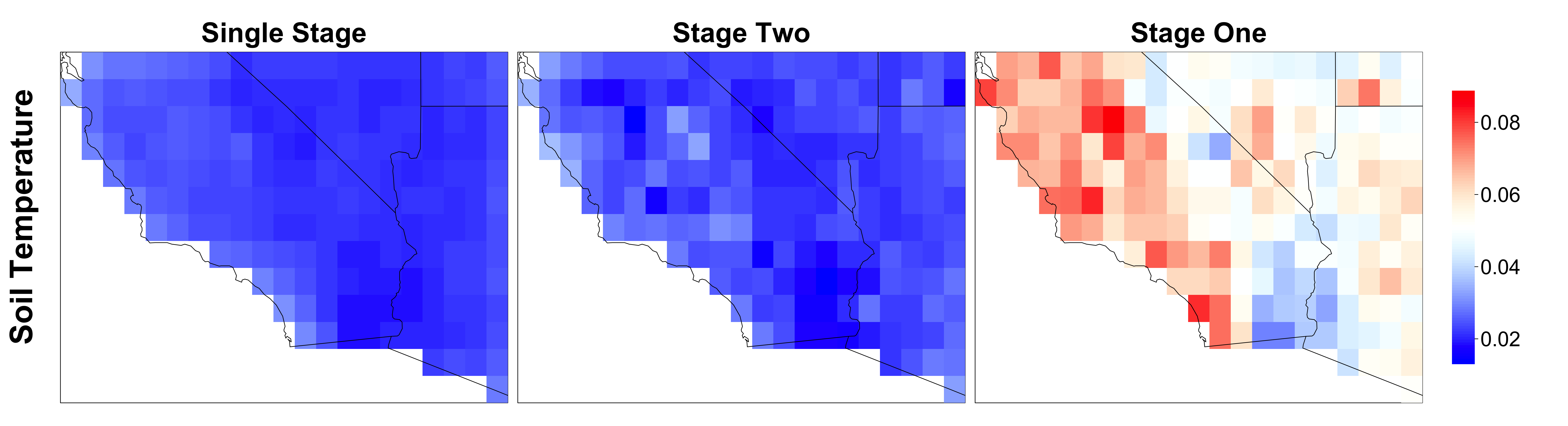}}
    \par\vspace{1mm}
    {\includegraphics[trim={0 0 0 85mm},clip,width=0.99\textwidth]{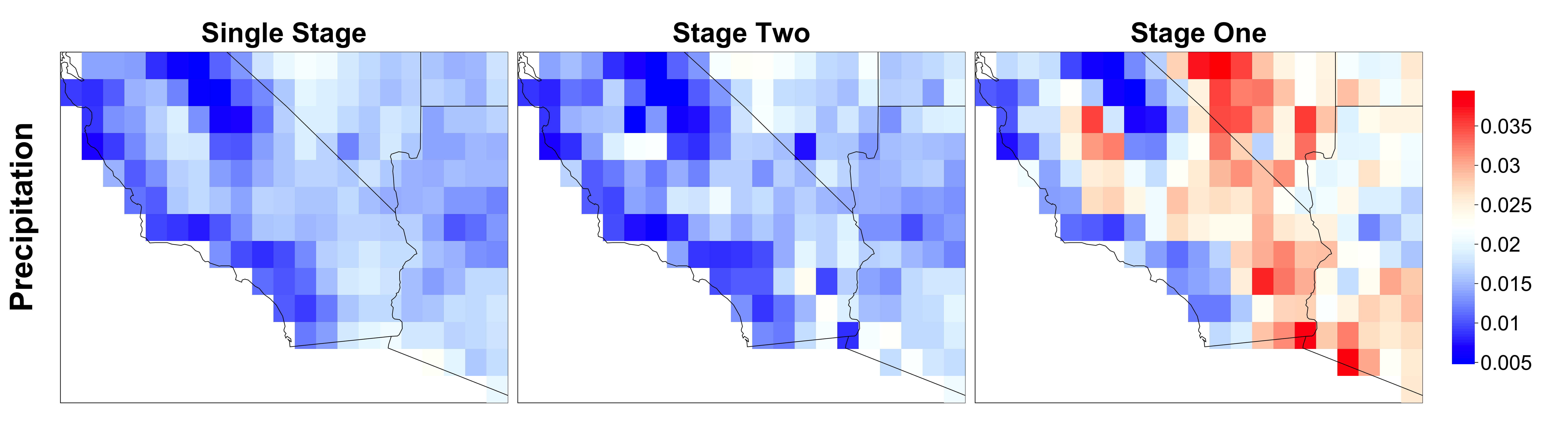}}
    \par\vspace{1mm}
    {\includegraphics[trim={0 0 0 85mm},clip,width=0.99\textwidth]{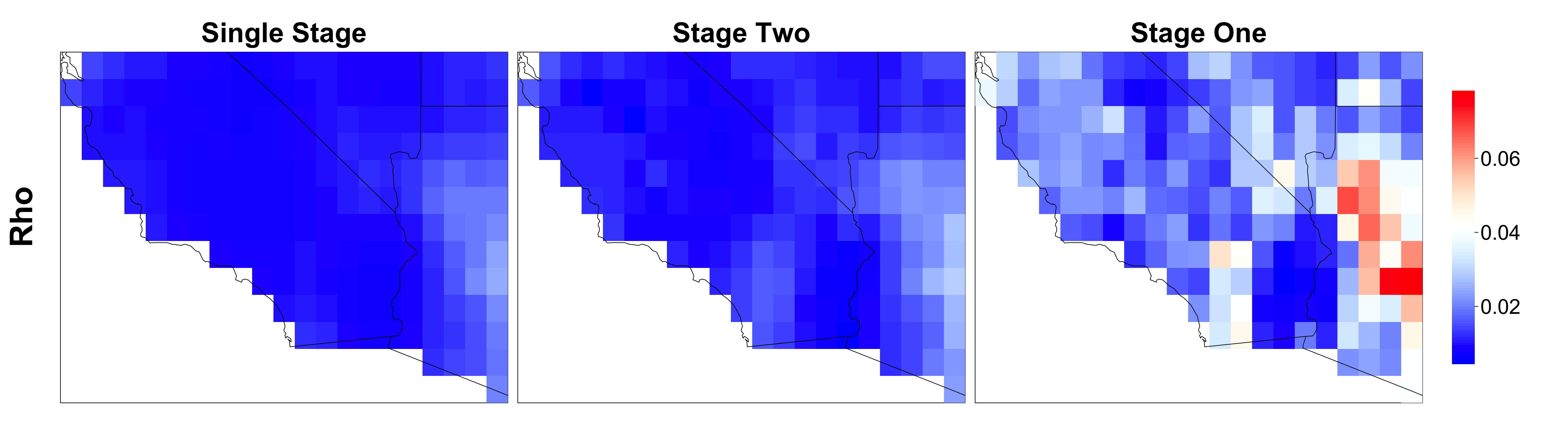}} 

    \caption{Posterior standard deviations of the location-specific parameters assigned spatial MVN priors in the continuous model under the single-stage method (left column), the two-stage method (center column), and after stage one only (right column). From top to bottom, the rows show the intercept $\beta_0$, potential evaporation effect $\beta_{\mathrm{pevap}}$, soil temperature effect $\beta_{\mathrm{tsoil}}$, precipitation effect $\beta_{\mathrm{apcp}}$, and autoregressive parameter $\rho$.}
    \label{fig:continuous-sd}
\end{figure}

\begin{figure}[H]
    \centering
    {\includegraphics[width=0.99\textwidth, height=4cm]{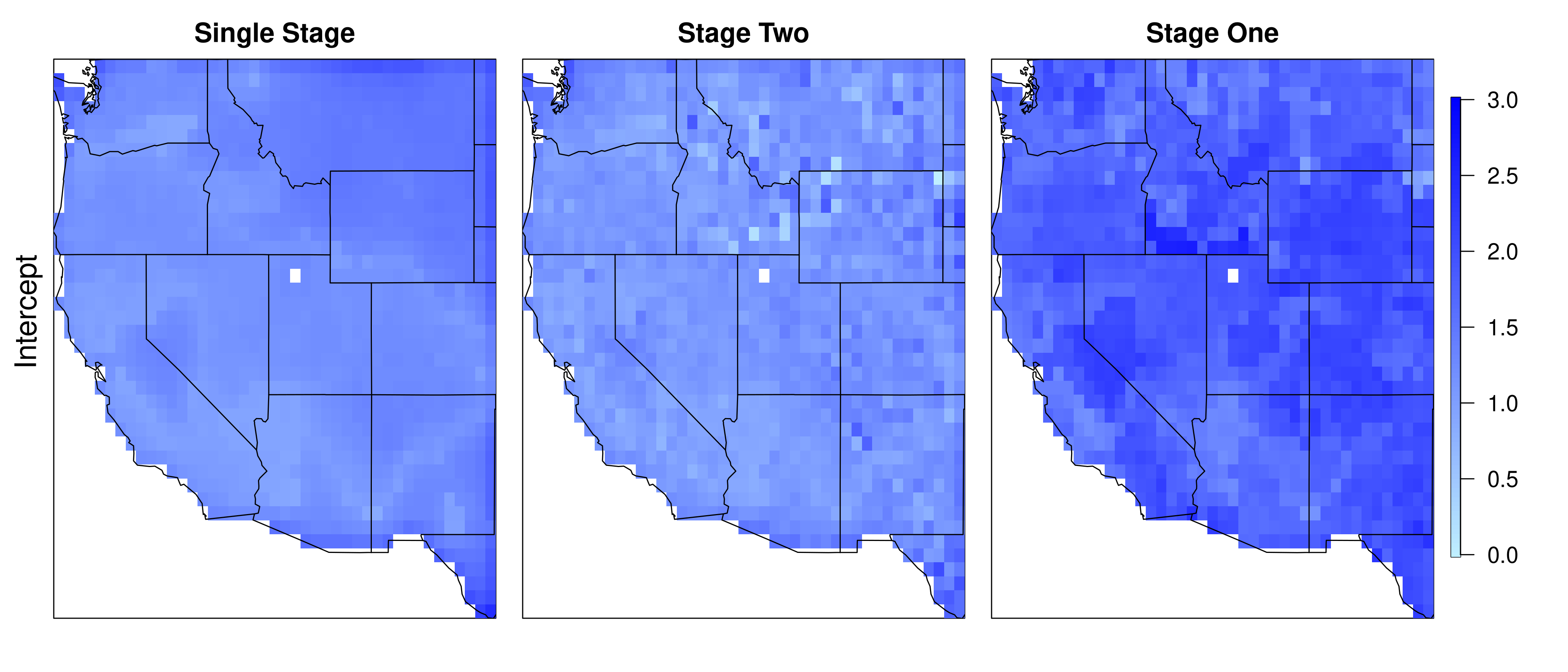}}  
    {\includegraphics[trim={0 0 0 11mm},clip, width=0.99\textwidth, height=4cm]{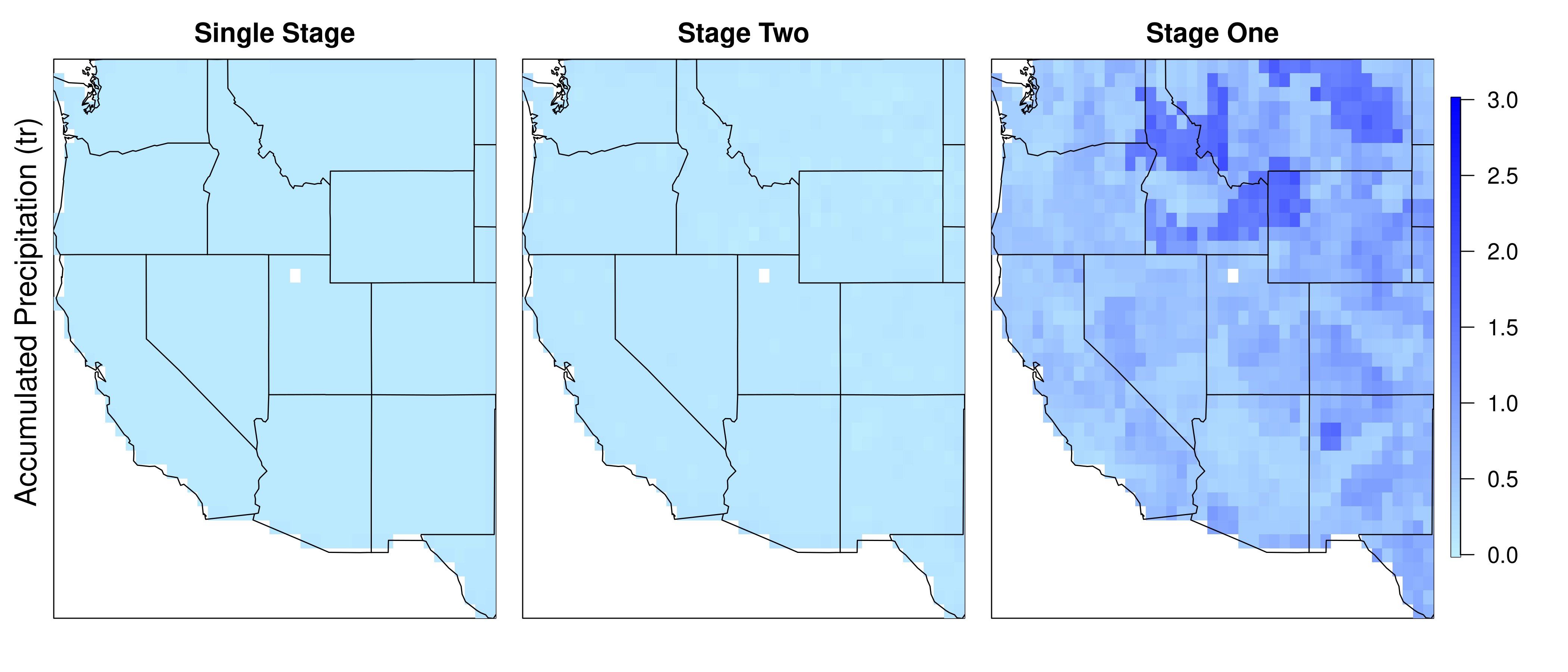}} 
    {\includegraphics[trim={0 0 0 11mm},clip,width=0.99\textwidth, height=4cm]{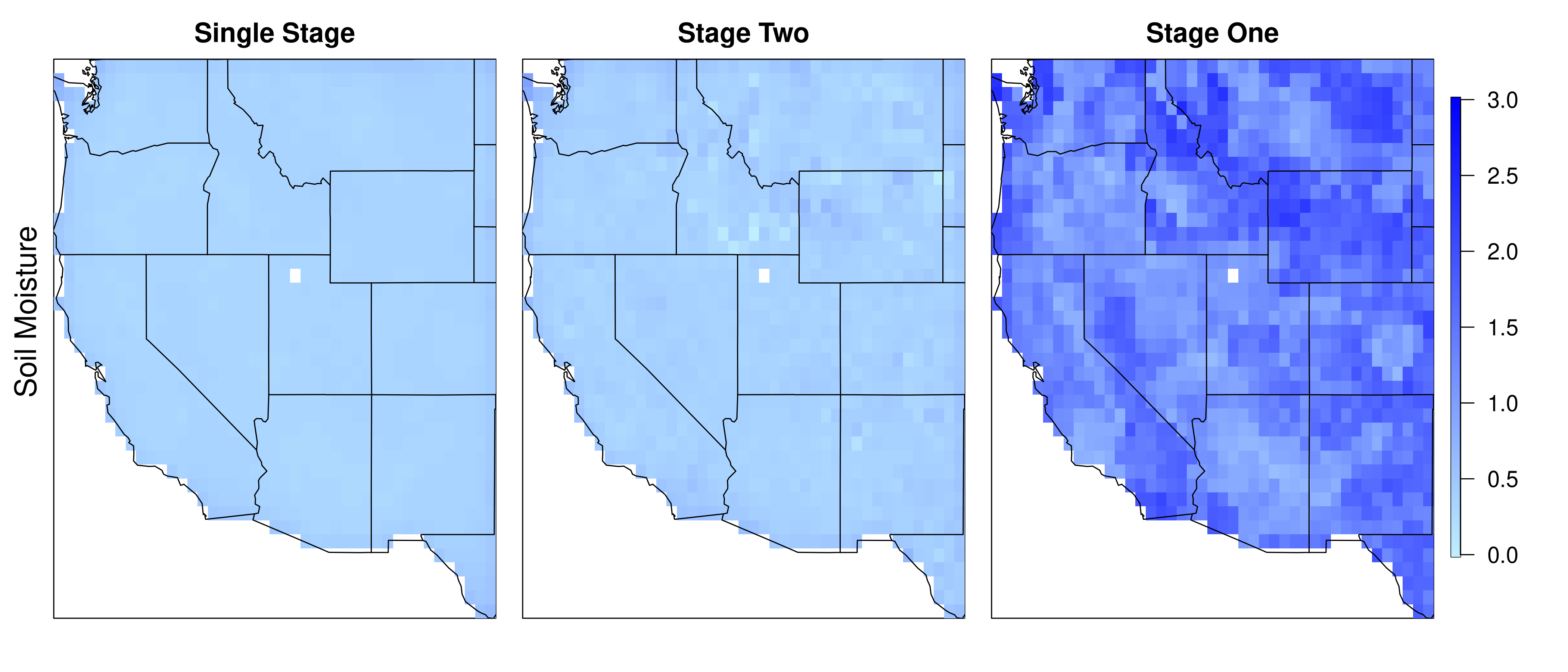}}
    {\includegraphics[trim={0 0 0 11mm},clip,width=0.99\textwidth, height=4cm]{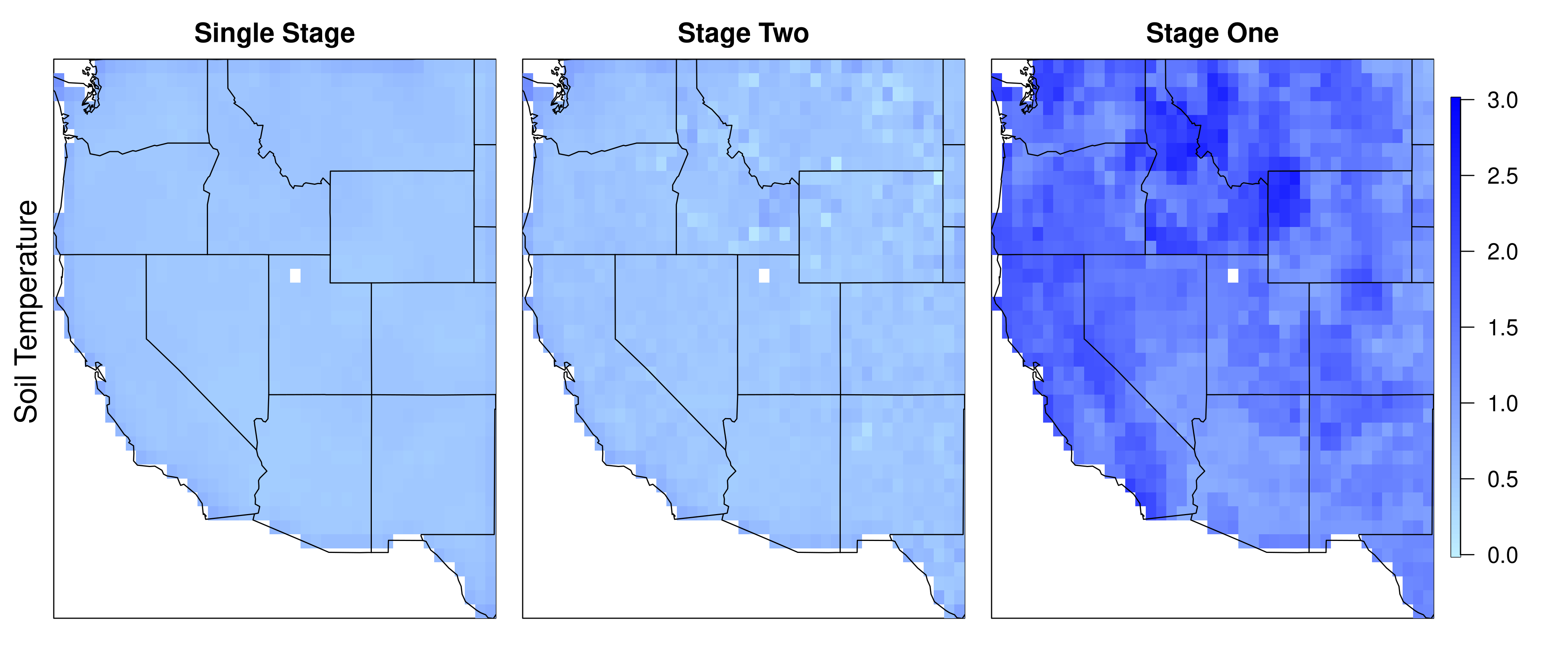}}
    {\includegraphics[trim={0 0 0 11mm},clip,width=0.99\textwidth, height=4cm]{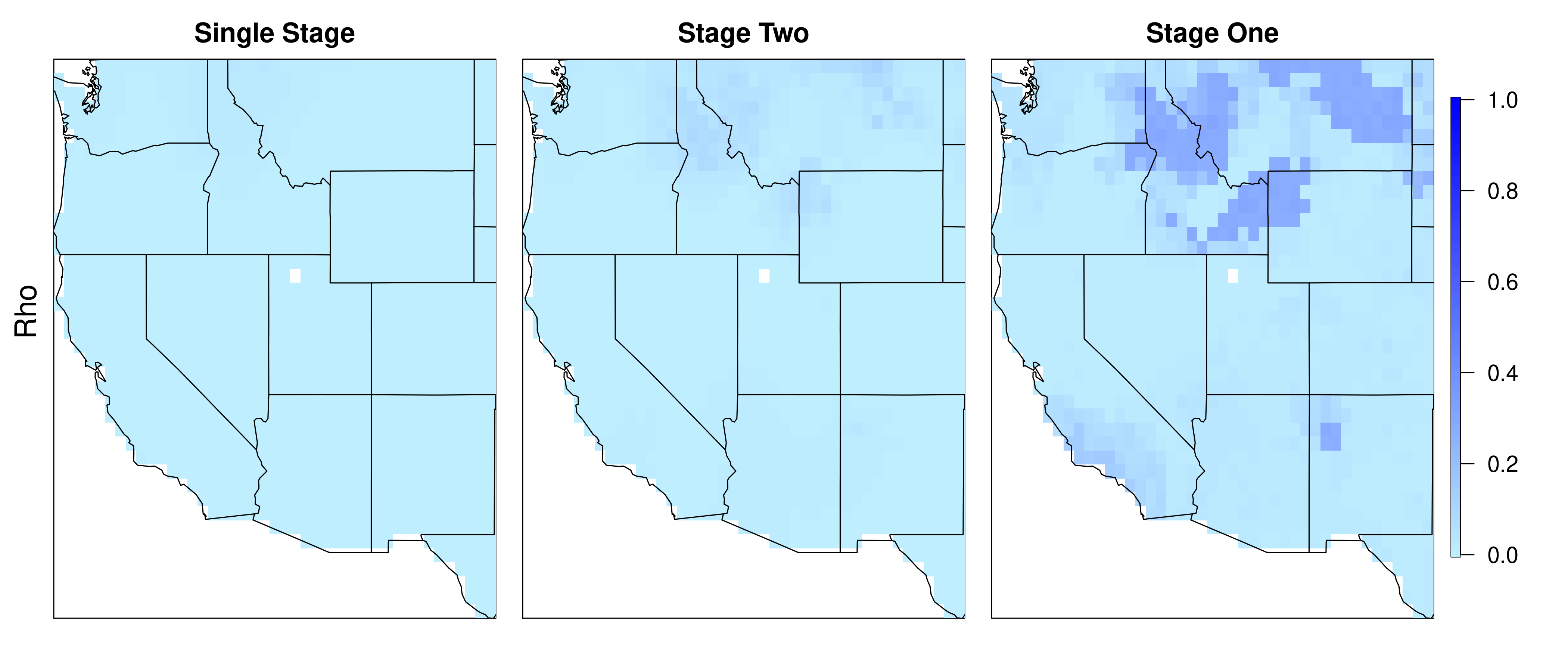}} 

    \caption{Posterior standard deviations of the location-specific parameters assigned spatial ICAR priors in the binary model taken from the single-stage method (left column), two-stage method (center column), and after stage one only (right column).  Parameters shown include: intercept $\beta_0$, effect of accumulated precipitation $\beta_{\mathrm{apcp}}$, effect of soil moisture $\beta_{\mathrm{soilm}}$, effect of soil temperature $\beta_{\mathrm{tsoil}}$, and  autoregression $\rho$.}.
    \label{fig:sd-binary}
\end{figure}

\clearpage
\subsection{Additional parameters}
\begin{figure}[H]
    \centering
    {\includegraphics[width=0.75\textwidth]{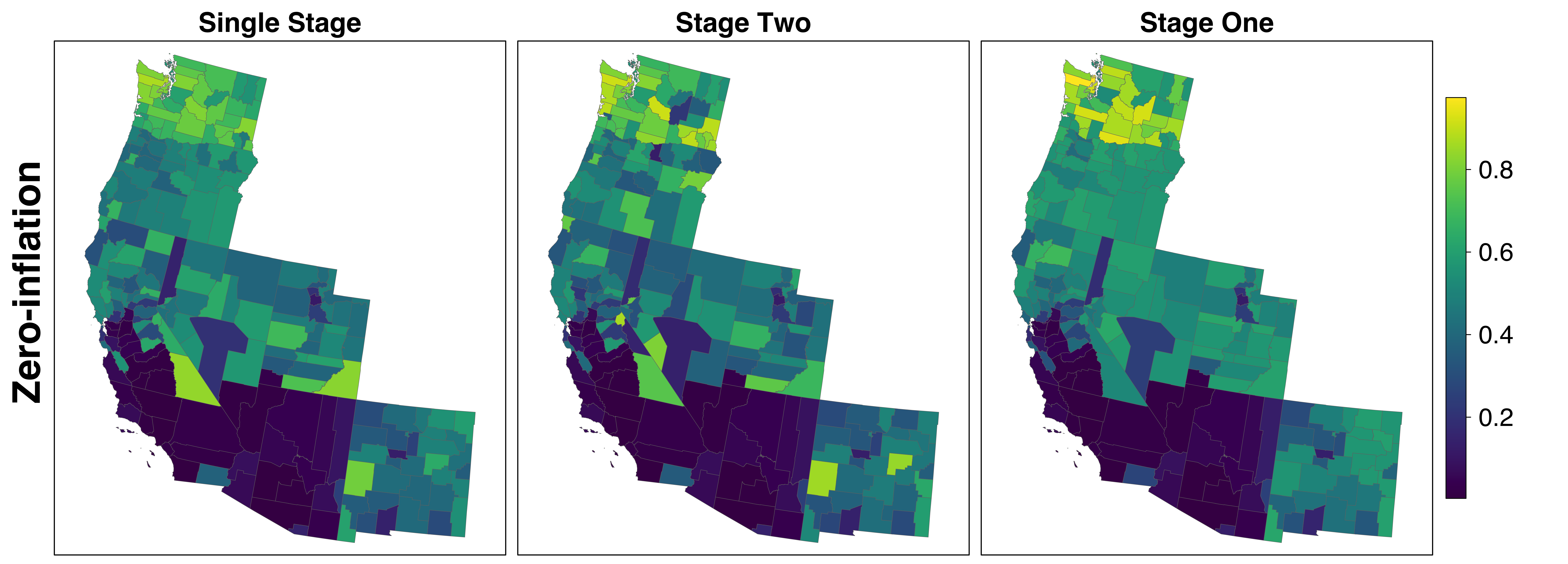}}  
    \par\vspace{1mm}
    {\includegraphics[trim={0 0 0 70mm},clip, width=0.75\textwidth]{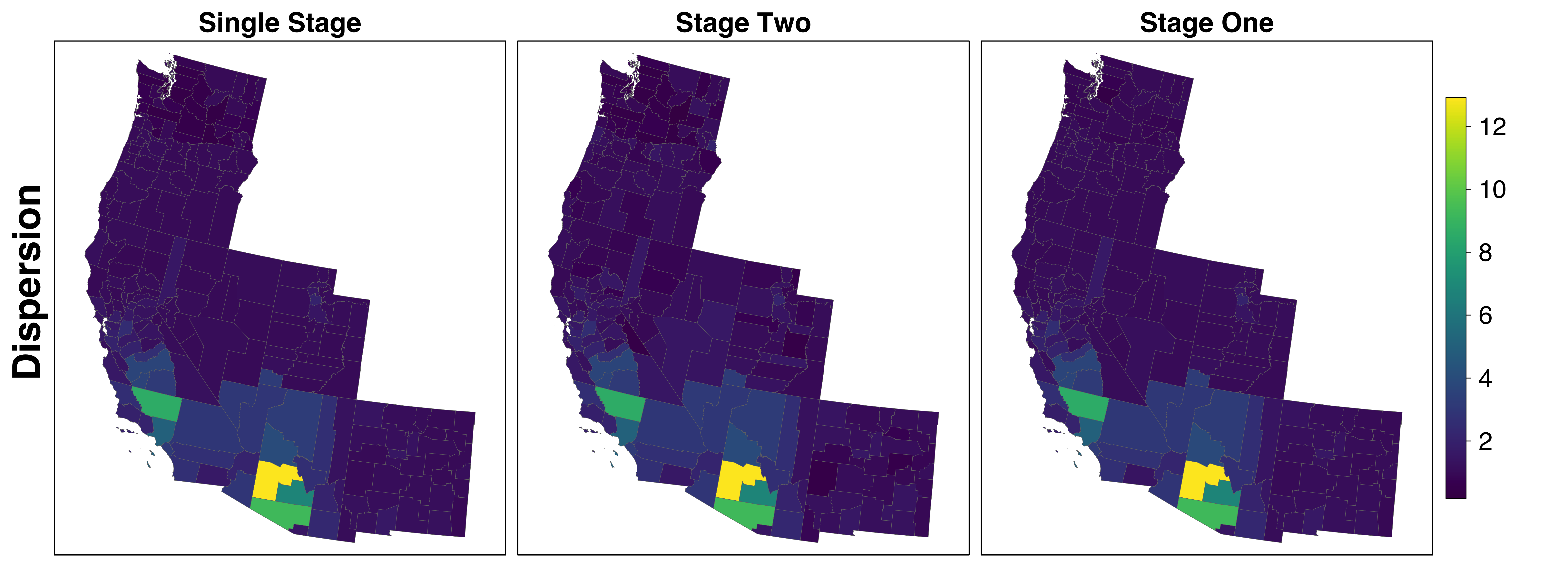}} 
    
    \caption{Posterior means of the zero-inflation and dispersion parameters in the count model under the single-stage method (left column), the two-stage method (center column), and after stage one only (right column).}
    \label{fig:count-mean-pi-r}
\end{figure}

\begin{figure}[H]
    \centering
    \par\vspace{1mm}
    {\includegraphics[width=0.73\textwidth]{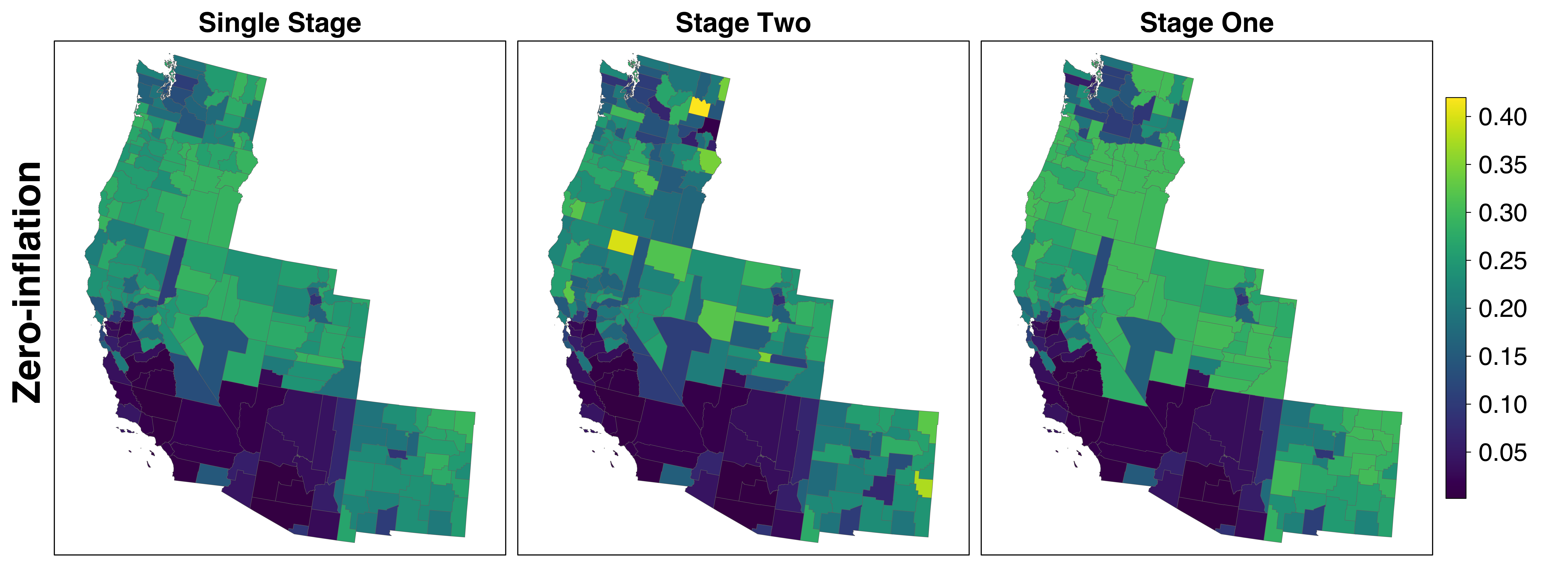}}  
    \par\vspace{1mm}
    {\includegraphics[trim={0 0 0 70mm},clip, width=0.73\textwidth]{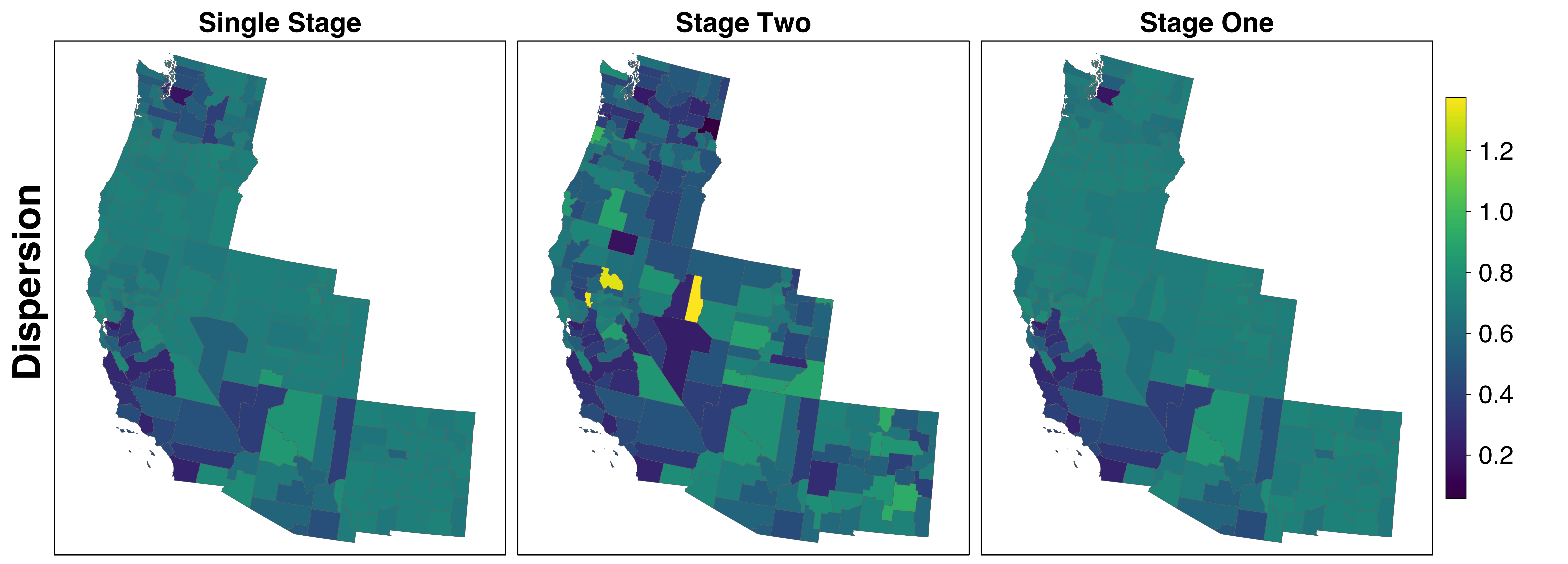}}  

    \caption{Posterior standard deviations of the zero-inflation and dispersion parameters parameters in the count model under the single-stage method (left column), the two-stage method (center column), and after stage one only (right column).}
    \label{fig:count-sd-pi-r}
\end{figure}

\begin{figure}[H]
    \centering
    {\includegraphics[width=0.99\textwidth]{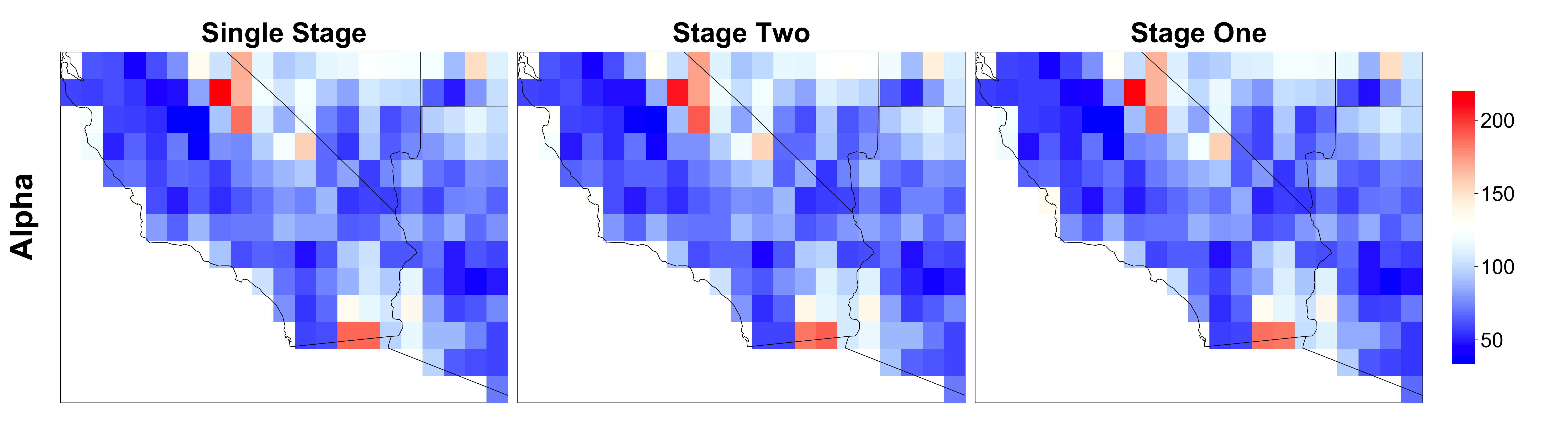}}   

    \caption{Posterior means of the gamma shape parameter in the continuous model under the single-stage method (left column), the two-stage method (center column), and after stage one only (right column).}
    \label{fig:continuous-mean-alpha}
\end{figure}

\begin{figure}[H]
    \centering
    {\includegraphics[width=0.99\textwidth]{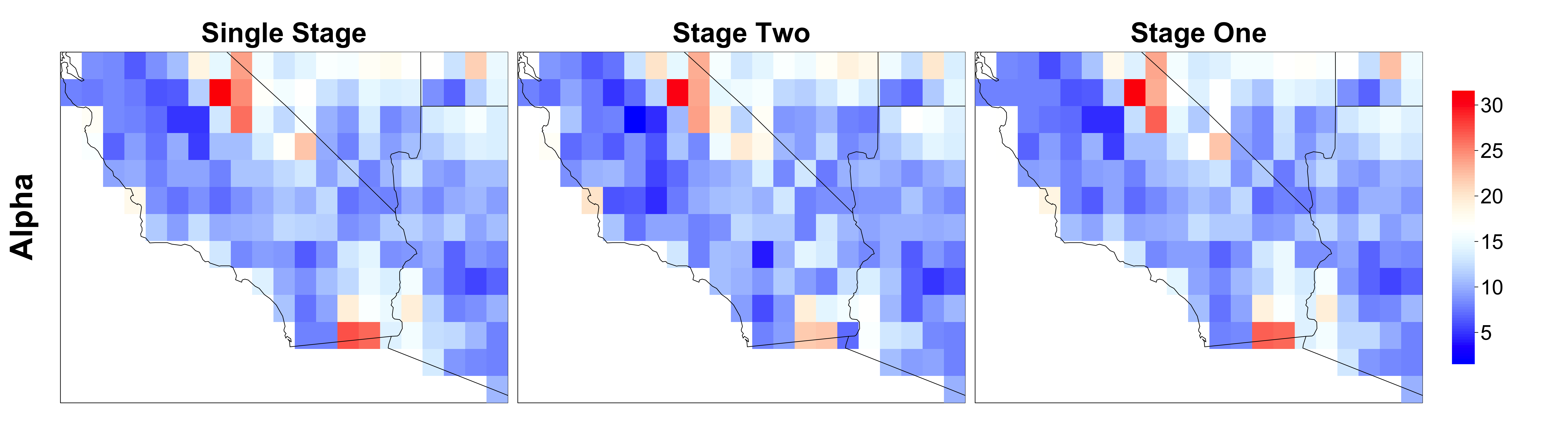}}  

    \caption{Posterior standard deviations of the gamma shape parameter in the continuous model under the single-stage method (left column), the two-stage method (center column), and after stage one only (right column).}
    \label{fig:continuous-sd-alpha}
\end{figure}

\clearpage
\subsection{Trace plots}

\begin{figure}[H]
    \centering
    {\includegraphics[width=0.99\textwidth]{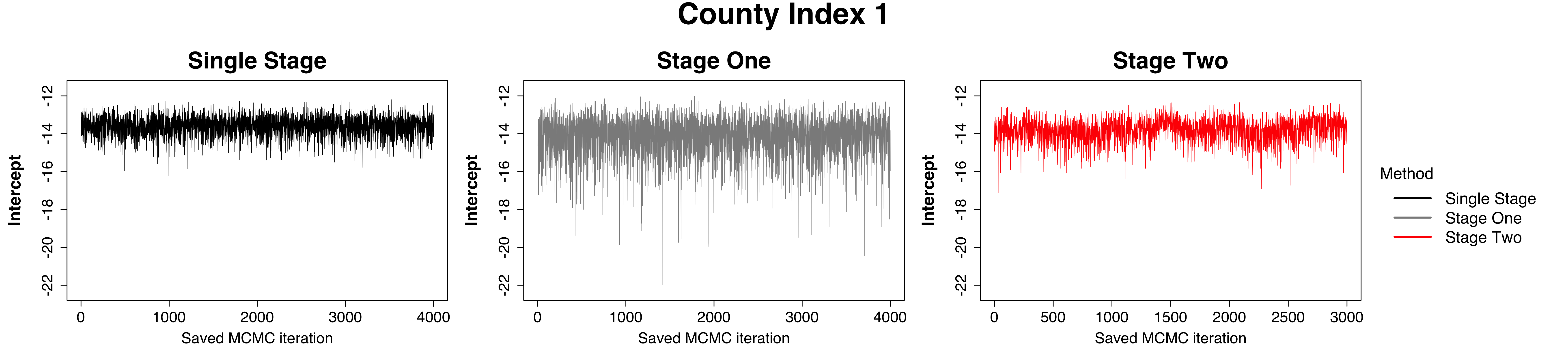}}  
    {\includegraphics[trim={0 0 0 120mm},clip, width=0.99\textwidth]{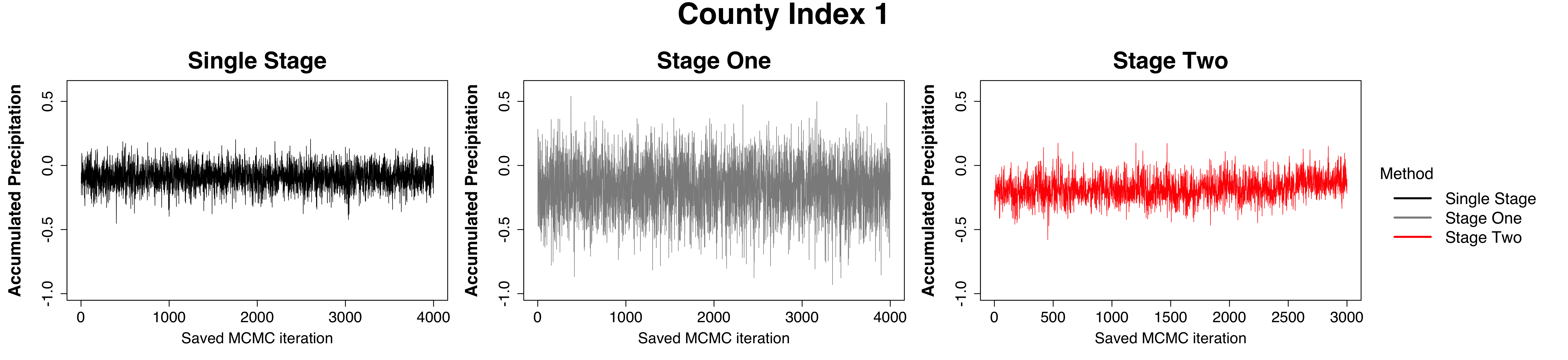}} 
    {\includegraphics[trim={0 0 0 120mm},clip,width=0.99\textwidth]{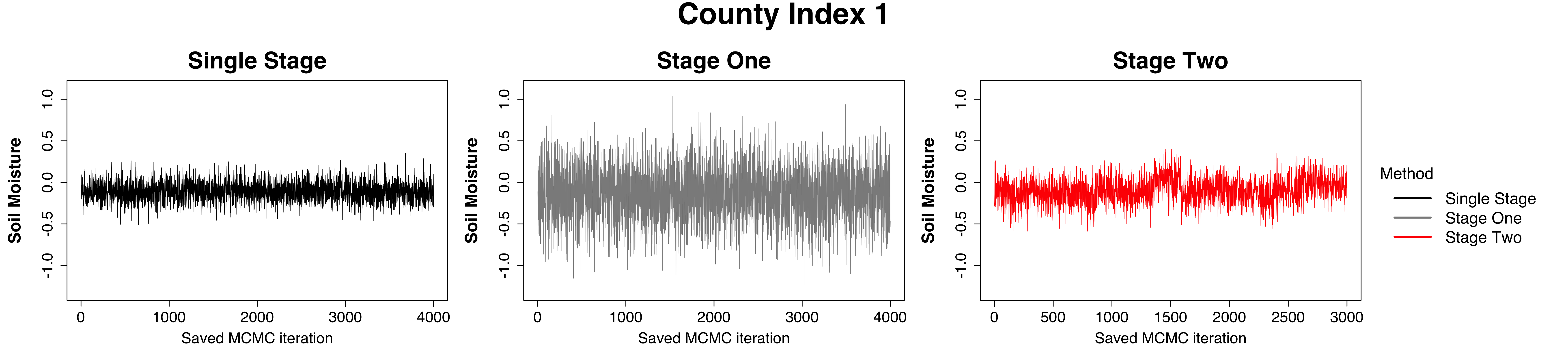}}
    {\includegraphics[trim={0 0 0 120mm},clip,width=0.99\textwidth]{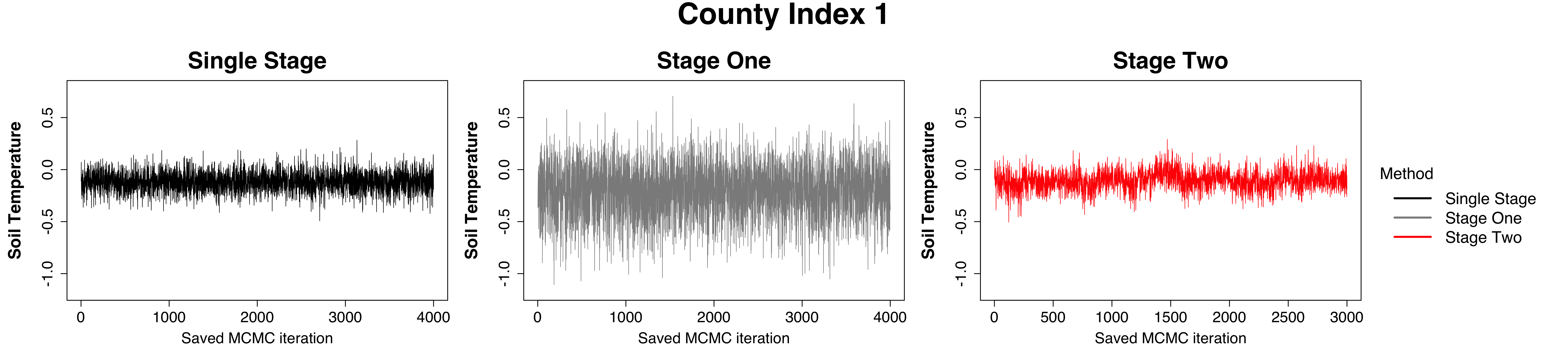}}
    {\includegraphics[trim={0 0 0 120mm},clip,width=0.99\textwidth]{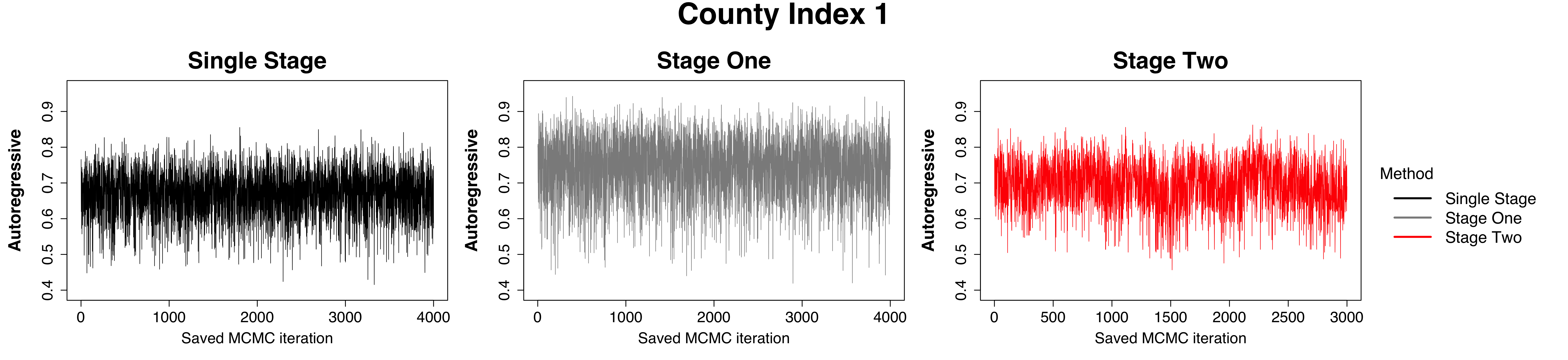}} 
    {\includegraphics[trim={0 0 0 120mm},clip,width=0.99\textwidth]{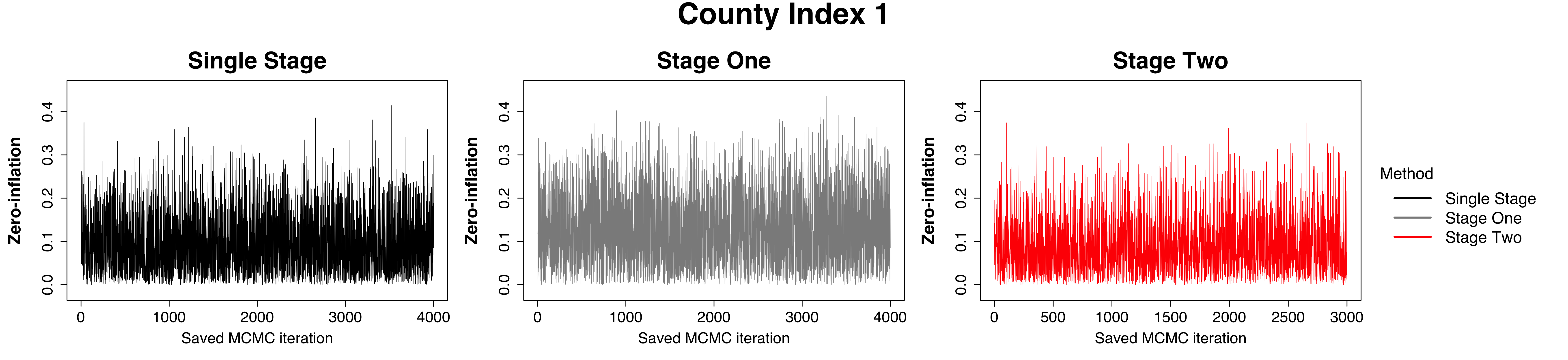}} 
    {\includegraphics[trim={0 0 0 120mm},clip,width=0.99\textwidth]{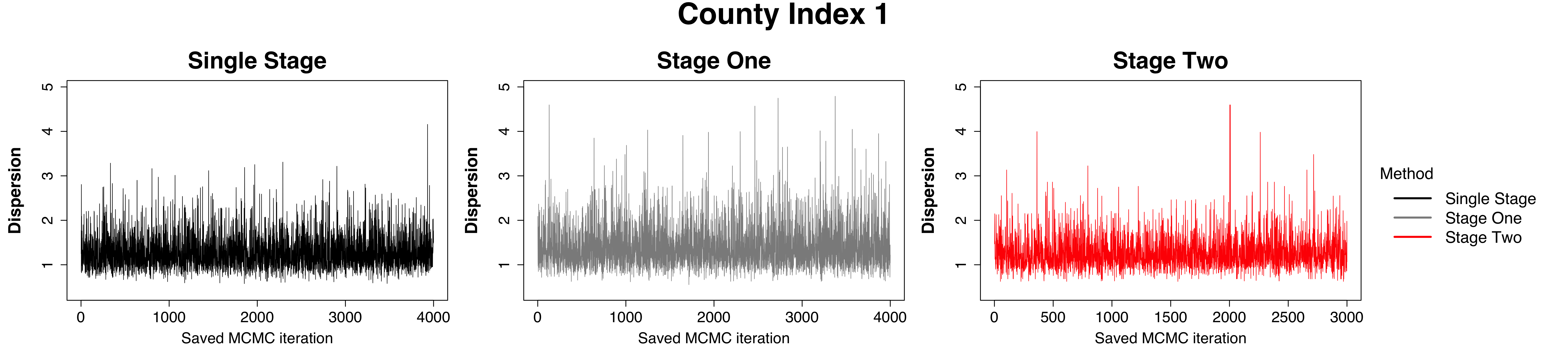}} 

    \caption{Trace plots of the county-specific parameters in the count model for one county.}
    \label{fig:count-trace}
\end{figure}

\clearpage
\begin{figure}[H]
    \centering
    {\includegraphics[width=0.99\textwidth]{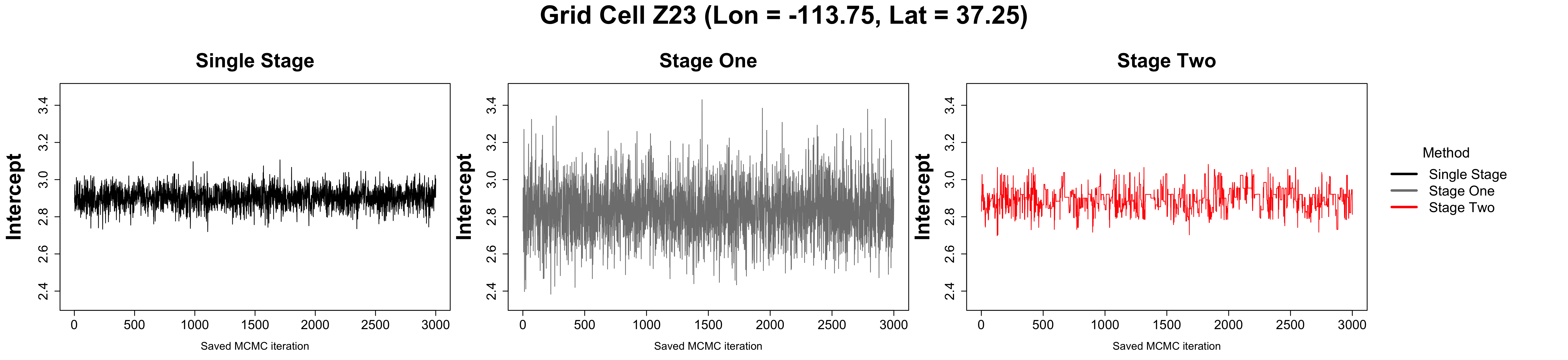}}  
    \par\vspace{1mm}
    {\includegraphics[trim={0 0 0 115mm},clip, width=0.99\textwidth]{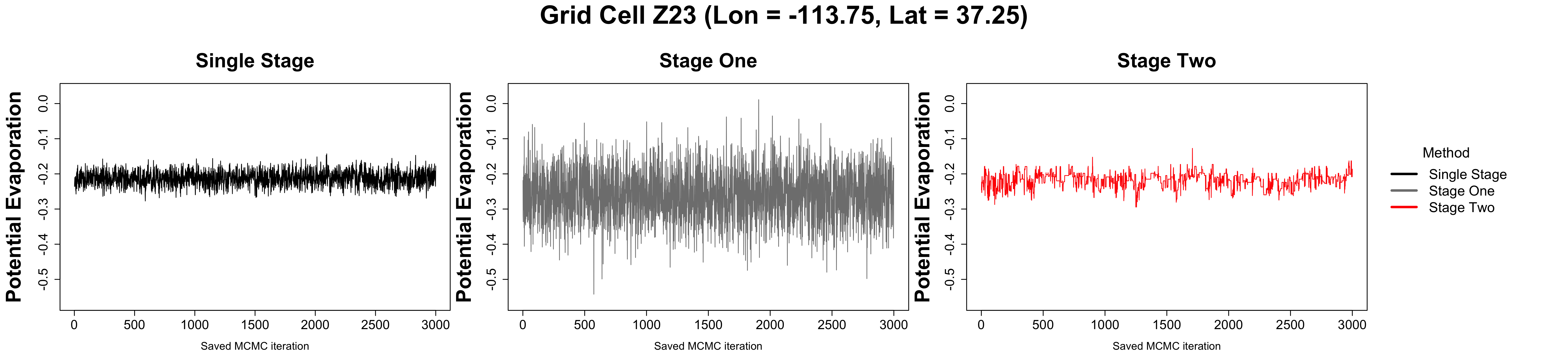}} 
    \par\vspace{1mm}
    {\includegraphics[trim={0 0 0 115mm},clip,width=0.99\textwidth]{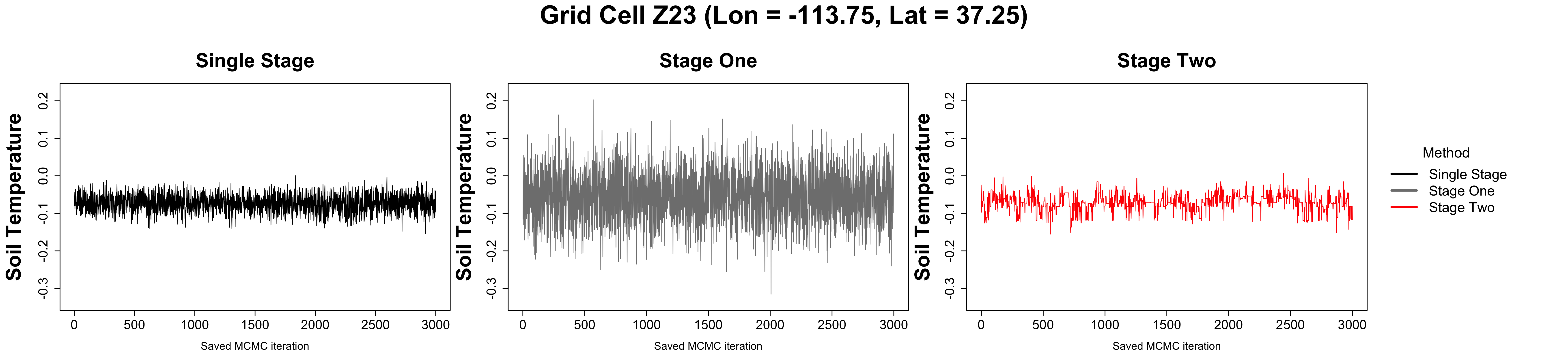}}
    \par\vspace{1mm}
    {\includegraphics[trim={0 0 0 115mm},clip,width=0.99\textwidth]{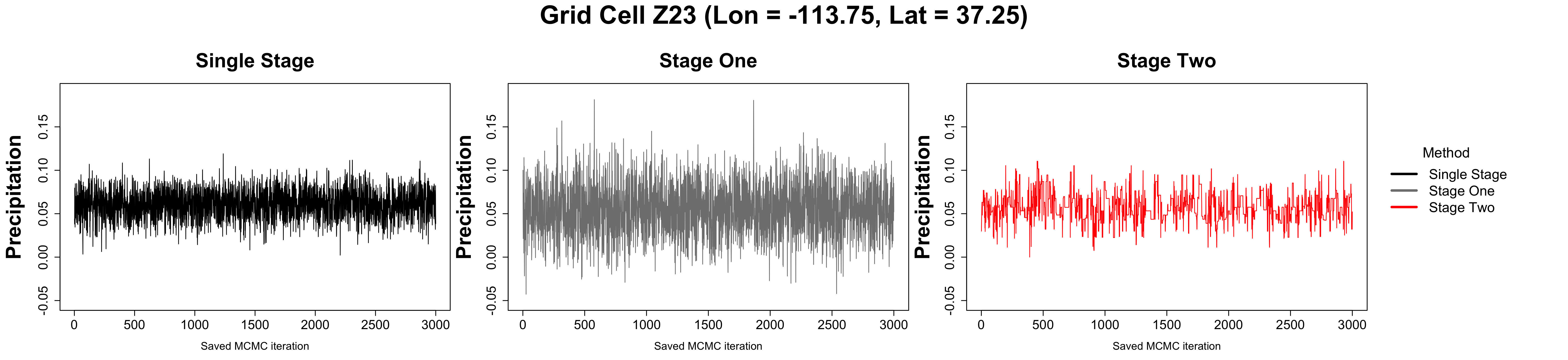}}
    \par\vspace{1mm}
    {\includegraphics[trim={0 0 0 115mm},clip,width=0.99\textwidth]{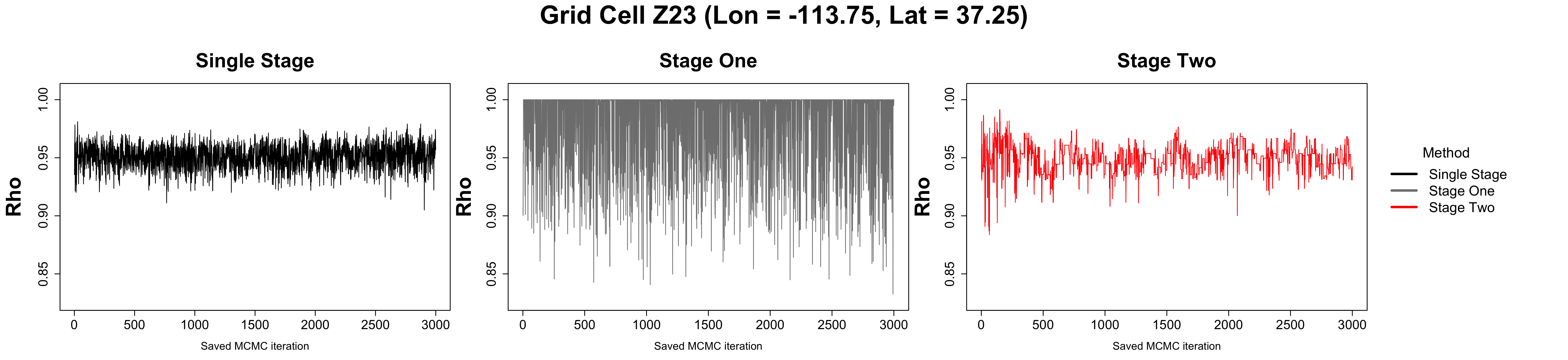}} 
    \par\vspace{1mm}
    {\includegraphics[trim={0 0 0 115mm},clip,width=0.99\textwidth]{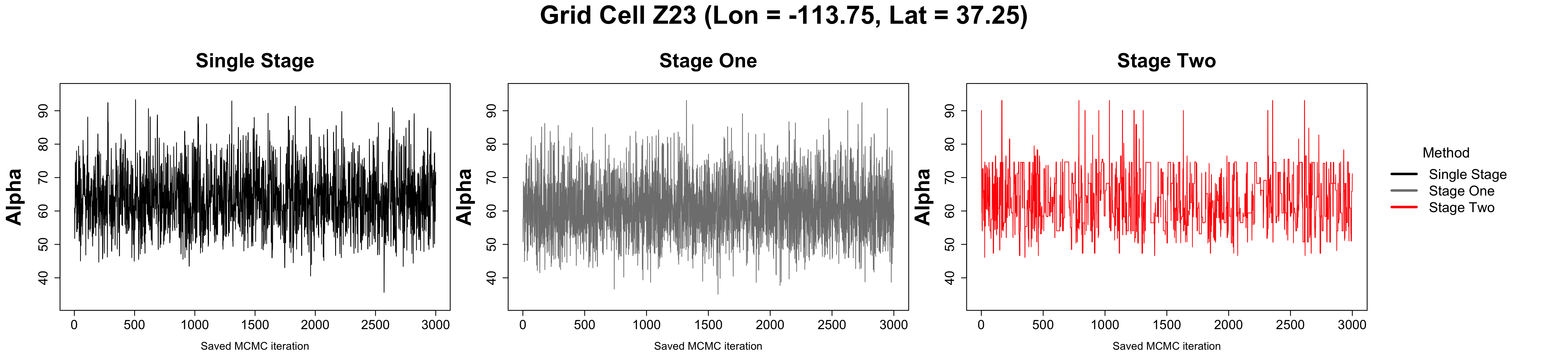}} 

    \caption{Trace plots of the location-specific parameters in the continuous model for one grid cell.}
    \label{fig:continuous-trace}
\end{figure}

\clearpage
\subsection{Trace plots for selected grid cells in the binary model} \label{sec:binary-grid-cells}
\begin{figure}[H]
\begin{center}
\includegraphics[width=4.6in]{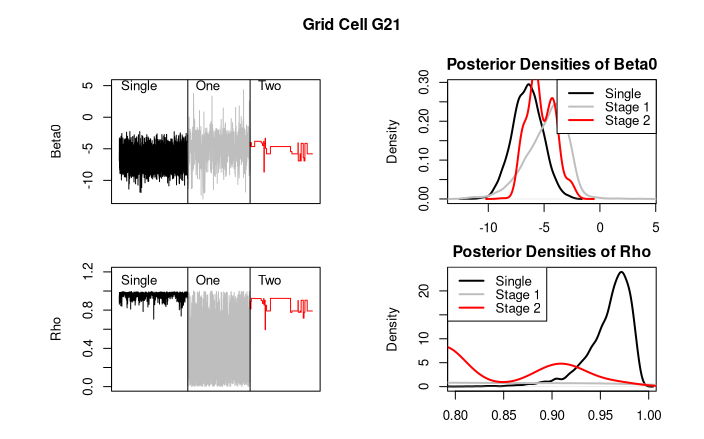}  
    \caption{Top left: trace plots of single-stage (black), stage one (gray), and stage two (red) posterior chains for $\beta_0$.  Top right: kernel density estimates of posterior distributions for the same three models.  Bottom left: trace plots of single-stage (black), stage one (gray), and stage two (red) posterior chains for $\rho$.  Bottom right: kernel density estimates of posterior distributions for the same three models. All results are shown for grid cell G21.}
        \label{fig:G21}
\end{center}

\end{figure}

This subsection presents trace plots and related results for the two representative grid cells which explore the source of mild disagreement between the posteriors from the single-stage and two-stage algorithms in the binary drought case study.  The grid cells shown below are G21 (in the problem area on the border between Idaho and Montana) and GG16 (in Southern California and not in the problem area). Over the training period, location G21 had no droughts, and accordingly the response variable $Y_{i=\mathrm{G21},\ t} = 0$ for all $t=1, \ldots, T$.  It is easy to imagine that the autocorrelation parameter $\rho$ is poorly informed in such a setting.  Figure \ref{fig:G21} expands on this.  Beginning with the lower left panel, we see that the trace plots of the single-stage, stage one, and stage two models deviate sharply from one another.  The stage one posterior is essentially the same as the stage one prior, uniform across (0,1), and so it is a poor approximation of the single-stage posterior.  Stage two mixes poorly, getting stuck as it repeatedly draws proposals from this poor stage one approximation and remains in place during the Metropolis-Hastings step.  This poor performance impacts other parameters, most notably the intercept $\beta_0$, whose stage one posterior approximation was much better but whose stage two posterior mixes poorly.  

\begin{figure}[H]
\begin{center}
\includegraphics[width=4.6in]{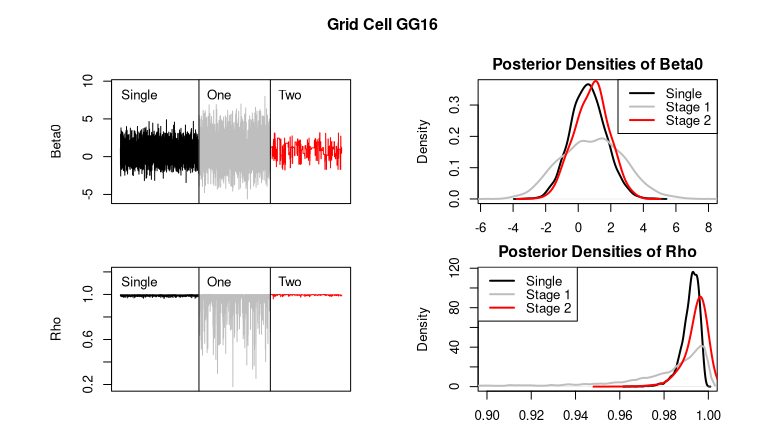}  
    \caption{Top left: trace plots of single-stage (black), stage one (gray), and stage two (red) posterior chains for $\beta_0$.  Top right: kernel density estimates of posterior distributions for the same three models.  Bottom left: trace plots of single-stage (black), stage one (gray), and stage two (red) posterior chains for $\rho$.  Bottom right: kernel density estimates of posterior distributions for the same three models. All results are shown for grid cell GG16.}
        \label{fig:GG16}
\end{center}
\end{figure}

Contrast this with grid cell GG16, which occurs in Southern California and has data with a mix of droughts and non-droughts.  Figure \ref{fig:GG16} shows the same trace plots and kernel density estimates of the posteriors for the intercept $\beta_0$ and autocorrelation parameter $\rho$.  While the stage one approximation for $\rho$ isn't perfect, it is close enough to permit much better mixing in stage two, which is able to re-orient the posterior much closer to the single-stage model output.

\end{document}